\documentclass[preprint,12pt]{elsarticle}

\usepackage{lineno}
\biboptions{sort&compress}
\usepackage[figuresright]{rotating}

\usepackage[USenglish]{babel}        
\usepackage{csquotes}
\usepackage{etoolbox}     

\usepackage{amsmath}        
\usepackage{amssymb}        
\usepackage{amsthm}         
\usepackage{dsfont}         
\usepackage{mathtools}      
\usepackage{bm}             

\usepackage[a4paper, left=32mm, right=26mm, bottom=41mm, top=28mm]{geometry}
\usepackage{lscape}         
\usepackage{microtype}      
\usepackage{fancyhdr}       

\usepackage{graphicx}       
\usepackage{subcaption}     
\usepackage{svg}            
\usepackage{adjustbox}      
\usepackage{float}          
\usepackage{tikz}           
\usetikzlibrary{shapes,arrows,positioning,fit,backgrounds,calc,shadows}

\usepackage{booktabs}       
\usepackage{multirow}       
\usepackage{tabularx}       
\usepackage{array}          

\usepackage[hyphens]{url}   
\usepackage{xurl} 
\usepackage{doi}            
\usepackage[capitalize]{cleveref} 
\usepackage{hyperref}       

\usepackage{sansmathfonts}  

\usepackage{caption}        
\usepackage[section]{placeins}       
\usepackage{lipsum}         
\usepackage{xspace}         
\usepackage{siunitx}        
\usepackage{enumitem}       
\usepackage{appendix}       
\usepackage{glossaries}     
\usepackage{changepage}

\usepackage{todonotes}
\newboolean{SHOWCOMMENTS}
\setboolean{SHOWCOMMENTS}{true}

\ifthenelse{\boolean{SHOWCOMMENTS}}{
\newcommand{\tdMK}[1]
{\todo[color=blue!25,inline]{\footnotesize{\bf Martin:} #1}}
}{
\newcommand{\tdMK}[1]{}
}
\newcommand{\MK}[1]{{\color{black}#1}}

\ifthenelse{\boolean{SHOWCOMMENTS}}{
\newcommand{\tdSK}[1]
{\todo[color=red,inline]{\footnotesize{\bf Sascha:} #1}}
}{
\newcommand{\tdSK}[1]{}
}
\newcommand{\SK}[1]{{\color{black}#1}}

\ifthenelse{\boolean{SHOWCOMMENTS}}{
\newcommand{\tdGK}[1]
{\todo[color=violet!50,inline]{\footnotesize{\bf Gerta:} #1}}
}{
\newcommand{\tdGK}[1]{}
}

\ifthenelse{\boolean{SHOWCOMMENTS}}{
\newcommand{\tdSW}[1]
{\todo[color=ForestGreen,inline]{\footnotesize{\bf Sascha:} #1}}
}{
\newcommand{\tdSW}[1]{}
}

\definecolor{Vadere}{RGB}{70, 130, 180}
\definecolor{memilio}{RGB}{220, 20, 60} 
\definecolor{degraded}{RGB}{255, 140, 0}
\definecolor{comparison}{RGB}{34, 139, 34}
\definecolor{lightgray}{RGB}{245, 245, 245}

\newcommand{\out}[1]{}

\begin{document}

\begin{frontmatter}

    \title{On the Effect of Missing Transmission Chain Information in Agent-Based Models: Outcomes of Superspreading Events and Workplace Transmission}

    \author[a,b]{Sascha Korf\fnref{ca}}
    \ead{sascha.korf@dlr.de}
    \author[c]{Sophia Johanna Wagner}
    \ead{sophia.wagner@hm.edu}
    \author[c]{Gerta Köster}
    \ead{gerta.koester@hm.edu}
    \author[a,d]{Martin J. Kühn}
    \ead{martin.kuehn@dlr.de}

    \address[a]{Institute of Software Technology, Department of High-Performance Computing,
        German Aerospace Center, Cologne, Germany}
    \address[b]{Institute of Bio- and Geosciences, IBG-1: Biotechnology, Forschungszentrum Jülich, Jülich, Germany}
    \address[c]{Department of Computer Science and Mathematics, Munich University of Applied Sciences HM, Munich, Germany}
    \address[d]{Bonn Center for Mathematical Life Sciences and Life and Medical Sciences Institute, University of Bonn, Bonn, Germany}

    \fntext[ca]{Corresponding author}

    \begin{abstract}
        Agent-based models (ABMs) have emerged as distinguished tools for epidemic modeling due to their ability to capture detailed human contact patterns \MK{and can, thus,} support decision-makers in times of outbreaks and epidemics. However, as a result of missing correspondingly resolved data transmission events are often modeled based on simplified assumptions. In this article, we present a framework to assess the impact of these simplifications on epidemic prediction outcomes, considering superspreading and workplace transmission events. We couple the VADERE microsimulation model with the large-scale MEmilio-ABM and compare the outcomes of four  outbreak events after 10 days of simulation in a synthetic city district generated from German census data. In a restaurant superspreading event, where up to four households share tables, we observe 17.2~\% more infections on day 10 after the outbreak. The difference increases to 46.0~\% more infections when using the simplified initialization in a setting where only two households share tables. We observe similar outcomes (41.3~\% vs. 9.3~\% more infections) for two workplace settings with different mixing patterns between teams at work. In addition to the aggregated difference, we show differences in spatial dynamics and transmission trees obtained with complete or reduced outbreak information. We observe differences between simplified and fully detailed initializations that become more pronounced when the subnetworks in the outbreak setting are mixing less. In consequence and aside from classical calibration of models, the significant outcome differences should drive us to develop a more profound understanding of how and where simplified assumptions about transmission events are adequate.
    \end{abstract}

    \begin{keyword}
        Agent-based models  \sep Epidemics \sep Outbreak \sep Superspreading \sep COVID-19 \sep Infectious Diseases \sep MEmilio software framework \sep Vadere software framework

        \MSC[2020] 37M05 \sep 92-04 \sep 92-08 \sep 92-10 \sep 92B05
    \end{keyword}

\end{frontmatter}



\section{Introduction}\label{sec:introduction}

The SARS-CoV-2 pandemic has revealed how infectious diseases can spread rapidly through a highly interconnected world and highlighted the importance of early-stage pandemic modeling in order to understand the disease dynamics and inform public-health decision making.
During emerging outbreaks, various mathematical models can be utilized to analyze disease spread, to forecast epidemic trajectories, to evaluate intervention strategies, and to guide resource allocation decisions.

Mathematical models for infectious disease dynamics span a wide spectrum of complexity and detail.
Models based on ordinary differential equations (ODEs) provide rapid assessment capabilities and have proven valuable for policy guidance during health emergencies; see~\cite{brauer_mathematical_2019}.
Using metapopulation approaches, these models can be leveraged to incorporate spatial and demographic stratification~\cite{balcan_multiscale_2009, liu_modelling_2022, zunker_novel_2024} and the problem of exponential distributions for disease state transition~\cite{wearing_appropriate_2005,donofrio_mixed_2004} can be addressed through linear chain tricks~\cite{Getz2017,ploetzke_lct_2025} or integro-differential~\cite{Wendler2025IDE} approaches.
However, all these models remain fundamentally aggregated and provide limited capacity for representing individual heterogeneity and studying highly resolved transmission properties and dynamics.

Agent-based models (ABMs) are perfectly suited to resolve individual heterogeneity and model transmission on a very detailed scale.
However, ABMs themselves can vary greatly in level of detail and computational complexity~\cite{collier_parallel_2013, willem_optimizing_2015, bershteyn_implementation_2018, ferguson2020report, ozik2021population, macal2018chisim, gaudou2020comokit, CUEVAS2020103827,bicher_evaluation_2021,hinch_openabm-covid19agent-based_2021,kerr_covasim_2021,muller_predicting_2021,NIEDZIELEWSKI2024100801,Ponge23,bicker_hybrid_2024}. \MK{During the COVID-19 pandemic, among others, Covasim~\cite{kerr_covasim_2021}, the population model GEPOC~\cite{bicher_evaluation_2021}, OpenABM~\cite{hinch_openabm-covid19agent-based_2021}, MatSim-EpiSim~\cite{muller_predicting_2021}, MEmilio~\cite{KERKMANN2025110269}, Cuevas' model~\cite{CUEVAS2020103827}, or pDyn~\cite{NIEDZIELEWSKI2024100801} have been used to evaluate epidemic outcomes and intervention strategies with an agent-based level.}
At the most detailed level, locomotion-based ABMs, which \MK{simulate} human walk \MK{behavior}, track individual positions and movements with high precision, enabling sophisticated calculation of transmission probabilities based on exact proximity and environmental factors~\cite{rahn_modelling_2022, atamer-balkan-2024-cdyn}.
While these models produce extensive and detailed simulation data for transmission events, the computational requirements are substantial and often prohibitive for large populations or simulations that span over multiple weeks or months. At a more granular level, ABMs let individuals move between defined locations with simplified contact patterns and population-level transmission parameters inside individual rooms such as schools or workplaces~\cite{ajelli_comparing_2010, KERKMANN2025110269}. 

However, all granular models face a fundamental challenge when being initialized and calibrated using incomplete surveillance data that is prone to underreporting and lacks detailed transmission chain information.
The problem of underreporting has already been studied in several places, in particular, integrating seroprevalence~\cite{gudina_seroepidemiology_2021,contento_integrative_2023} or wastewater data~\cite{rubio-acero_spatially_2021,SCHMID2025100836,mohring_estimating_2024,BTW25}.
Nevertheless, we generally face the issue that transmission chains are often either unknown, partially known, or cannot be incorporated, on time, in detailed models.

During outbreak situations, surveillance systems typically detect aggregate case counts but rarely capture complete transmission networks.
Contact tracing efforts often provide incomplete coverage and may lag significantly behind case detection~\cite{bayly_looking_2024,eva_why_2020,FETZNER_MEASURING_2021}.
Consequently, modelers must make assumptions about how detected infections cluster within households, workplaces, or communities without knowing the underlying transmission chains that generated the observed case distribution.

Studies comparing different modeling approaches typically evaluate final epidemic outcomes such as peak timing, attack rates, and intervention effectiveness rather than isolating the specific contribution of initialization assumptions or missing transmission chain information~\cite{dautel_validation_2023, tabataba_framework_2017, ajelli_comparing_2010, willem_lessons_2017}.

Fitting models to aggregated case counts delivers an essential understanding of ongoing dynamics. However, it is likely that simulated transmission chains deviate from the ground truth. When the disease spreads predominantly within particular subnetworks, it is more likely to encounter people who are already immune or infected than if it spread across the global network. 
This raises the question of the magnitude of the models' prediction error after days, weeks, or even months. Ultimately, the lack of detailed transmission chain information may compromise prediction reliability and, in turn, leave practitioners without evidence-based guidance on the accuracy of predictions over a longer time frame.

\MK{Through graph theory and network-based infectious disease models, we already know that epidemic uptake and development substantially differ for various types of networks; see, e.g.,~\cite{newman_random_2001,newman_spread_2002,keeling_networks_2005,colizza_invasion_2007,pastor-satorras_epidemic_2015} and the references therein. For instance, the authors of~\cite{newman_random_2001} considered random graphs with arbitrary degree distributions to study the size of the giant component and the realized clustering on the graph. This study then also led to consider the spread of epidemics on a network~\cite{newman_spread_2002}. In~\cite{pastor-satorras_epidemic_2015}, numerical simulations for an \textit{SIS} model with scale-free and power-law distributed networks were considered with respect to the network size; in addition, many different network types have been presented and effects of clustering have been discussed. In~\cite{keeling_networks_2005}, typical \textit{SIR} epidemics on random, lattice, small world, spatial and scale-free network types were considered, showing maxima between approximately 10~\% and 70~\% for the proportion of infectious individuals throughout the epidemic spread. In this study, we go beyond these learnings and quantify the variation in the corresponding outcomes for different initial settings in a complex, multilayered simulation with a realistic city district and mixing pattern.}

Microscopic simulation modeling, including realistic human locomotion, enables the generation of detailed transmission data that can serve as reference data for systematic validation studies. Vadere~\cite{kleinmeier_vadere_2019}, originally developed as a pedestrian crowd simulation framework, has been extended to model disease transmission with individual-level precision. 
Detailed transmission events are captured with explicit distances and aerosol output that results in individual agents' exposure to the pathogen.  
Introducing an exposure threshold for infection results in an implicit dose-response model; see~\cite{kleinmeier_vadere_2019,mayr_social_2021,rahn_modelling_2022,rahn_toward_2024}.
This creates an opportunity to directly measure information loss when initializing ABMs from detailed transmission chain or aggregated case data.

In this work, we analyze the differences in the outcomes of the large-scale MEmilio-ABM~\cite{KERKMANN2025110269} when either initialized with aggregate case counts or detailed microsimulation data, which is considered our \MK{reference} data, of particular (superspreading) outbreak events.
With aggregated case data, we use a uniform distribution of infected individuals. 
For the comparison, we initialize the model with exact transmission chain information of the simulated (superspreading) outbreak event.
We hypothesize that detailed transmission chains that include information on clustering patterns—such as colleagues infecting colleagues in workplace scenarios and household members infecting household members in restaurant scenarios—will result in epidemic trajectories that simplified initializations will not mirror.

\section{Methods}\label{sec:methods}
To analyze the effect of different levels of detail in ABM simulations, we focused on two different initializations for an otherwise identically set-up large-scale model. We developed a methodological framework coupling the detailed microsimulation tool Vadere with realistic human walk, distancing, and aerosol transmission with the large-scale MEmilio-ABM. In~\cref{sec:framework}, we describe the novel approach, while~\cref{sec:model-description} provides the descriptions of the already available models. Eventually, we describe our synthetic population in~\cref{sec:city_setup} and the outbreak scenarios in~\cref{sec:outbreak}. 

\subsection{A Framework for Coupling Detailed Microsimulation with Large-scale ABM}\label{sec:framework}
In order to generate reference transmission data, we use Vadere, which was already calibrated to a realistic restaurant superspreading event in~\cite{rahn_modelling_2022}. \MK{We then calibrate the MEmilio-ABM to generate the same number of infections as in the reference setting. Then, we }
use forward simulations of the MEmilio-ABM~\cite{KERKMANN2025110269} with different initializations, either integrating full transmission chain information or only using aggregated case numbers to initialize infected individuals.
We will systematically describe the process as well as the models in the following sections, in particular describing the different restaurant and workplace outbreak scenarios in~\cref{sec:outbreak}.

Through systematic comparison of outbreak scenarios, we quantify initialization-driven prediction error and establish evidence-based guidance for epidemic modeling when surveillance systems provide aggregated outbreak information but not transmission details.

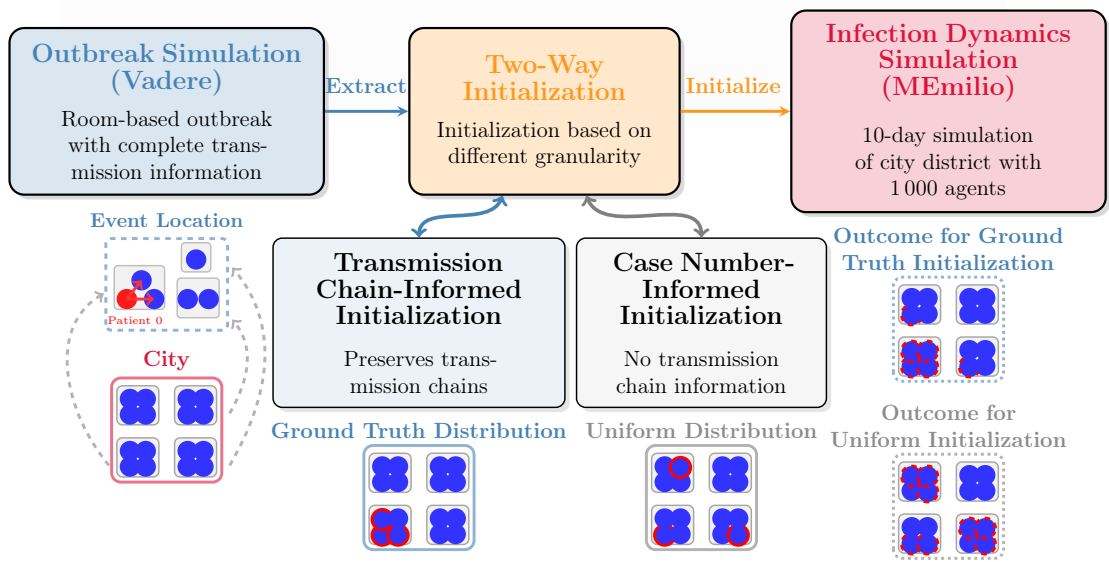
\begin{figure}[!h]
    \begin{adjustwidth}{-1in}{-1in}
        \centering
        \resizebox{\textwidth}{!}{
            
\begin{tikzpicture}[
    node distance=3cm,
    box/.style={rectangle, rounded corners=8pt, minimum width=3.5cm, minimum height=3.2cm, 
                text centered, draw, very thick, inner sep=10pt, drop shadow={opacity=0.3}},
    smallbox/.style={rectangle, rounded corners=5pt, minimum width=3.2cm, minimum height=2cm, 
                     text centered, draw, thick, inner sep=8pt, drop shadow={opacity=0.2}},
    agent/.style={circle, fill, minimum size=6pt},
    infected/.style={agent, fill=red!90},
    susceptible/.style={agent, fill=blue!80},
    arrow/.style={->, very thick, >=stealth},
    bigarrow/.style={->, ultra thick, >=stealth},
    title/.style={font=\Large\bfseries},
    stagenum/.style={font=\LARGE\bfseries},
]

\fill[lightgray, rounded corners=10pt, opacity=0.3] (-2,0) rectangle (16,2.5);

\node[box, fill=Vadere!20, text width=5.5cm, inner sep = 7pt] (stage1) at (0,0) {
    {\textcolor{Vadere}{\textbf{\large{Outbreak Simulation \\(Vadere)}}}}\\[0.2cm]
    {\normalsize Room-based outbreak with complete transmission information}
};

        
        
        

\node[below=0.05cm of stage1] (outbreak) {
    \begin{tikzpicture}[scale=0.8]
        \draw[ultra thick, dashed, Vadere!60, rounded corners=3pt] (-0.7,-1.0) rectangle (2.2,1.2);
        \node[above, font=\small, Vadere] at (0.75,1.3) {\textbf{Event Location}};
        
        \draw[thick, gray!50, fill=gray!10, rounded corners=2pt] (-0.5,-0.5) rectangle (0.7,0.55);
        \draw[thick, gray!50, fill=gray!10, rounded corners=2pt] (1.0,-0.7) rectangle (2.0,0.25);
        \draw[thick, gray!50, fill=gray!10, rounded corners=2pt] (1.1,0.4) rectangle (1.8,1.1);
        
        \node[infected] (agent1) at (-0.2,0) {};      
        \node[susceptible] (agent2) at (0.45,0) {}; 
        \node[susceptible] (agent4) at (0.15,0.45) {}; 
        
        \node[susceptible] (agent3) at (1.25,0) {};
        \node[susceptible] (agent6) at (1.75,0) {};
        
        \node[susceptible] (agent5) at (1.45,0.95) {};
        
        \draw[red!80, ultra thick, ->] (agent1.center) -- (agent2.center);
        \draw[red!80, ultra thick, ->] (agent1.center) -- (agent4.center);
        
        \node[below, font=\tiny, gray] at (0.15,-0.35) {};
        \node[below, font=\tiny, gray] at (1.5,-0.35) {};
        \node[right, font=\tiny, gray] at (1.85,0.85) {};
        
        \node[below, font=\tiny, red] at (0,-0.5) {Patient 0};
    \end{tikzpicture}
};
\node[box, fill=degraded!20, text width=4.8cm, inner sep=5pt] (stage2) at (7.2,0) {
    {\textcolor{degraded!90}{\textbf{\large{\mbox{Two-Way} \mbox{Initialization}}}}}\\[0.2cm]
    {\normalsize Initialization based on different granularity}
};

\node[box, fill=memilio!20, text width=5.5cm, inner sep=7pt] (stage3) at (14.9,0) {
    {\textcolor{memilio}{\textbf{\large \mbox{Infection Dynamics} Simulation\linebreak (MEmilio)}}}\\[0.4cm]
    {\normalsize  10-day simulation of
         city district with \\1\,000 agents
    } 
};

\node[below=2.7cm and -0.5cm of stage1] (population) {
    \begin{tikzpicture}[scale=1.0]
        \draw[ultra thick, memilio!60, rounded corners=5pt] (-0.6,-0.5) rectangle (1.5,1.5);
        \node[above, font=\normalsize, memilio] at (0.45,1.5) {\textbf{City}};
        
        \draw[thick, gray!70, fill=gray!10, rounded corners=3pt] (-0.5,-0.4) rectangle (0.3,0.35);
        \draw[thick, gray!70, fill=gray!10, rounded corners=3pt] (0.6,-0.4) rectangle (1.4,0.35);
        \draw[thick, gray!70, fill=gray!10, rounded corners=3pt] (-0.5,0.6) rectangle (0.3,1.35);
        \draw[thick, gray!70, fill=gray!10, rounded corners=3pt] (0.6,0.6) rectangle (1.4,1.35);
        
        \node[susceptible] at (-0.25,0) {};
        \node[susceptible] at (0.05,0) {};
        \node[susceptible] at (-0.25,0.3) {};
        \node[susceptible] at (0.05,0.3) {};
        
        \node[susceptible] at (0.85,0) {};
        \node[susceptible] at (1.15,0) {};
        \node[susceptible] at (0.85,0.3) {};
        \node[susceptible] at (1.15,0.3) {};
        
        \node[susceptible] at (-0.25,1.0) {};
        \node[susceptible] at (0.05,1.0) {};
        \node[susceptible] at (-0.25,1.3) {};
        \node[susceptible] at (0.05,1.3) {};
        
        \node[susceptible] at (0.85,1.0) {};
        \node[susceptible] at (1.15,1.0) {};
        \node[susceptible] at (0.85,1.3) {};
        \node[susceptible] at (1.15,1.3) {};
    \end{tikzpicture}
};

\node[smallbox, fill=Vadere!10, text width=5cm, inner sep = 8pt, below left = 0.8cm and -3.0cm of stage2] (detailed_branch) {
    \textbf{\large Transmission Chain-Informed Initialization}\\[0.3cm]
    {\normalsize Preserves transmission chains}
};

\node[below =0.0cm and 0.0cm of detailed_branch] (detailed_branch_example) {
    \begin{tikzpicture}[scale=1.0]
        \draw[ultra thick, Vadere!60, rounded corners=5pt] (-0.6,-0.5) rectangle (1.5,1.5);
        \node[above, font=\normalsize, Vadere] at (0.45,1.5) {\textbf{Ground Truth Distribution}};
        
        \draw[thick, gray!70, fill=gray!10, rounded corners=3pt] (-0.5,-0.4) rectangle (0.3,0.35);
        \draw[thick, gray!70, fill=gray!10, rounded corners=3pt] (0.6,-0.4) rectangle (1.4,0.35);
        \draw[thick, gray!70, fill=gray!10, rounded corners=3pt] (-0.5,0.6) rectangle (0.3,1.35);
        \draw[thick, gray!70, fill=gray!10, rounded corners=3pt] (0.6,0.6) rectangle (1.4,1.35);
        
        \node[susceptible, draw=red, ultra thick] at (-0.25,0) {};
        \node[susceptible, draw=red, ultra thick] at (0.05,0) {};
        \node[susceptible, draw=red, ultra thick] at (-0.25,0.3) {};
        \node[susceptible] at (0.05,0.3) {};
        
        \node[susceptible] at (0.85,0) {};
        \node[susceptible] at (1.15,0) {};
        \node[susceptible] at (0.85,0.3) {};
        \node[susceptible] at (1.15,0.3) {};
        
        \node[susceptible] at (-0.25,1.0) {};
        \node[susceptible] at (0.05,1.0) {};
        \node[susceptible] at (-0.25,1.3) {};
        \node[susceptible] at (0.05,1.3) {};
        
        \node[susceptible] at (0.85,1.0) {};
        \node[susceptible] at (1.15,1.0) {};
        \node[susceptible] at (0.85,1.3) {};
        \node[susceptible] at (1.15,1.3) {};
        
    \end{tikzpicture}
};

\node[smallbox, fill=lightgray!90, text width=4.2cm, below right=0.8cm and -2.0cm of stage2] (aggregate_branch) {
    \textbf{\large \mbox{Case Number}-Informed \mbox{Initialization}}\\[0.3cm]
    {\normalsize No transmission chain information}\\
};

\node[below =0.0cm and 0.0cm of aggregate_branch] (aggregate_branch_Example) {
    \begin{tikzpicture}[scale=1.0]
        \draw[ultra thick, gray!60, rounded corners=5pt] (-0.6,-0.5) rectangle (1.5,1.5);
        \node[above, font=\normalsize, gray!90] at (0.45,1.5) {\textbf{Uniform Distribution}};
        
        \draw[thick, gray!70, fill=gray!10, rounded corners=3pt] (-0.5,-0.4) rectangle (0.3,0.35);
        \draw[thick, gray!70, fill=gray!10, rounded corners=3pt] (0.6,-0.4) rectangle (1.4,0.35);
        \draw[thick, gray!70, fill=gray!10, rounded corners=3pt] (-0.5,0.6) rectangle (0.3,1.35);
        \draw[thick, gray!70, fill=gray!10, rounded corners=3pt] (0.6,0.6) rectangle (1.4,1.35);
        
        \node[susceptible, draw=red, ultra thick] at (-0.25,0) {};
        \node[susceptible] at (0.05,0) {};
        \node[susceptible] at (-0.25,0.3) {};
        \node[susceptible] at (0.05,0.3) {};
        
        \node[susceptible] at (0.85,0) {};
        \node[susceptible, draw=red, ultra thick] at (1.15,0) {};
        \node[susceptible] at (0.85,0.3) {};
        \node[susceptible] at (1.15,0.3) {};
        
        \node[susceptible] at (-0.25,1.0) {};
        \node[susceptible] at (0.05,1.0) {};
        \node[susceptible] at (-0.25,1.3) {};
        \node[susceptible, draw=red, ultra thick] at (0.05,1.3) {};
        
        \node[susceptible] at (0.85,1.0) {};
        \node[susceptible] at (1.15,1.0) {};
        \node[susceptible] at (0.85,1.3) {};
        \node[susceptible] at (1.15,1.3) {};
    \end{tikzpicture}
};

\node[below=0.0cm and 0.0cm of stage3] (gt_result) {
    \begin{tikzpicture}[scale=1.0]
        \draw[ultra thick,dotted, Vadere!60, rounded corners=5pt] (-0.6,-0.5) rectangle (1.5,1.5);
        \node[above, font=\normalsize, Vadere, align=center, text width=5cm] at (0.45,1.5) {\textbf{Outcome for Ground Truth Initialization}};
        
        \draw[thick, gray!70, fill=gray!10, rounded corners=3pt] (-0.5,-0.4) rectangle (0.3,0.35);
        \draw[thick, gray!70, fill=gray!10, rounded corners=3pt] (0.6,-0.4) rectangle (1.4,0.35);
        \draw[thick, gray!70, fill=gray!10, rounded corners=3pt] (-0.5,0.6) rectangle (0.3,1.35);
        \draw[thick, gray!70, fill=gray!10, rounded corners=3pt] (0.6,0.6) rectangle (1.4,1.35);
        
        \node[susceptible, draw=red, ultra thick, dashed] at (-0.25,0) {};
        \node[susceptible, draw=red, ultra thick, dashed] at (0.05,0) {};
        \node[susceptible, draw=red, ultra thick, dashed] at (-0.25,0.3) {};
        \node[susceptible, draw=red, ultra thick, dashed] at (0.05,0.3) {};
        
        \node[susceptible, draw=red, ultra thick, dashed] at (0.85,0) {};
        \node[susceptible] at (1.15,0) {};
        \node[susceptible] at (0.85,0.3) {};
        \node[susceptible] at (1.15,0.3) {};
        
        \node[susceptible, draw=red, ultra thick, dashed] at (-0.25,1.0) {};
        \node[susceptible] at (0.05,1.0) {};
        \node[susceptible] at (-0.25,1.3) {};
        \node[susceptible] at (0.05,1.3) {};
        
        \node[susceptible] at (0.85,1.0) {};
        \node[susceptible] at (1.15,1.0) {};
        \node[susceptible] at (0.85,1.3) {};
        \node[susceptible] at (1.15,1.3) {};
        
    \end{tikzpicture}
};

\node[below=0.0cm and 0.0cm of gt_result] (random_result) {
    \begin{tikzpicture}[scale=1.0]
        \draw[ultra thick, dotted, gray!60, rounded corners=5pt] (-0.6,-0.5) rectangle (1.5,1.5);
        \node[above, font=\normalsize, gray!90, align=center, text width=7.1cm] at (0.45,1.5) {\textbf{\mbox{Outcome for} \mbox{Uniform Initialization}}};

        \draw[thick, gray!70, fill=gray!10, rounded corners=3pt] (-0.5,-0.4) rectangle (0.3,0.35);
        \draw[thick, gray!70, fill=gray!10, rounded corners=3pt] (0.6,-0.4) rectangle (1.4,0.35);
        \draw[thick, gray!70, fill=gray!10, rounded corners=3pt] (-0.5,0.6) rectangle (0.3,1.35);
        \draw[thick, gray!70, fill=gray!10, rounded corners=3pt] (0.6,0.6) rectangle (1.4,1.35);
        
        \node[susceptible, draw=red, ultra thick, dashed] at (-0.25,0) {};
        \node[susceptible, draw=red, ultra thick, dashed] at (0.05,0) {};
        \node[susceptible] at (-0.25,0.3) {};
        \node[susceptible] at (0.05,0.3) {};
        
        \node[susceptible] at (0.85,0) {};
        \node[susceptible, draw=red, ultra thick, dashed] at (1.15,0) {};
        \node[susceptible, draw=red, ultra thick, dashed] at (0.85,0.3) {};
        \node[susceptible, draw=red, ultra thick, dashed] at (1.15,0.3) {};
        
        \node[susceptible] at (-0.25,1.0) {};
        \node[susceptible, draw=red, ultra thick, dashed] at (0.05,1.0) {};
        \node[susceptible, draw=red, ultra thick, dashed] at (-0.25,1.3) {};
        \node[susceptible, draw=red, ultra thick, dashed] at (0.05,1.3) {};
        
        \node[susceptible] at (0.85,1.0) {};
        \node[susceptible] at (1.15,1.0) {};
        \node[susceptible] at (0.85,1.3) {};
        \node[susceptible] at (1.15,1.3) {};
        
    \end{tikzpicture}
};

\begin{scope}[shift={(0,-7)}] 
    \draw[ultra thick, dashed, gray!60, ->] (-1.1,0.2) to[out=130, in=210] (-1.2,3.5);
    
     \draw[ultra thick,dashed, gray!60, ->] (1.2,0.2) to[out=60, in=310] (1.25,4.0);
    
     \draw[ultra thick, dashed,gray!60, ->] (1.2,1.2) to[out=60, in=310] (1.25,3.0);
\end{scope}

\draw[bigarrow, Vadere] (stage1.east) -- (stage2.west);
\draw[bigarrow, degraded!90] (stage2.east) -- (stage3.west);

\draw[Vadere, line width=2pt, <->] (stage2.south) +(-0.8,0) to[out=-110,in=90] (detailed_branch.north);
\draw[gray, line width=2pt, <->] (stage2.south) +(0.8,0) to[out=-70,in=90] (aggregate_branch.north);


\node[above, font=\normalsize, Vadere] at (3.8,0.2) {\textbf{Extract}};
\node[above, font=\normalsize, degraded!90] at (10.8,0.2) {\textbf{Initialize}};



\end{tikzpicture}

        }
    \end{adjustwidth}
    \caption{\textbf{Simulating and Comparing the Effect of Detailed Transmission Chain Initialization.} The workflow depicts how we analyze the effect of missing transmission chain information in epidemic predictions through three stages.
        The outbreak simulation (left) provides a realistic data set that serves as \MK{reference outbreak data} with full information on the initial transmission chains. Using either this full set of information or aggregations thereof to case numbers (center), we initialize the city district ABM, which is then run for 10 days (right). The outbreak event location is depicted below the outbreak simulation and shows agents positioned according to proximity, with one initially infectious agent (in red) spreading the disease to others. The smaller boxes represent a symbolic excerpt of the city district with four locations, such as households with four persons each.}
    \label{fig:framework}
\end{figure}

We investigate the initializations' impact through a three-stage process (Figure~\ref{fig:framework}) that quantifies how missing transmission chain information from outbreak events affects epidemic spread predictions.
Our approach compares a transmission chain-, or simply denoted transmission-informed initialization (using detailed Vadere microsimulation data), with case number-informed initialization (using only aggregate case counts and event attendance) for subsequent MEmilio-ABM epidemic simulations. \MK{With the Vadere simulation generating synthetic but realistic outbreak events, the model prediction error of the framework consists of the model error through the assumptions made in the MEmilio model~\cite{KERKMANN2025110269} and the potentially uninformed initialization of primary infections; as analyzed extensively in the results sections hereafter.}

\paragraph{Stage 1: Detailed Outbreak Simulation with Vadere}
Vadere generates detailed outbreak and superspreading scenarios with complete information on transmission events, capturing who-infected-whom relationships together with social clustering patterns.
Agents represent individuals from the synthetic city population (generated in advance and used in its entirety in Stage 3), preserving social structures such as household members sitting together at restaurant tables or colleagues sharing office spaces.

\paragraph{Stage 2: Two-way Initialization}
We initialize the MEmilio-ABM epidemic simulations using two different approaches that isolate the effects of missing transmission chain information: (i) The \textit{transmission-informed initialization} uses complete Vadere transmission data, preserving the exact clustering of infected individuals within households or workplaces; (ii) The \textit{case number-informed initialization} uses only aggregate case counts and uniformly distributes infections among event participants, simulating typical reporting limitations where only the plain number of outbreak cases is available. Thus, in the subsequent chapters, we'll refer to the case number-informed initialization also as \textit{uniform initialization}.

\paragraph{Stage 3: Epidemic Simulation Protocol and Comparison through MEmilio}
In Stage 3, both initialization approaches are used for conducting a 10-day epidemic simulation in a 1000-person synthetic German city using the MEmilio-ABM with consistent parameters and population structure as described in Sections~\ref{sec:memilio-abm} and~\ref{sec:city_setup}. The 10-day observation window captures critical early outbreak dynamics while avoiding epidemic saturation effects that could mask the initialization's event impact. Each scenario is run for 100 independent simulations using the identical city setup but allowing stochastic variation in agents' mobility and transmission events.
Hourly data collection captures epidemic state transitions and spatial infection distribution, enabling detailed comparative analyses of how transmission chain information loss impacts transmission dynamics over the 10-day simulation period.
By comparing epidemic outcomes between transmission-informed and case number-informed  initialization while controlling for all other parameters, we isolate prediction differences attributable solely to missing transmission chain information from the outbreak event.

We evaluate this framework across four outbreak and superspreading scenarios: two restaurant settings (with varying household clustering at dining tables) and two workplace environments (with different colleague meeting patterns), enabling a systematic assessment of how missing or disregarding detailed transmission data affects epidemic transmission.

\subsection{Model Descriptions}\label{sec:model-description}
To make the paper self-contained, we give a short overview of the two models used in this work. More details on the two models can be found in the original works~\cite{rahn_modelling_2022} and~\cite{KERKMANN2025110269,Bicker_MEmilio_v2_0_0} for the Vadere epidemic model and the MEmilio-ABM, respectively.

\subsubsection{Vadere epidemic model}\label{sec:Vadere}

The Vadere epidemic model, introduced by~\cite{mayr_social_2021,rahn_modelling_2022, rahn_toward_2024}, extends the microscopic crowd simulation software Vadere~\cite{kleinmeier_vadere_2019} with an exposure model and social distancing as a preventive measure.
In Vadere's microscopic simulation, small-scale scenarios such as restaurants or offices are modeled.
The model computes the exact movement of each agent, from their individual starting point to their target location. 
Each agent moves according to its individual speed and reduces the distance to its target while keeping a distance from other agents and obstacles.
In the exposure model, each agent has a health status and a breathing cycle. Agents are either infectious or susceptible.
Infectious agents release aerosol clouds containing pathogens that are released at the point of exhalation and remain suspended in that location.
These clouds can diffuse over time and their pathogen load decreases exponentially. When clouds overlap, their pathogen load adds up.
Susceptible agents inhale pathogens whenever they are positioned inside an aerosol cloud, and their individual exposure is quantified as the total number of pathogens they inhale over time.
In this work, we adopt the default parameter values of the exposure model as reported by Rahn et al.~\cite{rahn_modelling_2022}, see A.1. In particular, the exposure model is updated every 0.4 seconds.

We introduce a dose-response model based on the reference scenario from~\cite{rahn_modelling_2022}. We argue that agents who
are positioned less than 1.5 meters apart from an infected person for over a period of 10 minutes face a high infection risk; see also~\cite{rki_kontaktpersonen_2022}. We use simulations to compute the number of pathogens an agent in such a situation inhales. 
%
%
Following this definition, the exposure model is calibrated to an inhaled dose of approximately $3.2\times10^3$ pathogen particles as the critical transmission threshold for infection. That is, we assume that all agents with an exposure exceeding this threshold become infected.

\subsubsection{MEmilio Agent-Based Model}\label{sec:memilio-abm}
The MEmilio-ABM employs a modular location-based approach where agents move between defined locations with transmission occurring between agents who share the same location.
Locations represent different facility types (households, workplaces, schools, social venues) with location-specific contact patterns between age groups.
The model advances by discrete time steps, which are by default set to one hour.
Agents maintain state information, including demographic attributes such as age group, behavioral parameters such as compliance with nonpharmaceutical interventions (NPIs), and infection history with past infections and vaccinations.
The design and implementation enable efficient simulation of large-scale populations with individual-level detail for disease spread analysis~\cite{KERKMANN2025110269,Bicker_MEmilio_v2_0_0}.
Agents' health states follow discrete transitions between epidemiological compartments \MK{as they are also formulated for ODE-based SEIR-type models; cf.~\cite{brauer_mathematical_2019}}.
Susceptible agents can become infected from infectious agents at the same location.
They either go through a nonsymptomatic infection course and recover or have a symptomatic infection course, where they can either recover or die after the sequence of aggravating states \textit{Severe} (requiring hospitalization) and \textit{Critical} (requiring ICU).
By default, state time durations for SARS-CoV-2 follow lognormal distributions and transition probabilities depend on age and susceptibility.
We use parameters specific to the Alpha variant of SARS-CoV-2 throughout the simulations.
For this study, we implement simplistic behavior and do not use further NPIs.
That means symptomatic agents self-quarantine at home after two days of symptoms, reflecting cautious and early pandemic behavior before widespread policy implementation.
Exact parameters can be found in Table A.2.

\SK{Individual infectiousness is determined by the viral shed which depends on the agent's viral load through a nonlinear relationship~\cite{KERKMANN2025110269,jones_estimating_2021}. Viral load dynamics follow a two-phase trajectory: exponential increase to peak viral load followed by exponential decline to clearance. Individual-level shedding intensity is modeled stochastically to reflect the substantial variability observed in empirical viral load data~\cite{ke_daily_2022}. Viral shedding of an agent happens throughout its presence at a location and can then reach other susceptible agents at the same location. Transmissions happens based on contacts between susceptible and infected agents.} As we don't utilize mask wearing, vaccinations, or assumed prior infections for our simulation\MK{s in this paper}, transmission probability is calculated mainly from three components:

\begin{enumerate}
    \item \textbf{Age-stratified contact patterns} per location type, following empirical data from~\cite{mossong_social_2008,fumanelli_inferring_2012,prem_projecting_2017}, combined in~\cite{kuhn_assessment_2021}.
    \item \textbf{Viral shed exposure rate} $e_p$ derived from infectious agents at the same location and their individual viral \SK{shed as well as the contact rate between corresponding age groups of the agents}, see~\cite{KERKMANN2025110269}.
    \item \textbf{Transmission parameter} $\lambda$ establishing the relationship between $e_p$  and an agents' probability of infection. We model this relationship as linear, such that the infection rate $\tau_p = \lambda e_p$. The resulting infection rate $\tau_p$ then serves as the rate parameter for an exponential distribution used to calculate the probability of infection occurrence; compare~\cite{KERKMANN2025110269}.
\end{enumerate}
\MK{We provide visualization on the above mentioned parameters in the appendix in A.1.} In this work, we calibrate $\lambda$ using workplace Scenario~W1 (see~\cref{sec:outbreak_work}) as our reference standard.
We replicate this scenario in the MEmilio-ABM with 26 agents (age group 35-59) in a 7-hour workplace setting, testing $\lambda$ values from 10 to 50 with 100 replications per value and a step size of 0.1.
We select $\lambda = 22.6$, which produces an average of two infections, matching the Vadere reference for Scenario~W1. This calibration ensures comparable baseline transmission rates between models while maintaining realistic epidemic dynamics for early pandemic conditions without interventions.

\MK{Generally, the value $\lambda = 22.6$ is chosen in line with previous calibrations of Vadere~\cite{rahn_modelling_2022} and such that it produces relevant (super)spreading scenarios over all settings. This means that we intend to generate epidemic uptakes which are neither too fast nor too slow within a 10-day time frame. The 10-day time frame is chosen such that an essential uptake can, first, be observed but which, second, is not driven by infections of the second or third generation after patient zero. After calibration of this parameter for Scenario~W1, we furthermore validated with Scenario~W2,~R1, and~R2 that comparable general conditions hold.}

\subsection{Simulation City Setup}\label{sec:city_setup}
\begin{figure}[!h]
    \centering
    \includegraphics[width=\textwidth]{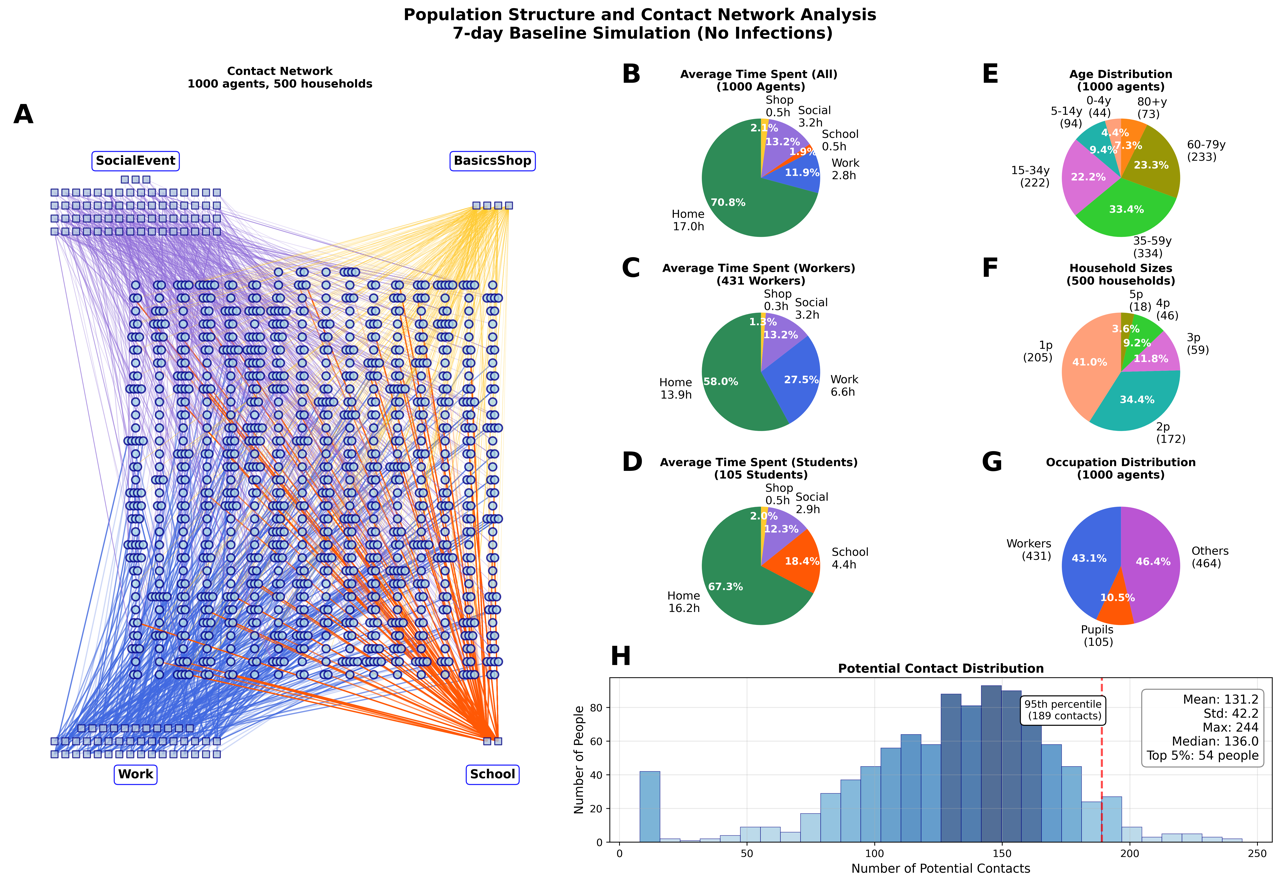}
    \caption{\textbf{Contact Network Analysis for Baseline MEmilio-ABM and Population Structure.}
        The figure presents an overview of a synthetic population comprising $N = 1000$ agents distributed across $H = 500$ households. Data is gathered from a 7-day MEmilio simulation with no infectious agents.
        \textit{Panel~A} displays the agent-based contact network where individual agents appear as light blue circles with dark blue outlines, organized within household groupings.
        Contact relationships (amount of hours spent at the same location) are visualized as colored edges connecting agents to location nodes (light blue squares) and back to the contacted agent, with colors corresponding to location types: blue for Work, orange for School, purple for SocialEvent, and yellow for BasicsShop.
        Edge thickness reflects contact intensity (more hours spent together).
        Location areas exclude Hospital/ICU types (as no infections occur and severe status is not reached) and household contacts to ensure visual clarity, as contact intensity is always high between household members.
        \textit{Panel~B} shows population-wide time allocation across all $N = 1000$ individuals.
        \textit{Panel~C} presents worker-specific patterns ($n_w = 431$ individuals).
        \textit{Panel~D} displays pupil allocation patterns ($n_s = 105$ individuals).
        \textit{Panel~E} reveals the demographic age structure, reflecting the German profile.
        \textit{Panel~F} shows household size distribution across $H = 500$ households: single-person, two-person, three-person, four-person, and five-person households, also reflecting the German profile.
        \textit{Panel~G} presents occupation distribution: Workers, Pupils, and Others; representing infants, retirees and non-employed agents.
        \textit{Panel~H} presents the potential contact distribution, which quantifies the number of distinct individuals with whom each individual spends at least one hour.  
        All temporal data represent average daily patterns, with percentages normalized relative to 24-hour totals.}
    \label{fig:population_structure}
\end{figure}

To study the aforementioned effects, the MEmilio-ABM operates on a synthetic population generated using a city builder module that creates realistic German demographic structures based on official statistical data~\cite{destatis_population_age}.
A general overview is given in~\cref{fig:population_structure}.
We create a city district with a population of 1000 agents following Germany's age group proportions: 4.4~\% aged 0–4 years, 9.4~\% aged 5–14 years, 22.2~\% aged 15–34 years, 33.4~\% aged 35–59 years, 23.3~\% aged 60–79 years, and 7.3~\% aged 80+ years.
Mobility of agents between locations follows priority-ordered rules where medical transitions (hospitalization, quarantine) take precedence over routine activities.
Normal mobility rules include prioritized age-dependent school attendance (ages 5-34) and work attendance (ages 15-59), along with stochastic shopping trips and social event participation.
In accordance with~\cite{noauthor_themenseite_nodate} we create 3.7 basic shops per 1000 agents.
Furthermore, we assume moderate gathering sizes with an average of 10 agents per workplace and 15 agents per social event.

\paragraph{Household Structure}
Household sizes reflect national distributions (41.5~\% single-person, 34.2~\% two-person households, 11.8~\% three-person households, 9.1~\% four-person households and 3.4~\% five-person households), with demographic constraints ensuring households containing children (ages 0-14) include at least one adult (age 15+ years)~\cite{noauthor_projected_nodate}.
Notably, maximum household size is limited to 5 persons.

\paragraph{Location assignment}
Agents receive assignments to multiple location types, creating the contact network topology.
The school location type will be divided into elementary and secondary schools with a 1:2 ratio and with an overall school rate of 10.5~\%~\cite{noauthor_pupils_nodate}.
Assignments prioritize elementary schools for individuals aged 5–14 years, assigning the remaining individuals of age 5–14 years to secondary schools and then filling the open slots of the secondary school with individuals of age group 15–34.
Workplace assignments target ages 15–59 years with a 77.5~\% employment rate for 15 to 65 years old~\cite{noauthor_erwerbstatigenquoten_nodate}.
All agents receive assignments to essential services (healthcare, shopping, social events), where 0-4 year olds just receive assignments to healthcare.

\paragraph{Mobility patterns}
Agents follow age-dependent mobility schedules, including school attendance (weekdays 8:00-15:00), workplace visits (weekdays with return around 17:00 and departure from 6:00 to 8:00), stochastic shopping trips (weekdays 7:00-22:00), and social events (weekends and evenings).
Here, agents only head shopping or to social events when they are at home, i.e., not directly from workplaces or after attending school.
Medical transitions or instructions override routine activities, i.e., when agents require hospitalization or ICU or when quarantine is deemed required.
The resulting time allocation patterns match empirical German time-use distributions~\cite{noauthor_zeitverwendungserhebung_nodate}.
The mobility structure in combination with the above population and infrastructure setup creates the resulting contact network through which epidemic spread occurs.

\paragraph{Seeds and Randomization}
Each simulation uses an advanced random number generator seed to ensure consistent and reconstructible runs (cf.~\cite{KERKMANN2025110269}).
These seeds are used to control stochastic elements during city initialization and maintain deterministic infrastructure scaling.
For the city construction and assignment of agents to locations, seeds influence the random assignment of individuals to specific households, workplaces, and schools, which creates different social mixing patterns between simulation runs.
To avoid mixing the studied effect with different city district setups, the overall district structure remains constant across different seeds, with infrastructure elements (number and size of households, workplaces, schools and stores) calculated deterministically using fixed population ratios, ensuring reproducible baseline conditions for comparative analysis.

\subsection{Outbreak Scenarios}\label{sec:outbreak}
We evaluate initialization effects across four outbreak scenarios, designed to represent real-world transmission or superspreading events, where detailed transmission chain information typically remains unavailable to surveillance systems.
Each scenario varies social grouping or contact patterns through controlled seating arrangements (restaurant) and meeting schedules (workplace), enabling systematic testing under different settings and conditions of how information loss and model error impact model outcomes.

In order to simulate the outbreak scenarios, a corresponding number of agents (with properties such as workplace or household assignment as defined in the scenario) are selected from the city district and handed over to the Vadere outbreak simulation.
In Vadere, persons get infected solely if they inhale a sufficient number of pathogens that an infectious agent left behind. Usually, that means that they step close to the infectious person.
Movements within the scenarios reflect realistic real-world scenarios.

Figure~\ref{fig:scenario_overview_static} provides an overview of all scenarios.
Furthermore, in~\cite{korf_2025_17048423} videos are available where the entire simulation for each scenario is shown.
A detailed description of every agent's movement is also given in Appendix B.

The scenarios represent typical social contact patterns that lead to a particular outbreak.
In the \textbf{Restaurant Scenarios}, multiple individuals (from different households) attend a restaurant.
The outbreak is driven by the social gathering with selected seating arrangements.
In the \textbf{Workplace Scenarios}, employees from multiple workplaces meet at a workplace setting in varying numbers of meetings. 

Vadere simulates each scenario with one initially infected agent, tracking exact transmission events, timing, and spatial locations throughout the event duration.
This produces detailed transmission chains identifying precisely which agents become infected, together with their social relationships (household and workplace affiliations), providing both granular transmission data and aggregate infection counts for subsequent MEmilio initialization.

For the two restaurant scenarios, we do not vary the total number of contacts and only seat individuals from different households differently. 
Note that this means that we end up with the same number of infected individuals from the outbreaks, as actual household assignment has no influence on the simulation. 
While in the two workplace scenarios, we simulate the same individuals and workplace assignments for the office rooms but change contact numbers and durations through different conference room meeting settings, resulting in different outcomes for the outbreak settings.

\subsubsection{Restaurant Scenarios}
Based on the documented superspreading event in Guangzhou, China~\cite{LI2021107788}, we simulate a 2-hour restaurant event with 89 agents, including one initially infectious individual, as shown in \cref{fig:scenario_overview_static}, Panel A.
Here, agents join and leave the restaurant room and go to their respective seats over the course of the two-hour simulation.
Around 70~\% of the agents are already in the restaurant at the start of the simulation, with the infectious agent coming in after one minute.
The agents remain in the restaurant for periods ranging between 60 and 90 minutes.
We assume two scenarios with different household distribution patterns.

\begin{itemize}
    \item \textbf{Scenario~R1 (Limited inter-household mixing):} In R1, each table with {five} or fewer seats exclusively seats agents from the same household.
          When table size exceeds maximum household capacity (more than five seats), an additional household is added to fill the remaining seats.
          This reflects typical family dining patterns where relatives sit together; compare the red outline in the \cref{fig:scenario_overview_static}.
          As this results in a maximum of two households sitting together at a table, it also reflects NPIs that were in place in, e.g., Germany at particular time points of the pandemic and where only individuals of two households were allowed to meet; see~\cite{seidel_treffen_2021}.
          The total number of infected agents from the Scenario~R1 outbreak is nine.
    \item \textbf{Scenario~R2 (More inter-household mixing):} In R2, similarly to R1, each table with {three} or fewer seats exclusively seats agents from the same household.
          When table size exceeds three seats, an additional household is added to fill the next three seats until the table size is reached.
          This represents mixed social dining with more inter-household contacts.
          The maximum number of households at a table is four; compare the cyan outline in \cref{fig:scenario_overview_static}.
          The total number of infected agents from the Scenario~R2 outbreak is also nine, only distributing the infected individuals differently across the households.
\end{itemize}

\subsubsection{Workplace Scenarios}\label{sec:outbreak_work}
\begin{figure}[!h]
    \centering
    \includegraphics[width=\textwidth]{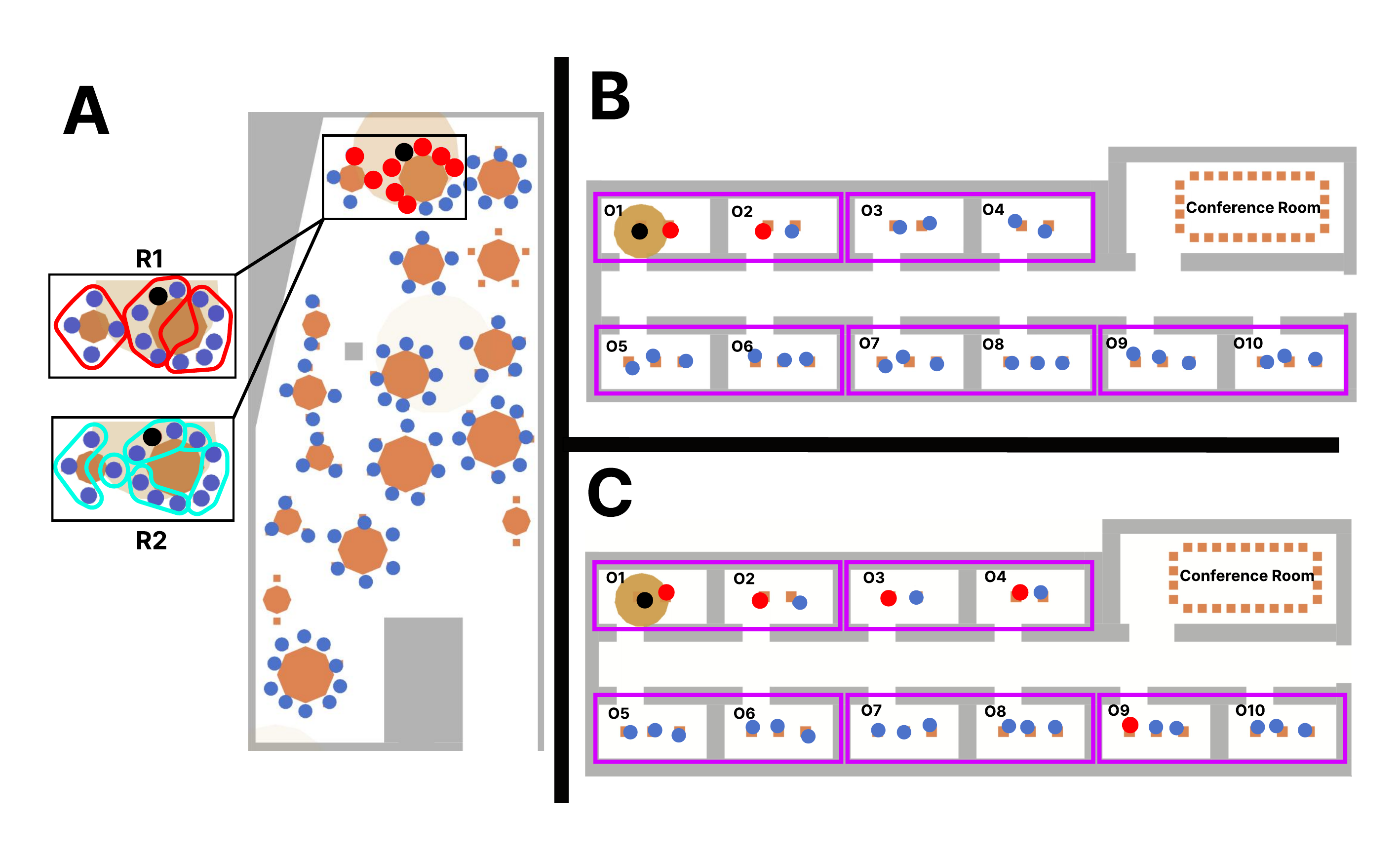}
    \caption{\textbf{Scenario Overview Showing Final Infection Spread Across Different Scenarios and Initial Conditions}. Panels show outbreak-related infected agents (red circles for newly infected and black circles for the initially infectious agent). Colored outlines reflect households (red and cyan outlined groups) for the restaurant setting and workplaces (violet outlined office pairs) for the work setting. Blue circles indicate susceptible individuals after the simulation. Initially, infectious agents emit virus particles, visualized as brown circles surrounding them. \textit{Panel A}: Restaurant outbreak scenarios showing spatial distribution of $n=89$ agents across dining tables with one initially infectious agent. The R1 excerpt shows household grouping at the two selected tables (top inset, red outlined groups) at the start of the simulation, while the R2 excerpt (bottom inset, cyan outlined groups) shows inter-household mixing at the start of the simulation. Households are shown only for the two tables where infections happen. \textit{Panel B:} Workplace Scenario~W1 featuring $n=26$ agents with few meetings and limited mixing.
        \textit{Panel C:} Workplace Scenario~W2 with identical office structure but more meetings and more mixing.}
    \label{fig:scenario_overview_static}
\end{figure}

For the two workplace scenarios, we simulate work settings with 26 agents. We designed a floor layout consisting of 10 offices in total that form five (small) workplaces. We group 4 to 6 (2-3 per office) individuals into different workplaces made up of two offices each; see \cref{fig:scenario_overview_static} Panels B and C. Agents from these workplaces meet each other during the outbreak settings as described below. Note that in the subsequent simulations, all of these (small) workplaces will be filled independently with additional agents from the city district to create workplaces of size 10 or 11 and then do not mix further with the workplaces created from the other office pairs. 

For both scenarios, we implement varying contact patterns through different meeting frequencies and configurations.
While agents maintain a certain distance in meetings when possible, aerosol transmission and remaining close(r) contacts create distinct transmission dynamics between scenarios.

In both scenarios, Person 1 from Office 1 is defined as the single initially infected patient zero, allowing for direct comparison of transmission patterns under different meeting structures.
For simplicity, agents only move between their assigned offices and the meeting room throughout the 8-hour simulation period.

\textbf{Scenario~W1 (Few meetings, limited mixing):} In Scenario~W1, we implement three small sequential one-hour meetings with limited mixing between workplaces, yielding reduced transmission potential.
\begin{itemize}
    \item \textbf{Meeting 1:} 120--180 minutes | Offices 1, 5 (Persons 1-2 and 9-11)
    \item \textbf{Meeting 2:} 180--240 minutes | Offices 2, 6 (Persons 3-4 and 12-14)
    \item \textbf{Meeting 3:} 240--300 minutes | Offices 9, 10 (Persons 21-26)
\end{itemize}
Notably, the sequential timing of Meeting 1 and 2 produces aerosol transmission events, where one agent from Office 2 becomes infected through lingering aerosols from the preceding meeting.
The final amount of infected agents is three.

\textbf{Scenario~W2 (More meetings, more mixing):} In Scenario~W2, we implement four meetings with more mixing between workplaces, increasing transmission potential.
\begin{itemize}
    \item \textbf{Meeting 1:} 0--60 minutes | All offices (All persons)
    \item \textbf{Meeting 2:} 60--120 minutes | Offices 3,4,7,8 (Persons 5-8 and 15-20)
    \item \textbf{Meeting 3:} 240--300 minutes | Offices 1,2,5,6 (Persons 1-4 and 9-14)
    \item \textbf{Meeting 4:} 300--360 minutes | Offices 9,10 (Persons 21-26)
\end{itemize}
The opening meeting creates a superspreading event where all 26 agents are present, resulting in four infections with transmissions happening across office boundaries.
The subsequent larger meetings maintain elevated mixing compared to W1, with an additional aerosol transmission occurring between the third and fourth meeting.
The final amount of infected agents is six, effectively more than doubling the number of new infections compared to W1.

\section{Results}\label{sec:results}
The results section is divided into the presentation of the different initializations for the four scenarios and three complementary analyses on the whole 10-day epidemic simulation. 
In these, first, we quantify aggregate epidemic dynamics through cumulative infection trajectories over the 10-day simulation period. Second, we examine spatial transmission patterns to understand how initialization affects disease spread across households and workplaces within the city district. 
Third, we analyze detailed transmission trees  to characterize the structural differences in epidemic networks that drive the observed macroscopic patterns. In order for the results to be most representative, we pre-ran the simulations after initialization and selected one city setup that returned average relative differences in final infections for all four scenarios.

\subsection{Initializations}
In this section, we briefly provide details on the different initializations to show properties of the outbreak at the beginning of the MEmilio-ABM simulation. 
Especially, we show information on the individuals and units (households or workplaces) that are infected. 
For uniform initializations, we distribute the infectious agents across different tables for the restaurant scenario and, for the workplace scenario, across different workplaces, resulting in minimal infection clustering compared to the transmission-informed initialization.
For the Scenarios R1 and R2, uniform initialization makes each infected agent belong to a different household. 
For Scenario~W1, uniform initialization makes all infected agents belong to different workplaces, whereas in Scenario~W2, only two infected agents belong to the same workplace. \cref{tab:initialization-details} provides the detailed initializations for one exemplary seed and summarizes the clustering and mixing characteristics of the different initializations.
Note that for infected agents of the outbreak scenario, we infect agents of the subsequent 10-day simulation with the MEmilio-ABM as follows:
Agents from the same household in the outbreak scenario are also from the same household in the MEmilio-ABM simulation (where for each of those households in the MEmilio-ABM simulation, up to five agents are assigned in total).
Agents from the same outbreak workplace are assigned to the same workplace in the MEmilio-ABM simulation (where for each workplace 10 to 11 agents are assigned in total).

\begin{table}[h]
    \centering
    \caption{\textbf{Detailed Initialization Specifications by Scenario for one Exemplary Seed.}}
    \begin{adjustwidth}{-1in}{-1in}
        \resizebox{1.3\textwidth}{!}{%
            \begin{tabular}{lcclc}
                \toprule
                \textbf{Scenario}   & \textbf{Total}     & \textbf{Initialization} & \textbf{Global IDs}                                                      & \textbf{Units Affected by at Least} \\
                                    & \textbf{Infected}  &                         & \textbf{(Agent ID, Household (R1, R2) or Workplace (W1,W2) ID)}                       & \textbf{one Infectious Agent}       \\
                \midrule
                \multirow{2}{*}{R1} & \multirow{2}{*}{9} & Transmission-Informed   & (3,5), (5,5), (45,28), (46,28), (47,28), (48,28), (49,28), (53,29), (54,29)           & 3 households                        \\
                                    &                    & Uniform                 & (3,5), (559,291), (8,6), (523,270),  (602,312), (273,142), (25,17), (45,28), (661,347) & 9 households                        \\
                \midrule
                \multirow{2}{*}{R2} & \multirow{2}{*}{9} & Transmission-Informed   & (7,6), (9,7), (3,5), (4,5), (5,5), (22,17), (23,17), (32,21), (0,3)                       & 6 households                        \\
                                    &                    & Uniform                 & (7,6), (124,65), (14,10), (100,52), (80,42), (51,29), (27,18), (3,5), (93,49)          & 9 households                        \\
                \midrule
                \multirow{2}{*}{W1} & \multirow{2}{*}{3} & Transmission-Informed   & (12,503), (96,503), (180,503)                                                            & 1 workplace                        \\
                                    &                    & Uniform                 & (12,503), (262,505), (189,507)                                                         & 3 workplaces                        \\
                \midrule
                \multirow{2}{*}{W2} & \multirow{2}{*}{6} & Transmission-Informed   & (12,503), (96,503), (180,503), (18,504), (184,504), (30,507)                            & 3 workplaces                        \\
                                    &                    & Uniform                 & (12,503), (18,504), (24,505), (397,505), (263,506), (104,507)                           & 5 workplaces                        \\
                \bottomrule
            \end{tabular}%
        }
    \end{adjustwidth}
    
    \label{tab:initialization-details}
\end{table}

The scenario initializations demonstrate differently strong infection clustering patterns, with infections concentrated closest among household members in Scenario~R1 and among the same workplace in Scenario~W1. 

\subsection{Epidemic Curve Analysis}

Epidemic trajectories are visualized using cumulative infection counts over the 10-day simulation period, with confidence intervals calculated from 100 independent simulation runs per scenario. 
The plots display median trajectories (solid lines) with 50~\% and 90~\% confidence intervals through the p5, p25, p75, and p95 percentiles (dark and light shaded regions) for both transmission-informed (red) and case number-informed (blue), in the following simply denoted uniform, initialization approaches. 
Vertical dashed lines indicate when each approach reaches 100 cumulative infections, enabling direct comparison of epidemic growth timing. 
Day-10 epidemic size differences are annotated with absolute case counts and percentage increases, providing quantitative measures of initialization effects across scenarios.

\begin{figure}[!h]
    \includegraphics[width=\textwidth]{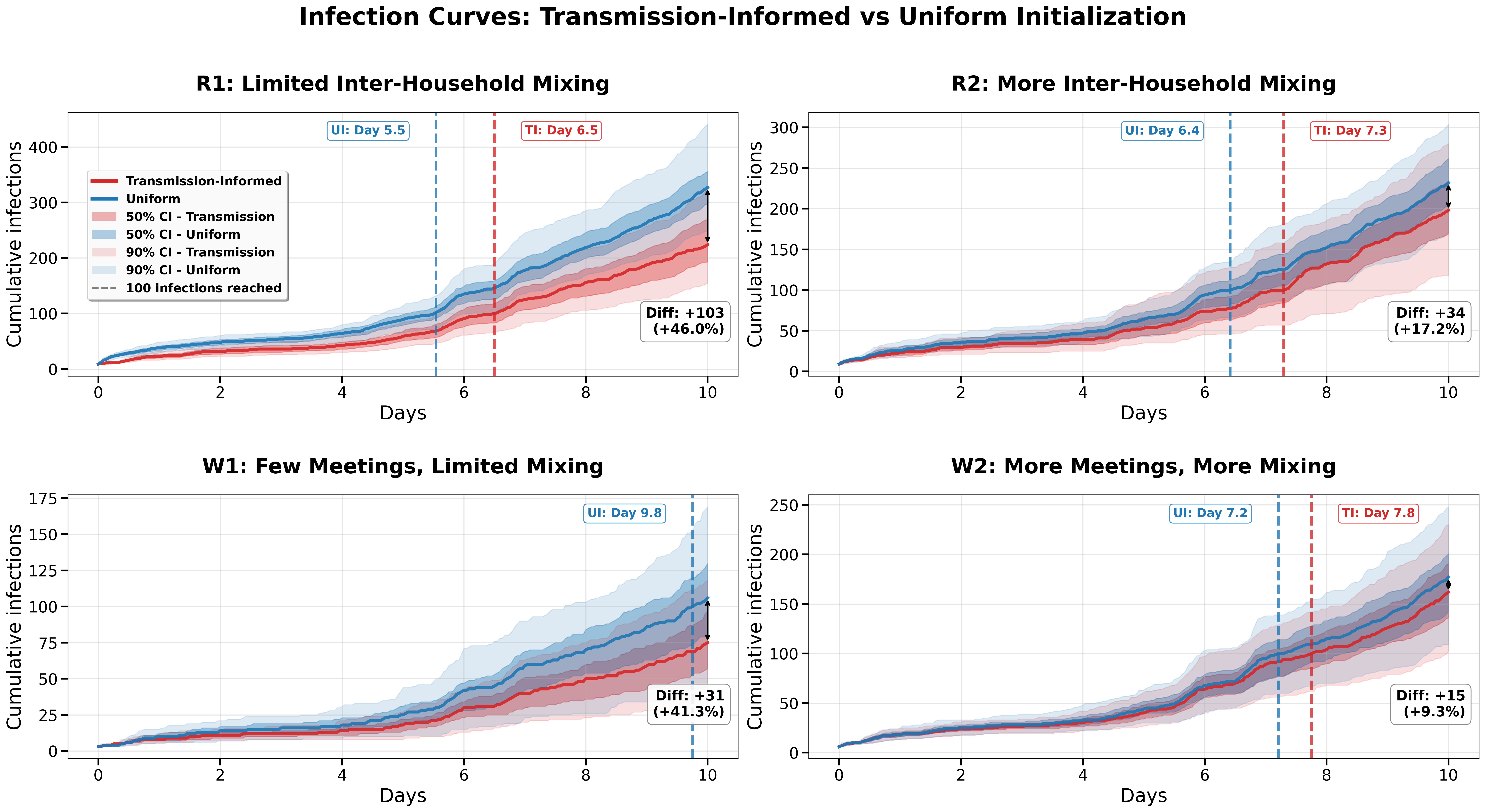}
    \caption{\textbf{Epidemic Trajectories comparing Transmission-Informed versus Uniform Initialization across four Outbreak Scenarios.} Each panel shows cumulative infections over 10 days with 50~\% (dark shaded) and 90~\% (light shaded) confidence intervals from 100 simulations. Red lines represent transmission-informed initialization preserving \MK{reference}  infection data from Vadere's microsimulation, while blue lines show uniform initialization distributing infections randomly across outbreak participants. Vertical dashed lines indicate when each approach reaches 100 infections (UI: uniform, TI: transmission-informed). Final day differences are shown in boxes with absolute case counts and percentage increases.}
    \label{fig:epidemic_curves}
\end{figure}

The epidemic trajectories reveal substantial and systematic differences between transmission-informed and uniform initialization approaches across the outbreak scenarios; see \cref{fig:epidemic_curves}. 
These differences are shown in both the epidemic growth and in total infection counts, with the magnitude of effects varying considerably by scenario type and initialization. 
Note that although Scenarios R1 and R2 are initialized with the same number of agents, the final infection outcome cannot be compared directly between the two scenarios due to different households participating in the outbreak event; compare~\cref{tab:initialization-details}.
Since households with weaker or stronger connections to the entire district are involved, only comparisons between the two initializations within the same scenario setting are meaningful.
In contrast, the workplace scenarios W1 and W2 are initialized with overlapping infected populations, as the agents infected in W1 represent a subset of those infected in W2.

Restaurant scenarios demonstrate slightly more pronounced differences between initializations. 
In Scenario~R1 (limited mixing), uniform initialization leads to 103 additional infections by day 10, representing a 46.0~\% increase over the transmission-informed baseline. 
Note, the trajectories begin diverging instantly, with uniform initialization consistently producing faster epidemic growth throughout the simulation period.
The uniform approach reaches 100 infections one day earlier than transmission-informed initialization (Day 5.5 vs. Day 6.5), underlining accelerated early-phase dynamics.

Scenario~R2 (more mixing) exhibits more modest but still significant differences, with uniform initialization producing 34 additional infections (+17.2~\%) by day 10. 
The temporal dynamics show similar patterns to R1, with uniform initialization reaching the 100-infection threshold 0.9 days earlier (day 6.4 vs. day 7.3).

\begin{figure}[!h]
    \centering
    \includegraphics[width=\textwidth]{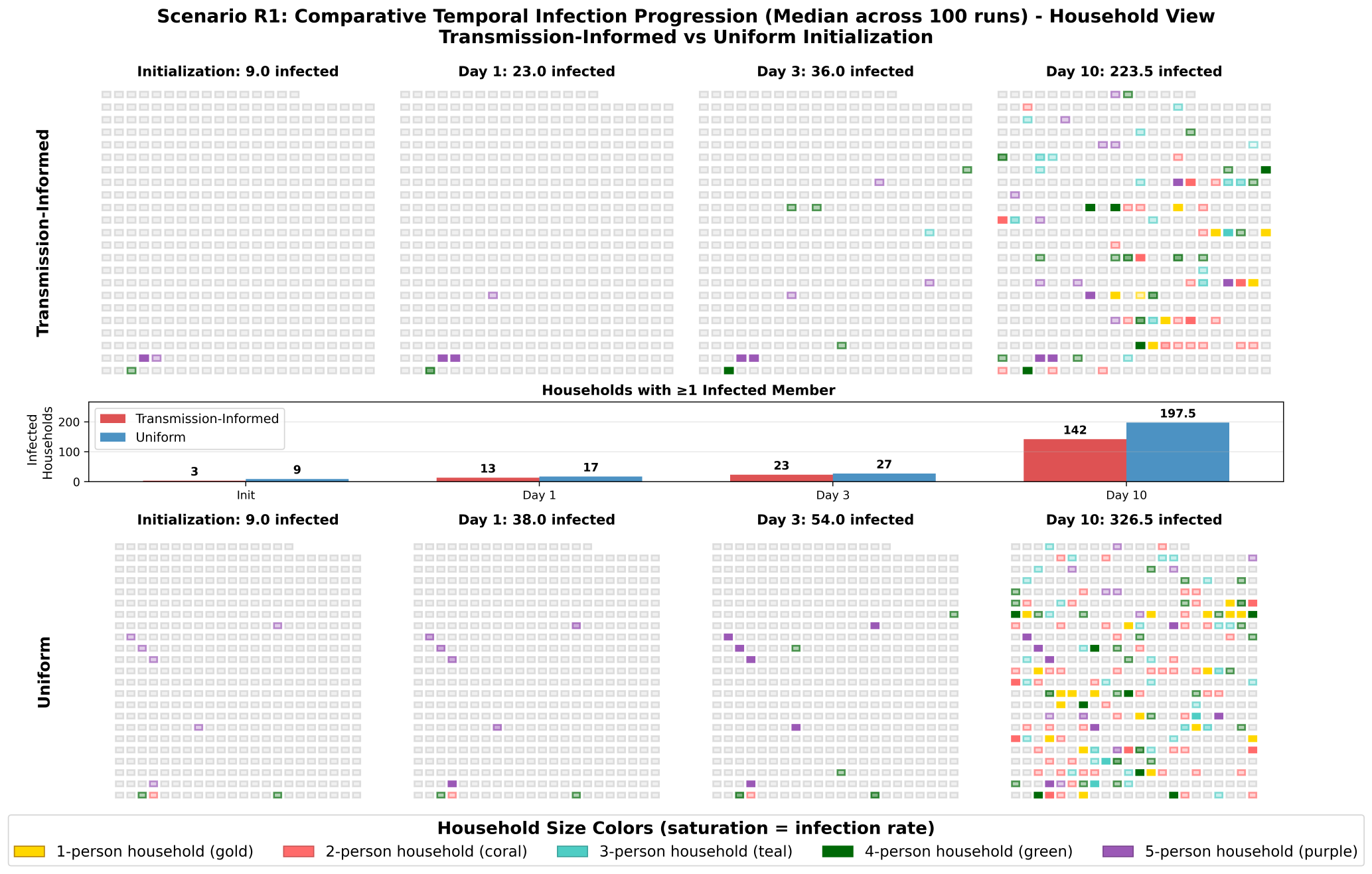}
    \caption{\textbf{Heatmap of Spatially Distributed Infections for Restaurant Scenario~R1 (Limited mixing)}: We compare transmission-informed initialization (top row) versus uniform initialization (bottom row) across four time points: initialization, day 1, day 3, and day 10. Rectangular shapes represent households, with color intensity indicating the median number of infected members per household across 100 simulation runs (one color per household size, light shading for a low percentage of infected, dark shading for a high percentage). The center panel shows the median number of affected households over time.}
    \label{fig:heatmap_r1}
\end{figure}

Workplace scenarios show similar initialization effects, which is also an effect of the reduced outbreak size. 
Scenario~W1 (limited mixing) demonstrates a 31-infection difference (+41.3~\%) choosing uniform initialization. 
The 100-infection milestone occurs on day 9.8 for the uniform initialization, while the transmission-informed initialization ends with 75 infectious agents.
Scenario~W2 (more mixing) produces a larger epidemic size on day 10 and similar patterns with a 15-infection difference (+9.3~\%). 
Here, 100 infections are reached at day 7.2 with uniform and day 7.8 with transmission-informed initialization.

The confidence intervals reveal that initialization effects exceed usual stochastic variation across all scenarios. 
The 50~\% and 90~\% confidence intervals for transmission-informed and uniform approaches show small overlap in limited mixing settings (R1 and W1), indicating that the observed differences are robust. Workplace scenarios show greater overlap but are still consistent.

\begin{figure}[!h]
    \centering
    \includegraphics[width=\textwidth]{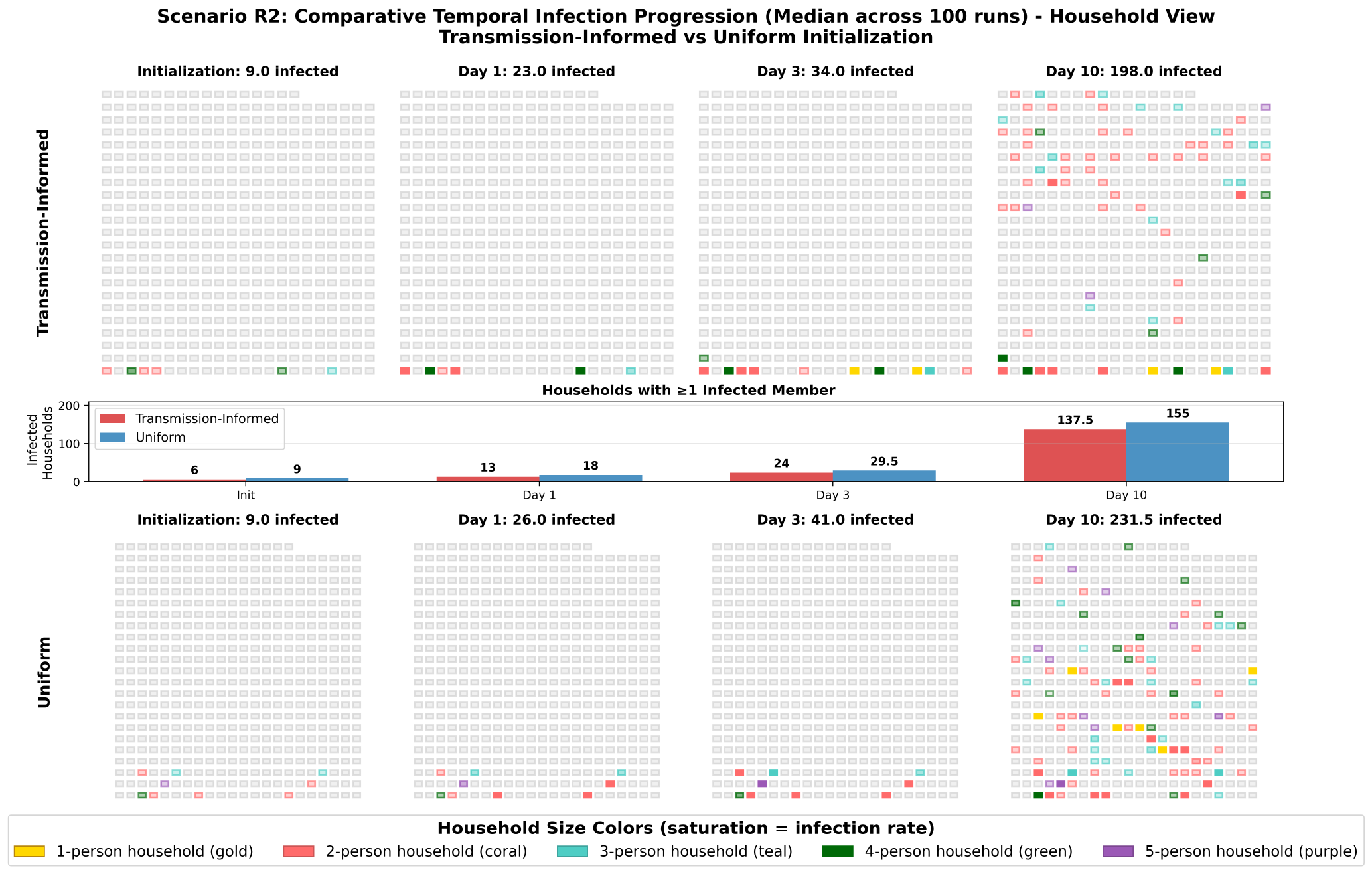}
    \caption{\textbf{Heatmap of Spatially Distributed Infections for Restaurant Scenario~R2 (More mixing)}: We compare transmission-informed initialization (top row) versus uniform initialization (bottom row) across four time points: initialization, day 1, day 3, and day 10. Rectangular shapes represent households, with color intensity indicating the median number of infected members per household across 100 simulation runs (one color per household size, light shading for a low percentage of infected, dark shading for a high percentage). The center panel shows the median number of affected households over time.}
    \label{fig:heatmap_r2}
\end{figure}

These results establish that missing transmission chain information creates systematic bias in epidemic predictions, with the magnitude of bias strongly dependent on the social clustering and mixing characteristics (of the initial outbreak). 
Limited mixing scenarios, which involve limited inter-household and inter-workplace mixing, show the largest sensitivity to initialization assumptions, while more inter-household and inter-workplace mixing scenarios demonstrate smaller but still noteworthy effects.
Here, in comparison to workplace scenarios, household scenarios show heightened relative differences.

Furthermore, multi-seed analysis across \SK{50} different city configurations confirms that these patterns are qualitatively robust to different household variations (Appendix Figures C.4--C.7). 
Again, the consistency of epidemic spread across all seeds demonstrates that initialization choice creates systematic rather than stochastic prediction errors.

\subsection{Spatial Disease Spread Patterns}

In this section, we consider spatial infection patterns, which are visualized through temporal heatmaps displaying household or workplace infection densities at four key time points: initialization, day 1, day 3, and day 10. 
Each rectangular (household) or hexagonal (workplace) unit is color-coded by the median number of infected members across 100 simulation runs, using different colors for different household sizes and deeper shades to indicate higher infection concentrations. 
It is important to note that these visualizations represent infected agents grouped by their assigned household or workplace locations, rather than infections occurring at those specific locations. 
This approach illustrates how disease propagates through distinct subnetworks within the broader contact network structure.
The visualization employs uniform grid layouts, which are ordered by the global index of the location in the city district, while bar charts quantify the median number of affected units with at least one infected agent at that time.

The spatial heatmaps reveal how the initialization assumptions create distinct patterns of household and workplace infection clustering, explaining the macroscopic epidemic differences observed in the aggregate trajectories.

\begin{figure}[!h]
    \centering
    \includegraphics[width=\textwidth]{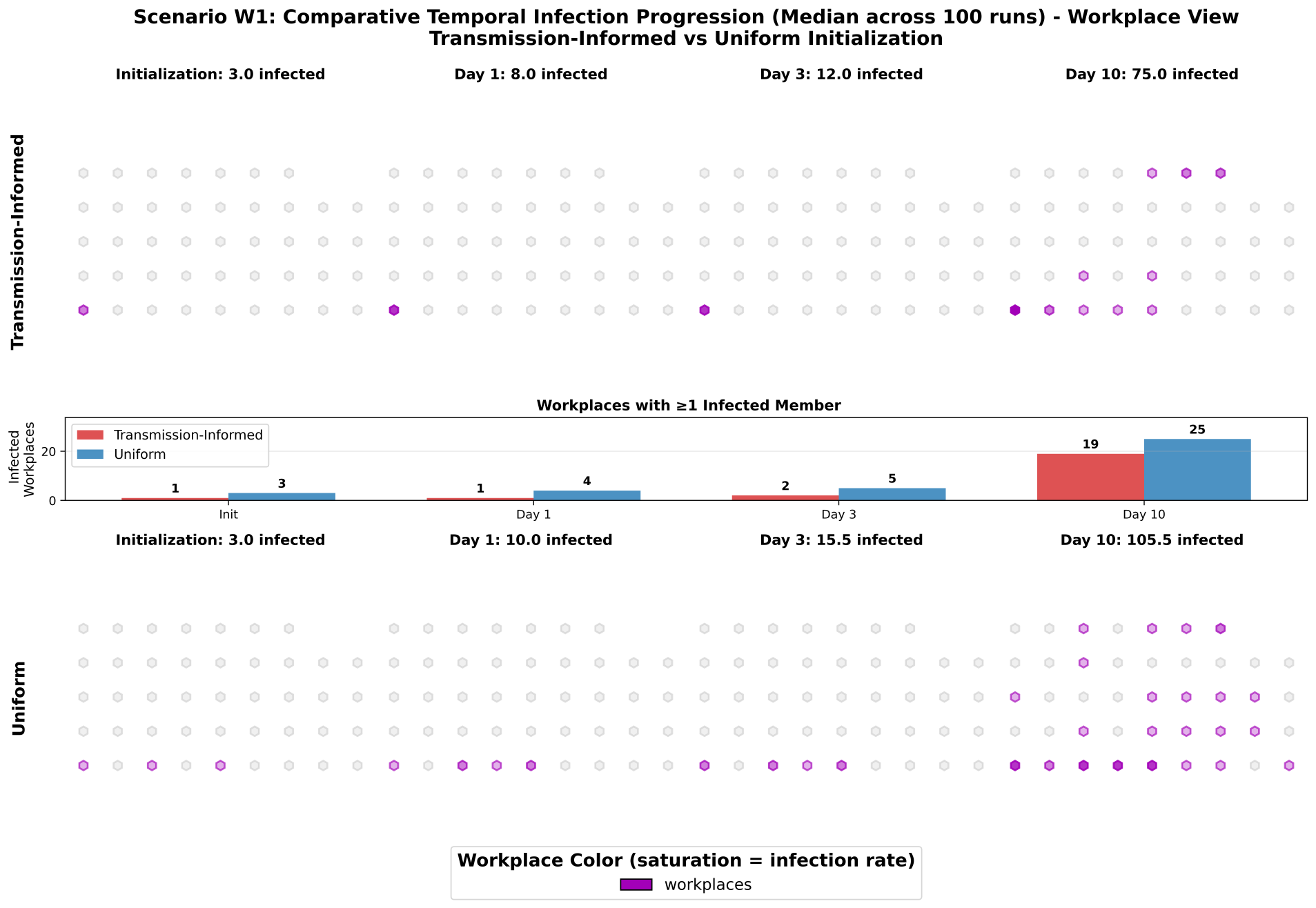}
    \caption{\textbf{Heatmap of Spatially Distributed Infections for Workplace Scenario~W1 (Limited mixing)}: We compare transmission-informed initialization (top row) versus uniform initialization (bottom row) across four time points: initialization, day 1, day 3, and day 10. Hexagonal shapes represent workplaces, with color intensity indicating the median number of infected members per workplace across 100 simulation runs (light shading for a low percentage of infected, dark shading for a high percentage). The center panel shows the number of affected workplaces over time.}
    \label{fig:heatmap_w1}
\end{figure}

\begin{figure}[!h]
    \centering
    \includegraphics[width=\textwidth]{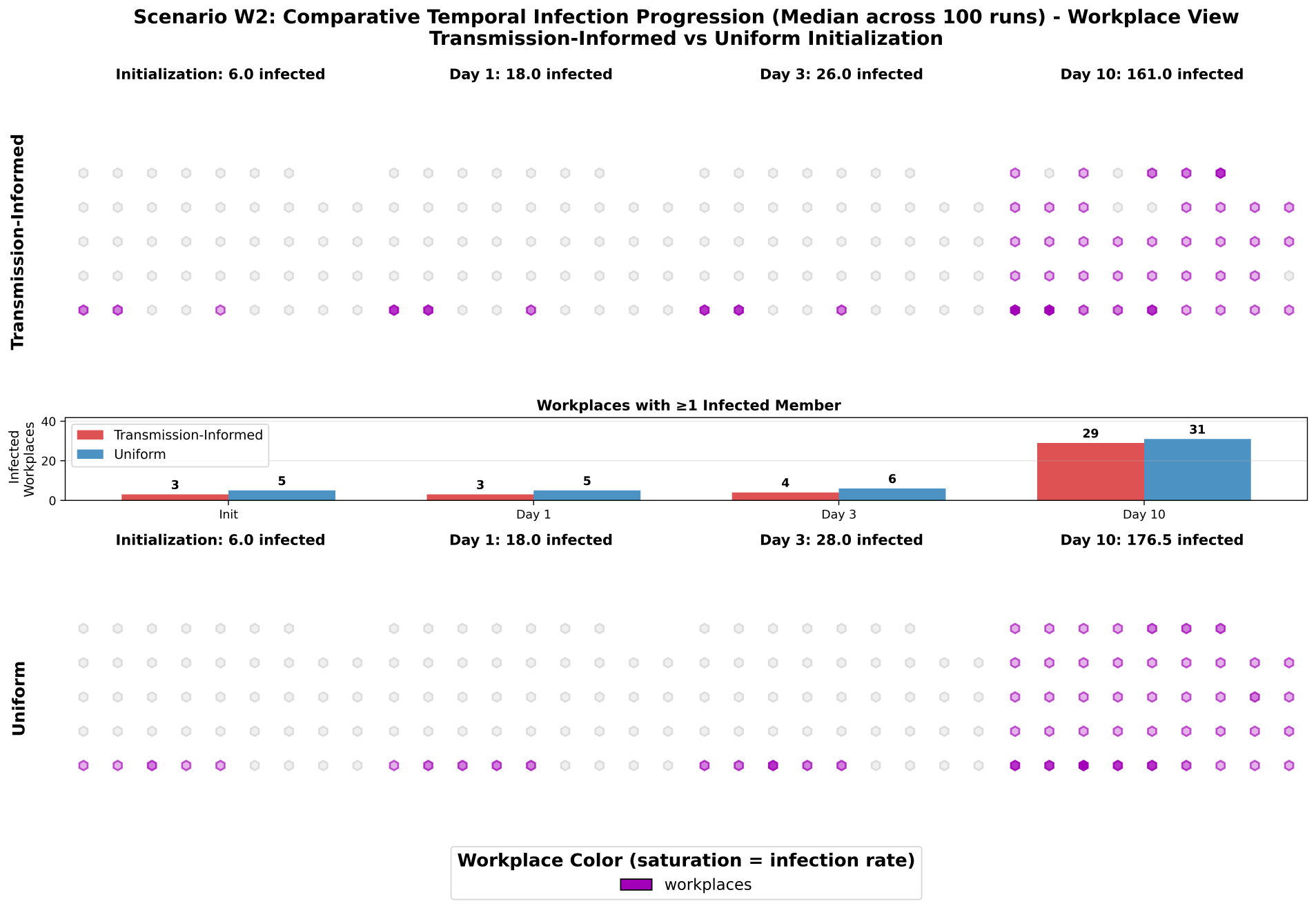}
    \caption{\textbf{Heatmap of Spatially Distributed Infections for Workplace Scenario~W2 (More mixing)}: We compare transmission-informed initialization (top row) versus uniform initialization (bottom row) across four time points: initialization, day 1, day 3, and day 10. Hexagonal shapes represent workplaces, with color intensity indicating the median number of infected members per workplace across 100 simulation runs (light shading for a low percentage of infected, dark shading for a high percentage). The center panel shows the number of affected workplaces over time.}
    \label{fig:heatmap_w2}
\end{figure}

The restaurant scenarios (Figures~\ref{fig:heatmap_r1} and~\ref{fig:heatmap_r2}) demonstrate the critical role of household clustering. 
In Scenario~R1, transmission-informed initialization begins with infections concentrated in just 3 households. While we see infection-saturated households at days 1 and 3 with transmission-informed initialization, uniform initialization shows more but much less saturated households at day 1, which, on the other hand, become more saturated by day 3. 
Under transmission-informed initialization, infections saturate the initially affected households but face natural barriers when attempting to spread to new household units.
By day 10, we see a substantial number of saturated households in both approaches. 
Notably, the proportion of infected agents divided by affected households is larger for uniform initialization (1.65 vs. 1.57  infected per infected household), showing a higher average saturation for uniform initialization. 
This is likely due to the persisting effects of the initialization in nine large households and early saturation effects of the small city district scale.

Scenario~R2 exhibits similar but attenuated patterns. 
Due to the weaker household clustering and more mixing at the tables, the difference in the effect of the two initializations is reduced while preserving the bias. 
From day 0 to day 3, we see the similar effect of more households getting saturated with uniform initialization. 
Furthermore, transmission-informed initialization affects 6 households initially, growing to 137.5 affected households by day 10, while uniform initialization spreads from 9 to 155 affected households. 
Again, we see that the uniform initialization yields a slightly higher average saturation (1.49 vs. 1.44).
Here, this is likely due to early saturation effects of the small city district scale, as initial infected households are not substantially bigger.

The workplace scenarios (Figures~\ref{fig:heatmap_w1} and~\ref{fig:heatmap_w2}) reveal more modest but consistent initialization effects. 
In contrast to the household simulation, we here show saturation and diffusion effects across different workplaces, where the office pairs and workplaces from the outbreak scenario are integrated into five of the 43 workplaces with 10-11 agents each.

In Scenario~W1, transmission-informed initialization results in concentrated infections within one workplace.
By day 10, affected workplaces grow from 1 to 19, while uniform initialization spreads from 3 initially affected departments to 25 total affected workplaces, showing a more widespread infection distribution.

As Scenario~W2 yields more infected agents than W1, this creates more opportunities for transmission, visible in the broader infection distribution throughout the time steps. 
Transmission-informed initialization begins with 3 affected workplaces and reaches 29 affected workplaces, while uniform initialization spreads from 5 initially affected workplaces to 31 total affected workplaces.

Given that workplaces are of size ten or eleven, transmission saturation occurs comparatively slower than in household settings, resulting in smaller infection differences between initializations relative to Scenario~R1.
However, the systematic bias toward broader spread under uniform initialization remains evident across both workplace scenarios, especially with the highly clustered Scenario~W1.

\begin{figure}[!h]
    \centering
    \includegraphics[width=\textwidth]{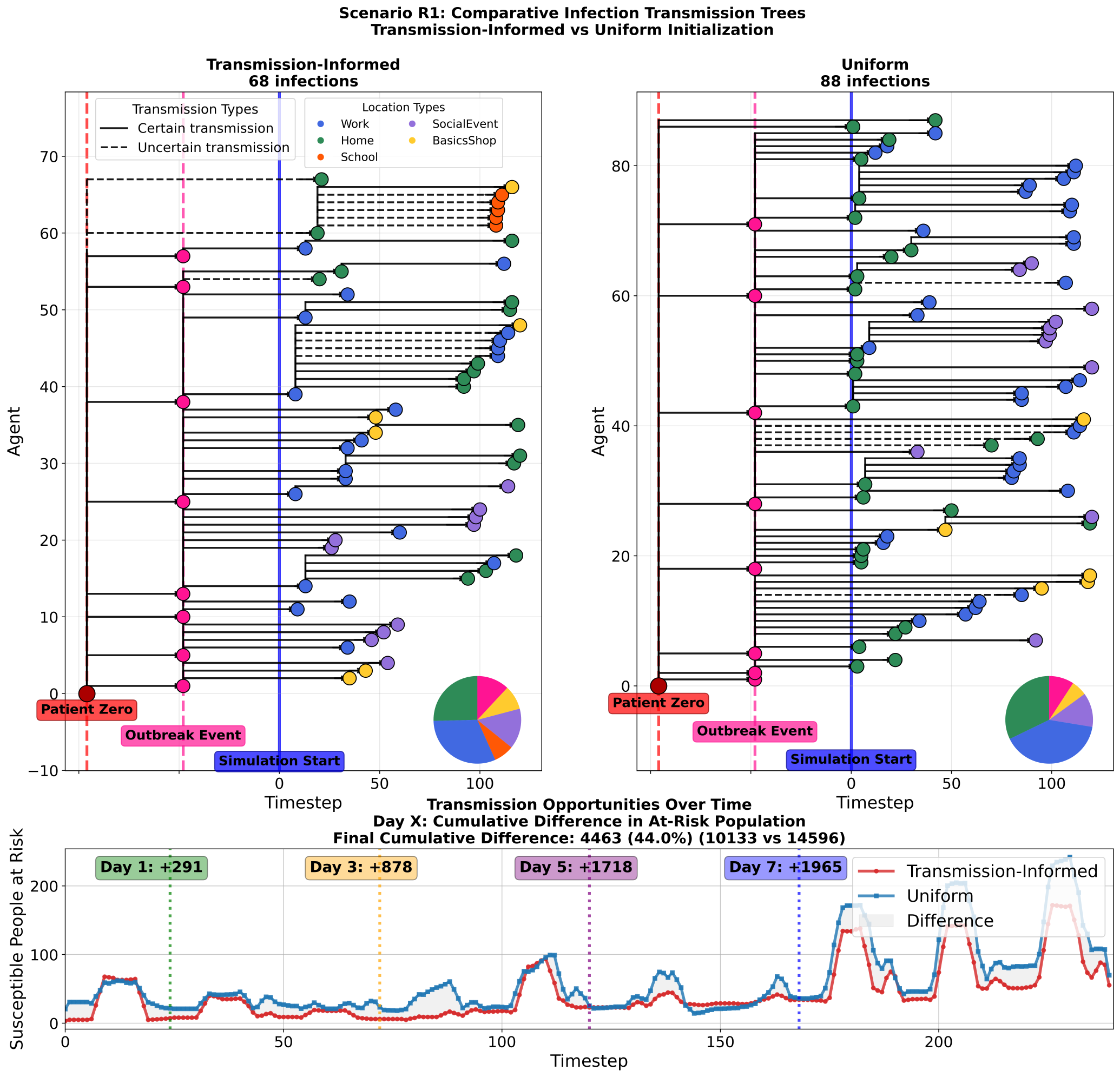}
    \caption{\textbf{Transmission Trees and Population at Risk for Restaurant Scenario~R1.} Comparison of transmission-informed initialization (left) versus uniform initialization (right). Top panels show transmission chains over five days, with colored dots representing infected individuals by location type (blue: Work, green: Home, orange: School, yellow: BasicsShop, purple: SocialEvent) plus pink for the outbreak event and red for patient zero (counted as part of the outbreak event). Solid lines indicate certain transmission chains, while dashed lines represent transmission events with uncertain infector relationships. Pie charts display the distribution of infections by location type. The bottom panel shows the 3-day moving average difference in susceptible population at risk over the whole 10 days. }
    \label{fig:transmission_tree_R1}
\end{figure}

\begin{figure}[!h]
    \centering
    \includegraphics[width=\textwidth]{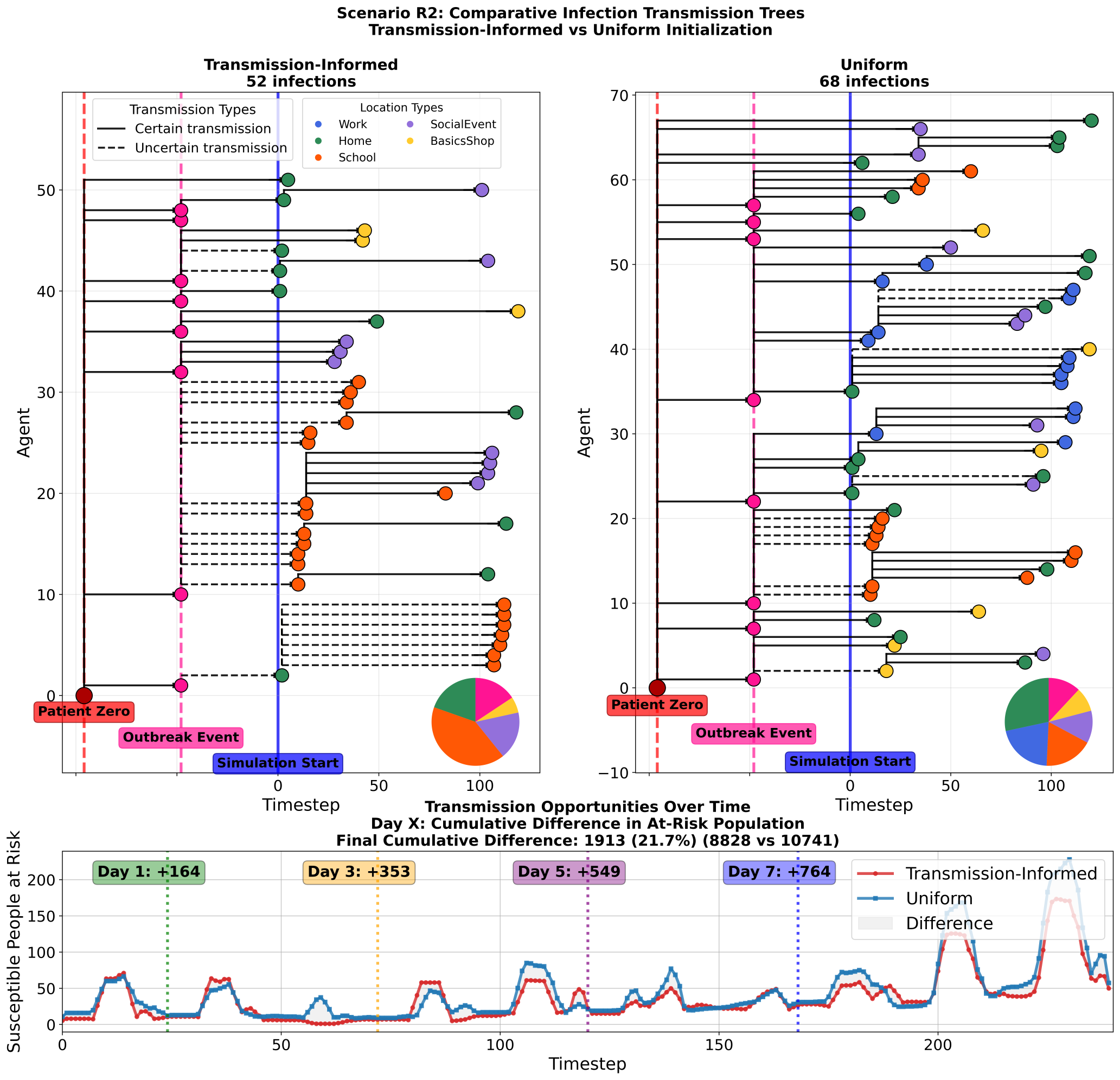}
    \caption{\textbf{Transmission Trees and Population at Risk for Restaurant Scenario~R2.} Comparison of transmission-informed initialization (left) versus uniform initialization (right). Top panels show transmission chains over five days, with colored dots representing infected individuals by location type (blue: Work, green: Home, orange: School, yellow: BasicsShop, purple: SocialEvent) plus pink for the outbreak event and red for patient zero (counted as part of the outbreak event). Solid lines indicate certain transmission chains, while dashed lines represent transmission events with uncertain infector relationships. The bottom panel shows the 3-day moving average difference in susceptible population at risk over the whole 10 days.}
    \label{fig:transmission_tree_R2}
\end{figure}

The heatmap analysis reveals the fundamental mechanism underlying initialization effects: infection clustering versus dispersion trade-offs. 
Uniform initialization distributes initial infections across more social units, creating multiple simultaneous transmission “sources” that collectively drive faster epidemic expansion. Transmission-informed initialization creates transmission “sinks” where infections become concentrated in highly connected social units (households or workplaces), leading to accelerated local spread but is limited by local saturation, resulting in, to some extent, slower spatial spread, i.e., fewer infected units throughout the simulation.

The spatial analysis demonstrates that the aggregate epidemic differences documented in Figure~\ref{fig:epidemic_curves} emerge from the systematic differences in the distribution of transmission events. 
The initialization choice alters not just the timing of epidemic growth but also the spatial pathway through which infections propagate across the city district's population.

\subsection{Transmission Trees and Population at High Risk}

In this section, transmission trees are constructed to get a more profound insight into evolving dynamics at the beginning of the simulation, revealing how the two different initialization choices alter transmission ways, location-based infection patterns, and the depletion of susceptible populations. 
The trees are constructed by tracking infection events and inferring transmission relationships based on agents being at the same location and being infectious at that time point. 
For the sake of a clear visualization, we only show the first five days of the simulation.
In addition, we provide an overview of the susceptible population being at risk of infection, i.e., being at the same location as infectious individuals.

\textbf{Transmission Trees.} As the MEmilio-ABM does not use direct contacts explicitly, we obtain situations where multiple infectious agents, present at the same location, create ambiguity regarding the actual source of transmission. 
With potential transmission by aerosols, this might, however, not be resolved in observations either.
In the transmission trees, we express this ambiguity through dashed lines where we assign the agent who was earliest infected the role of the infector (as the origin of the connecting lines).
Each node represents an infected individual, colored by the location type where infection occurred (red: patient zero, pink: outbreak event, blue: workplace, green: home, orange: school, yellow: shopping, purple: social events), with y-axis positions reordered to minimize visual crossing of transmission chains. 
Solid lines indicate certain transmission events.
Pie charts summarize the distribution of infections by location type. The bottom panels track the amount of susceptible populations that have at least one infected agent at the same location at the corresponding time step of the simulation, providing insights into transmission opportunities and epidemic potential.
This data is gathered from one of the 100 runs, which has the minimum distance to the median output. 
That means we take the run where the Euclidean norm of cumulative infected agents subtracted from the median amount of cumulative infected agents from all 100 runs becomes minimal. 
Note that while this minimizes the distance to the median output and therefore shows a, to some extent, representative output, individual transmission chains remain highly stochastic and cannot be considered representative for all runs.

\begin{figure}[!h]
    \centering
    \includegraphics[width=\textwidth]{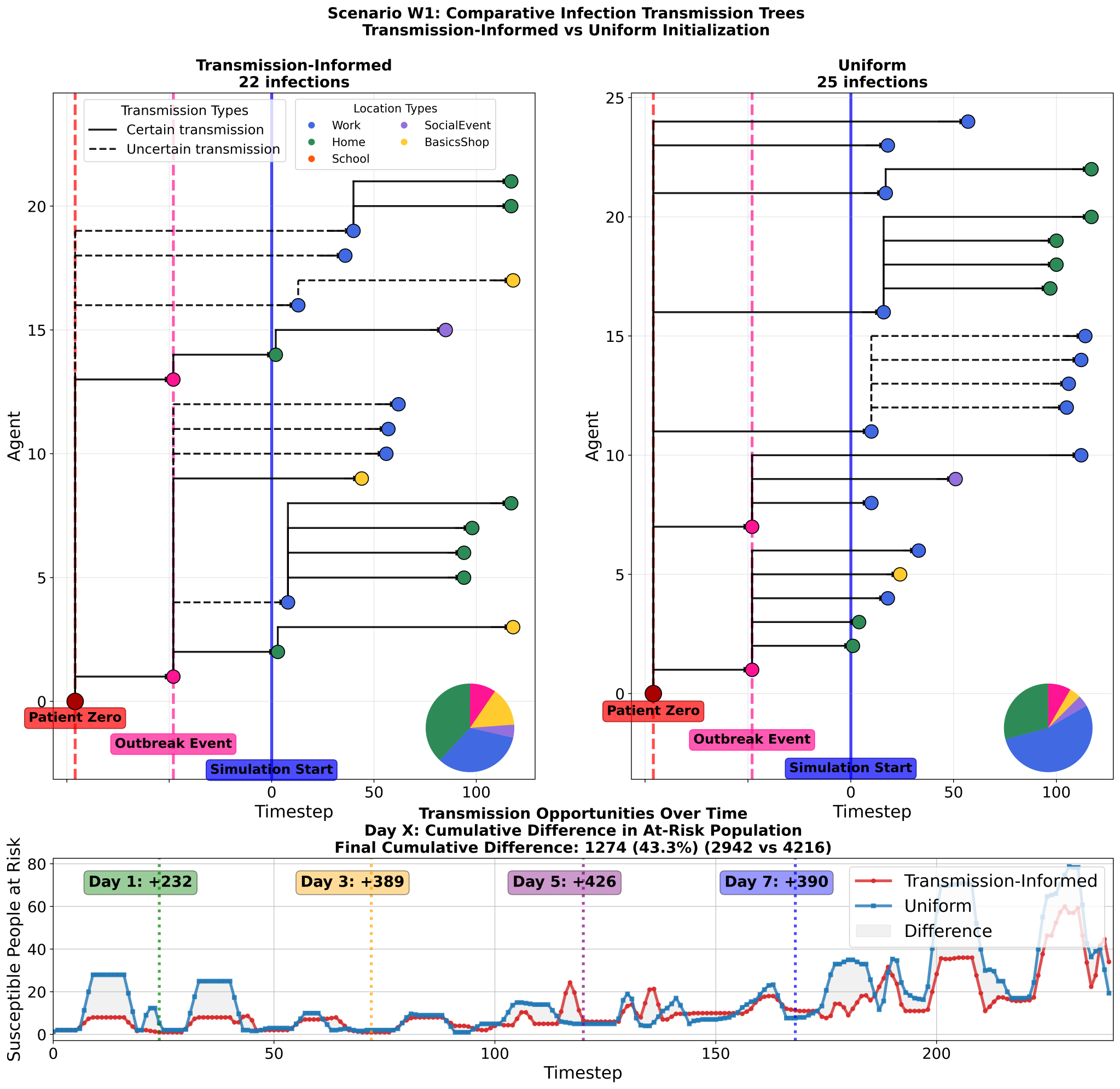}
    \caption{\textbf{Transmission Trees and Population at Risk for Workplace Scenario~W1.} Comparison of transmission-informed initialization (left) versus uniform initialization (right). Top panels show transmission chains over five days, with colored dots representing infected individuals by location type (blue: Work, green: Home, orange: School, yellow: BasicsShop, purple: SocialEvent) plus pink for the outbreak event and red for patient zero (counted as part of the outbreak event). Solid lines indicate certain transmission chains, while dashed lines represent transmission events with uncertain infector relationships. Pie charts display the final distribution of infections by location type. The bottom panel shows the 3-day moving average difference in susceptible population at risk over the whole 10 days. }
    \label{fig:transmission_tree_W1}
\end{figure}

\begin{figure}[!h]
    \centering
    \includegraphics[width=\textwidth]{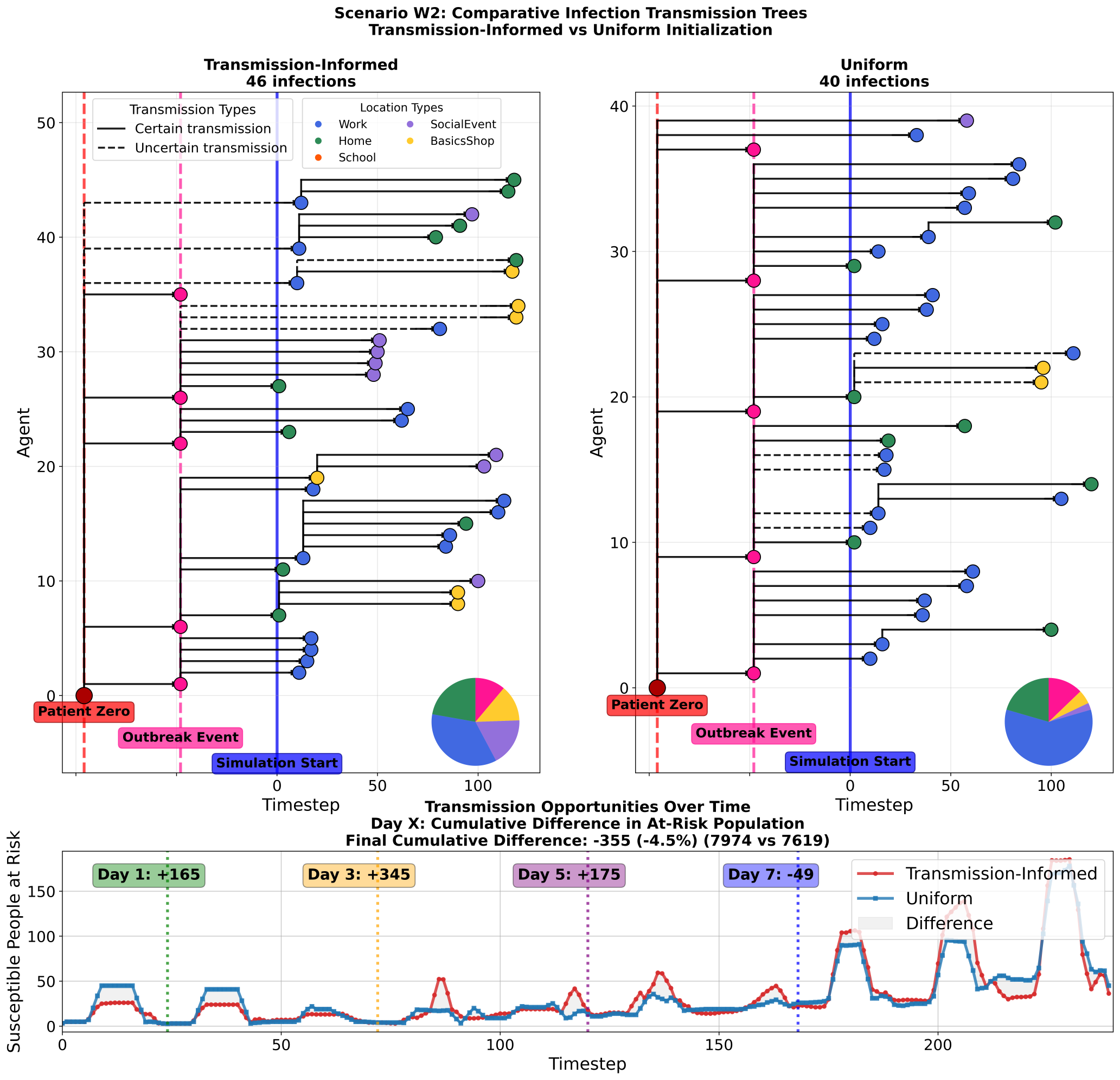}
    \caption{\textbf{Transmission Trees and Population at Risk for Workplace Scenario~W2.} Comparison of transmission-informed initialization (left) versus uniform initialization (right). Top panels show transmission chains over five days, with colored dots representing infected individuals by location type (blue: Work, green: Home, orange: School, yellow: BasicsShop, purple: SocialEvent) plus pink for the outbreak event and red for patient zero (counted as part of the outbreak event). Solid lines indicate certain transmission chains, while dashed lines represent transmission events with uncertain infector relationships. Pie charts display the final distribution of infections by location type. The bottom panel shows the 3-day moving average difference in susceptible population at risk over the whole 10 days. }
    \label{fig:transmission_tree_W2}
\end{figure}

The transmission trees of the median-minimizing run for the restaurant scenarios (Figures~\ref{fig:transmission_tree_R1} and~\ref{fig:transmission_tree_R2}) demonstrate markedly different network topologies between initialization approaches. 

In Scenario~R1, transmission-informed initialization creates a smaller transmission network with 68 infections after five days. 
The network shows distinct transmission generations, with clear temporal separation between infection generations with concentrated household-based transmission chains in the second generation after the initial outbreak, underlining the saturation. 
In contrast, uniform initialization generates a larger network with 88 infections (+29.4~\% increase), exhibiting multiple simultaneous transmission chains that propagate more rapidly through the population.
The transmission tree structure reveals broader temporal overlap between transmission generations, indicating accelerated epidemic dynamics consistent with the aggregate trajectory analysis. 
The household infection potential of uniform initialization is clearly visible through numerous simultaneous within-household infections occurring shortly after the simulation start.
The location-based infection distribution, shown in pie charts, still reveals certain similarities in overall transmission dynamics. 
Under both initializations, workplace transmission dominates (blue segments). 
Household transmissions (green segments) remain secondary, with limited basic shop (yellow) and social event (purple) involvement. 
However, note that this information cannot be considered representative if it is driven by stochastic processes and, e.g., if the initial outbreak contained infected school children or not, as demonstrated in Scenario~R2. 
For a more representative analysis, see Fig. D.8.. 
Again, let us also note that Scenarios R1 and R2 cannot be compared against each other directly, as different households and individuals participate in the outbreak event.

Scenario~R2 exhibits similar but toned-down patterns, with transmission-informed initialization producing 52 infections compared to 68 under uniform initialization (+30.7~\% increase). 
The transmission networks show less pronounced structural differences in immediate household infections, consistent with the weaker household clustering and stronger overall mixing in the initial outbreak configuration.
In transmission-informed initialization, school spread dominates, due to several agents initialized as pupils while with uniform initialization, household and work transmission are more or equally important.

The workplace transmission trees (Figures~\ref{fig:transmission_tree_W1} and~\ref{fig:transmission_tree_W2}) reveal distinct patterns. 
Scenario~W1 shows transmission-informed initialization producing 22 infections, while uniform initialization creates 25 infections (+13~\% increase). 
The location distribution shows workplace and household transmission (blue and green, respectively) dominating under both approaches, but with increased workplace-based secondary transmission under uniform initialization.
Clustered transmission patterns become evident as the three initial infectious agents under transmission-informed initialization rapidly infect 7 agents within the same workplace shortly after the simulation starts. 
Conversely, uniform initialization produces 8 infections distributed across 3 distinct workplaces in the first transmission generation, demonstrating reduced local transmission intensity and risk but enhanced spatial dispersion of epidemic potential across the workplace contact network.

Scenario~W2 shows transmission-informed initialization having accelerated early infection dynamics. 
Transmission-informed initialization produces 46 infections compared to 40 under uniform initialization, representing a rare case where the initialization effects are turned around. As remarked before, note that confidence intervals for the total epidemic size overlap closest with Scenario~W2, meaning that such a situation is more likely to appear in this setting.
This shows that sometimes increased local transmission intensity and risk from multiple infectious agents at the same location can overcome initial dispersion effects when only minimally limited by local saturation effects.

\noindent\textbf{Susceptible Population at Risk.} In the bottom panels of the Figures~\ref{fig:transmission_tree_R1}--\ref{fig:transmission_tree_W2}), we reveal the temporal dynamics of the susceptible population at high risk, meaning that these individuals are at the same location as infectious individuals. 
Thus, providing insights into the epidemic potential over time. 
Note that this metric is just one indicator of infection potential, as multiple infectious agents at the same location do not lead to a higher at-risk population but increase the local infection risk.
All scenarios but W2 show consistent patterns where uniform initialization creates sustained higher transmission potential, as evidenced by consistently high at-risk population levels.
The cumulative differences grow progressively larger over the 10-day simulation period, reaching maximum divergence by day 7-10 across all scenarios.

The W2 scenario exhibits more variable at-risk population dynamics.
Here, early uniform initialization produces a higher at-risk population, overtaken at day 7 of the simulation.

\MK{

\subsection{Sensitivity with transmission parameter and population size}

While our parameter $\lambda = 22.6$ is chosen in line with previous calibrations of Vadere~\cite{rahn_modelling_2022} and carefully calibrated such that it produces relevant (super)spreading scenarios over all settings which should reproduce epidemic uptakes that are neither too fast nor too slow within a 10-day time frame, we here provide an additional sensitivity analysis for $\lambda$ scaling it with a quarter, a half and three quarters.

In~\cref{tab:percentual_difference_lambda}, we provide the difference between the median outcomes for both initialization approaches and Scenarios~R1,~R2,~W1, and~W2. While we see quantitative differences for Scenarios~R1,~R2, and~W1, we still obtain qualitatively similar behavior for all scenarios and between the different initialization schemes, supporting the outcome of systematic rather than stochastic prediction errors. The increasing difference between Scenarios R1 and R2 with smaller $\lambda$ values can be explained by relocated transmission dynamics. While overall transmission rates decrease, household transmission becomes even more important; cf. Fig. F.9. In transmission-informed initialization, infection-saturated households have limited opportunities to infect new agents outside their cluster, constraining epidemic growth meaning that the difference from the initialization effect becomes more pronounced at lower transmission rates. In contrast, Scenario W1 exhibits decreasing differences with $\lambda$ decreasing. This is due to a limited saturation effect at the workplaces where the (10 or 11) workplace members only become entirely infected around simulation day 10. As before, epidemic progression becomes more driven by households which are not aligned with the designed clustering.

\begin{table}[!h]
\centering
\caption{\MK{\textbf{Difference between Transmission-Informed and Uniform Initialization with varying $\lambda$ Values}. For Scenarios~R1,~R2,~W1, and~W2, we provide the difference between the median outcomes for both initialization approaches.}}
\begin{tabular}{lcccc}
\toprule
\textbf{Scenario} & $\boldsymbol{\lambda}$ & $\frac{\bm{3}}{\bm{4}}\boldsymbol{\lambda}$ & $\frac{\bm{2}}{\bm{4}}\boldsymbol{\lambda}$ & $\frac{\bm{1}}{\bm{4}}\boldsymbol{\lambda}$  \\
\midrule
R1 & 41.0~\%& 45.5~\% & 54.1~\% & 63.9~\% \\
R2 & 12.4~\%& 14.2~\% & 18.0~\% & 21.7~\% \\
W1 & 27.1~\%& 23.4~\% & 19.1~\% & 10.8~\% \\
W2 & 6.5~\%& 7.0~\% & 7.0~\% & 6.2~\% \\
\bottomrule
\end{tabular}
\label{tab:percentual_difference_lambda}
\end{table}

Furthermore, we consider our model's and model setup's sensitivity with respect to the chosen population size. In~\cref{tab:percentual_difference_agents}, we provide the outcomes of the differences between the median simulations for our default setup with 1\,000 agents and for scaled setups between 2\,000 and 50\,000 agents in a city (district). While with more agents, the initial infections have a smaller share on the total population and, thus, less saturation effects appear, initializations with larger number of agents are also stochastic such that slightly different networks and transmission dynamics can realize. In the appendix, in Table E.3, we provide the average household sizes of the infected individuals of the first initial outbreak, for instance providing reasons for the slight outliers such as in Scenario~W1 with 10\,000 individuals. Given the overall complexity of the model and corresponding parameters, a difference of 10-15~\% should be considered with care while differences around 30, 40, or even up to almost 80 percent show consistently the systematic biases from the uniform initialization.\\

\begin{table}[!h]
\caption{\MK{\textbf{Difference between Transmission-Informed and Uniform Initialization for the four Different Scenarios with Varying Population Sizes.}}}
 \vspace{2mm} 
\begin{adjustwidth}{-1in}{-1in}\centering
        \resizebox{1.3\textwidth}{!}{%

\begin{tabular}{lccccccccc}
\toprule
\textbf{Scenario} & 1k & 2k& 5k & 7k & 10k &12k & 15k & 20k & 50k \\
\midrule
R1 & 41.0~\% &57.6~\%&64.8~\%&63.7~\%&78.8~\% &66.0~\%&72.8~\%&78.5~\%&71.1~\% \\
R2 & 12.4~\% &15.2~\%&13.1~\%&11.0~\%& 25.0~\% &13.0~\%&17.0~\%&25.2~\%&19.4~\%\\
W1 & 27.1~\% &23~\%&29.9~\%&29.4~\%& 41.7~\% &25.7~\%&27.3~\%&24.6~\%&28.9~\% \\
W2 & 6.5~\% &5.4~\%&7.5~\%&11.1 ~\%& 10.0~\% &10.1~\%&9.8~\%&7.6 ~\%&6.8~\% \\
\bottomrule
\end{tabular}
}

\end{adjustwidth}
\label{tab:percentual_difference_agents}
\end{table}

\subsection{Application to other household and contact structures}

\begin{figure}[!t]
    \centering
    \begin{minipage}{0.68\textwidth}
       \includegraphics[width=1\linewidth]{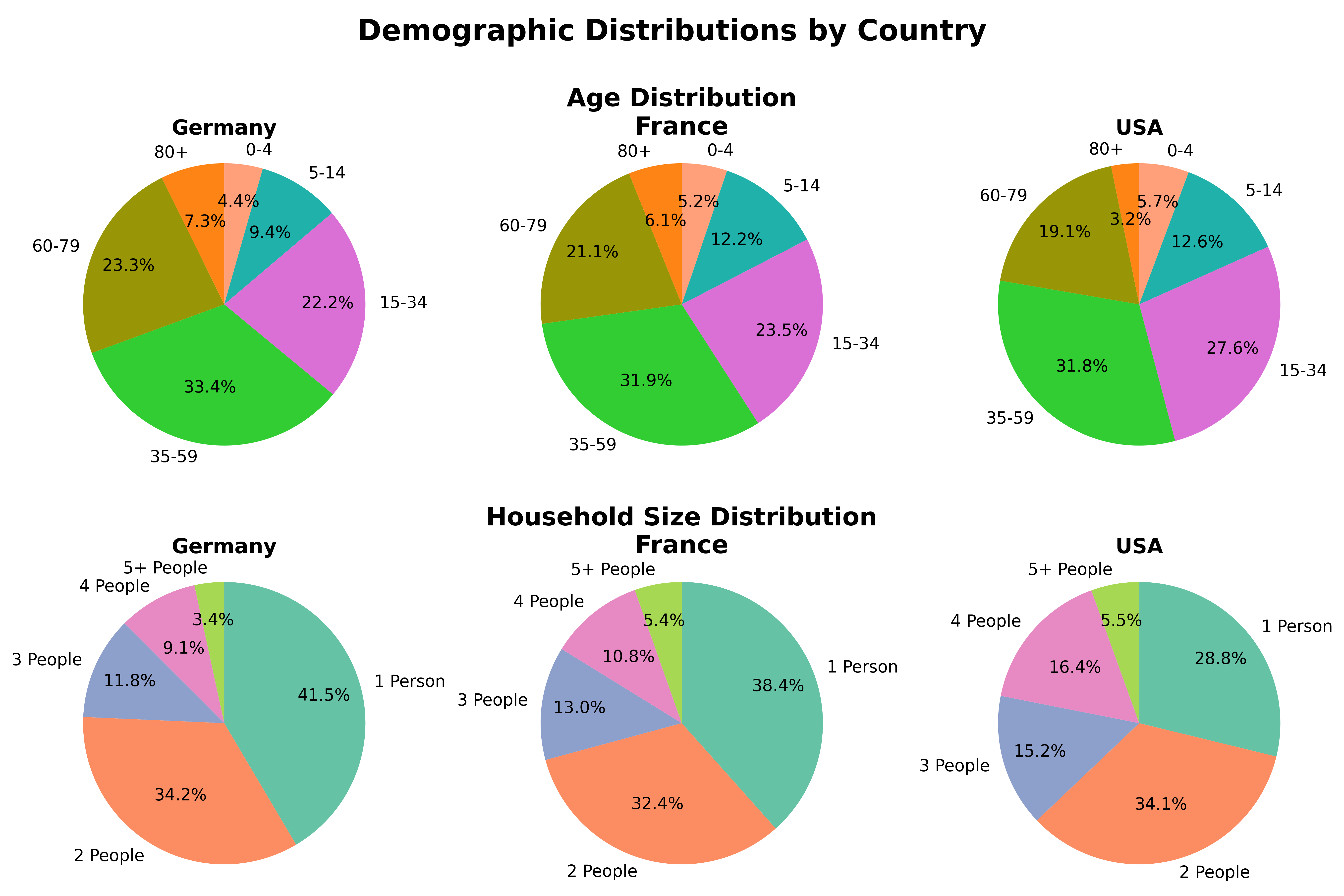}     
    \end{minipage}
\hspace*{0.2cm}
\begin{minipage}{0.28\textwidth}
  \includegraphics[width=0.8\linewidth]{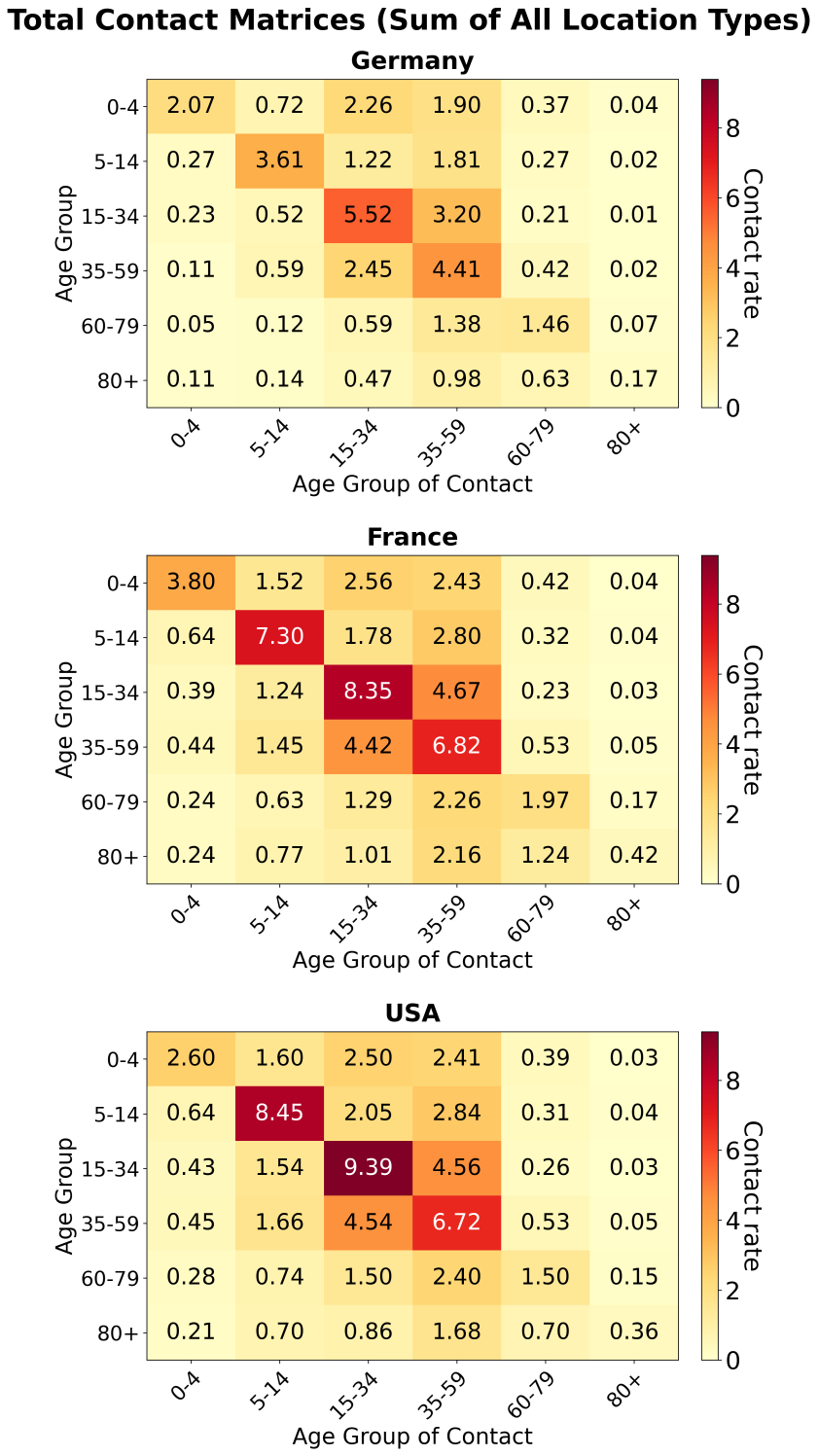}     
\end{minipage}
    \caption{\MK{\textbf{Comparison between household and contact structures in Germany, the USA, and France.} \textit{Left:} Share of households of different sizes in the three different countries. \textit{Right:} Contact matrices summed up over all locations (Home, School, Work, Other) and for the three different countries.}}
    \label{fig:usa_fr}
\end{figure}

In the previous subsections, we have provided detailed insights into four different scenarios, also varying transmission probability and population sizes. In order to support our findings more generally, we here consider city setups with different population, household, and contact structures. Therefore, instead of building upon German data, we here consider household and contact structures from France and the USA. Population and household data is obtained from~\cite{fra_household_size,fra_population_age,usa_household_size,usa_population_age} and contact matrices are built upon~\cite{prem_projecting_2017}, stratified into the six age groups as before. In~\cref{fig:usa_fr}, we provide the different household structures of Germany, the USA, and France (left) as well as the contact matrices (right). While both countries have a younger population, with the USA even younger than France, the USA also has a substantially smaller number of single person households and a substantially larger number of households with three, four, or five persons. The corresponding initialization thus leads to different, implicitly realized networks. In addition, the reported number of contacts is substantially larger for both countries (France and USA).

\begin{figure}[!h]
    \centering
    \includegraphics[width=1.0\textwidth]{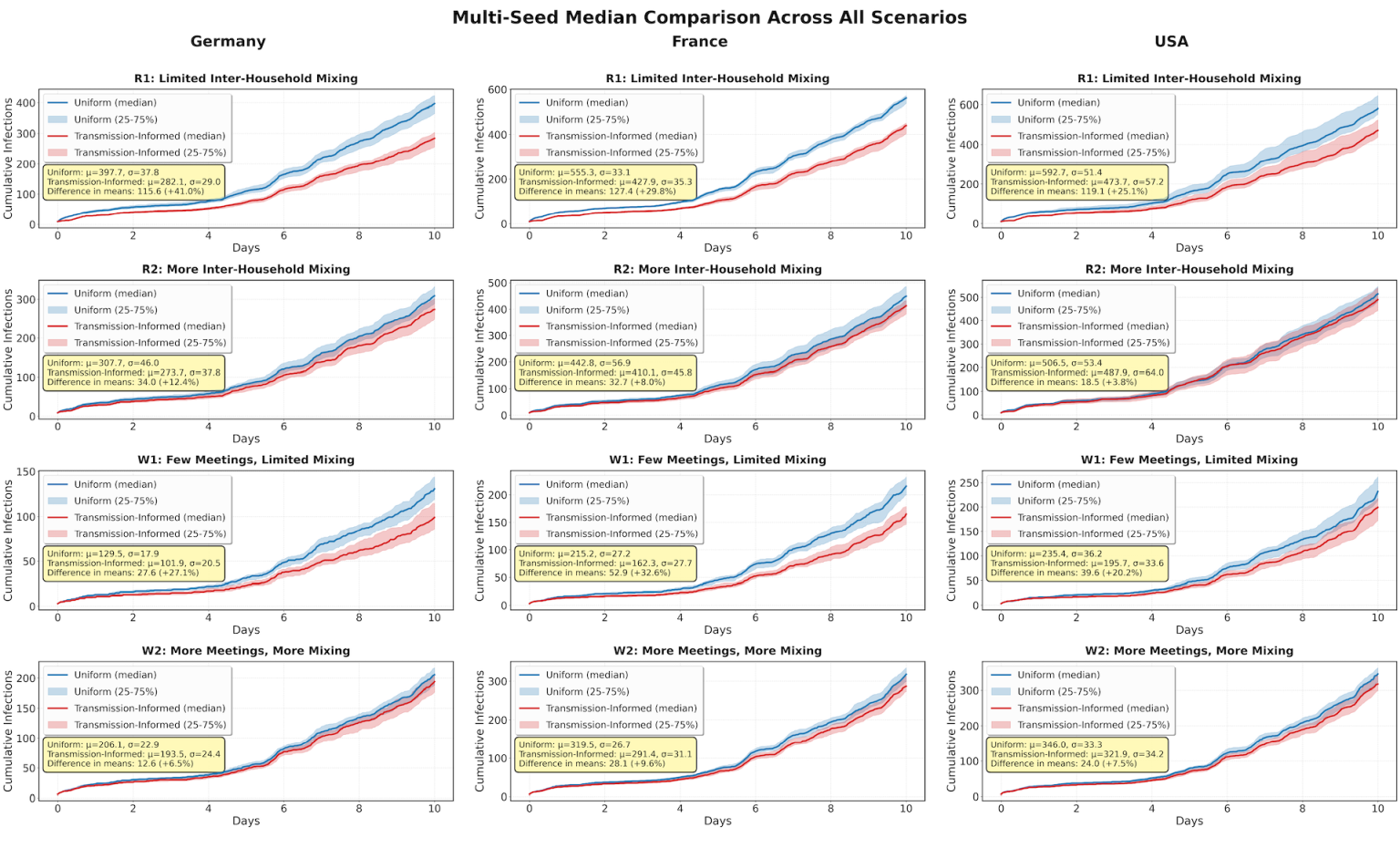}
    \caption{\MK{\textbf{Multi-Seed Simulation Comparison for the four Scenarios Initialized with Demographic Data from Germany, France and the USA, each with Uniform (blue) and Transmission-Informed (red) Initialization.} Scenarios~R1,~R2,~W1,~and~W2 are shown in rows while countries are shown in columns with Germany (left), France (center), and the USA (right). Each subfigure presents the median cumulative infections with 25th-75th percentile confidence intervals aggregated across all seed runs.}}
    \label{fig:three_counties_multi_seed}
\end{figure}

In~\cref{tab:percentual_difference_counties}, we provide the differences between the median outcomes for the city district setups with 1\,000 agents representatively drawn from the household and population structure of the USA and France. We see that the simulation results for France are closer to those of Germany which is due to the fact of the population and household structure which also have more similarities. However, as individuals in both countries have, on average, a substantially larger number of daily contacts, epidemic uptake is faster and saturation effects arrive earlier. In~\cref{fig:three_counties_multi_seed}, we also provide the median outcomes for all scenarios together with the percentiles as in lower left-hand side panels of Figs.~C.4 to C.7. While we see quantitative differences for the three country demographics, transmission-informed initialization stays in median always below uniform initialization. In addition, differences are consistently more pronounced in the Scenarios~R1 and~W1 with more pronounced clusterings. A summary on the median outcomes is also given in~\cref{tab:percentual_difference_counties}.

\begin{table}[!t]
\centering
\caption{\MK{\textbf{Difference between Transmission-Informed and Uniform Initializations for the four Different Scenarios and three Countries: Germany, France and the USA.}}}

\begin{tabular}{lccc}
\toprule
\textbf{Scenario} & GER & FRA & USA \\
\midrule
R1 & 41.0~\% &29.8~\% & 25.1~\% \\
R2 & 12.4~\% & 8.0~\% & 3.8~\% \\
W1 & 27.1~\% & 32.6~\% & 20.2~\% \\
W2 & 6.5~\% & 9.6~\% & 7.5~\% \\
\bottomrule
\end{tabular}

\label{tab:percentual_difference_counties}
\end{table}

}

\section{Discussion}\label{sec:discussion}

In this work, we shed light on the question of how simplifying assumptions about transmission events alter disease dynamics and epidemic outcomes.

Standard surveillance and reporting approaches providing case counts and detailed models using this data together with contact diary information to model human contact behavior can prove insufficient. 
This is since they do not capture precise seating arrangements, conversation patterns, or the subtle social clustering that drives transmission chains. More detailed contact tracing typically records event attendance but often cannot forward this information on time to detailed models, thus making these models miss the spatial and temporal patterns of disease dynamics that determine actual transmission risk.

In this work, we already modeled the outbreak events with a lot of detail and realism, using realistic human locomotion models, but simplified the setting by not considering additional mixing locations like washing rooms, office kitchens, or a canteen. The outcomes presented here can thus be understood under the assumption that transmission in these places is well prevented through hygiene measures and protective interventions to avoid further transmission between individuals that otherwise do not mix. 

The synthetic city district employed here represents a globally unclustered contact network with many small clusters through, e.g., households. Further studies will be needed to understand the effects of different contact networks leading to different saturation, locally and globally, better. In particular, for local saturation, it is still an open research question why secondary attack rates in households were observed to be substantially below 50~\%, see~\cite{shah_secondary_2020,koh_what_2020}, and how this can be represented realistically for a completely immune-naive population. 
In addition, while we generated the synthetic population with realism according to German Census data, future work will also contain modeling inter-household contacts and family members sharing an extended contact network, e.g., the same friends, more realistically.

Microscale simulators such as Vadere are intended to be used with single events and large-scale ABMs generally do not model transmission events with full detail. To study the model error of the latter, we proposed the framework as described in the beginning of the paper. To avoid the limitation of the large-scale ABM washing out the initialization's impact, we kept the simulation interval and number of infection generations short. While the outcomes for a fully coupled simulation with a large-scale ABM and a microsimulation for all transmission events is out-of-scope of realization (in the near future), we intuitively assume that a correspondingly developed model would further reduce the model error and, additionally, increase the difference regarding the simplified uniform transmission.

Finally, these results are specific to the MEmilio-ABM framework, which incorporates multiple transmission parameters beyond simple contact frequency. While local saturation effects in contact networks should be model-independent phenomena, the quantitative differences between initialization methods reflect this particular model's assumptions about viral load dynamics, age-stratified contact rates, and specific infection rates derived from infectious agents at the same locations. Nevertheless, in~\cite{KERKMANN2025110269}, it was shown that these assumptions could realistically represent real city-size disease dynamics.

\section{Conclusion}\label{sec:conclusion}

In this study, we presented a framework for investigating the importance of detailed outbreak information to reliably forecast epidemic evolution with its spatial distributions. The use of highly detailed microsimulations enabled us to isolate \MK{and quantify} initialization effects that would otherwise remain confounded with \MK{the model's} structural differences. 

Our study demonstrates that missing transmission chain information in agent-based epidemic models can result in substantial and systematic prediction errors that persist throughout epidemic trajectories, with clustering patterns and social mixing being the critical determinants of prediction accuracy. Our findings reveal that the magnitude of prediction error is strongly dependent on the social clustering and mixing characteristics of the outbreak and transmission events. \MK{With respect to prior findings from network epidemic models, we quantified these differences for a realistic demographic setting and mixing pattern in a complex multilayer model and well selected outbreak (superspreading) events.}
\MK{With a representative setting of 1\,000 individuals from German demographic data,} Restaurant scenarios, which involve the strongest inter-household transmission patterns, show the most pronounced effects, with uniform initialization producing up to 46.0~\% more infections than transmission-informed initialization by day 10. The consistency of these effects was validated across 100 simulations and \SK{50} different city configurations, confirming that missing transmission chain information creates systematic rather than stochastic prediction errors. \MK{We could also further quantify these effects for larger city district setups with up to 50\,000 individuals. Here, less saturation effects appeared but we could qualitatively confirm our findings for these varying population sizes as well as for contact and household structures from French and US demographic data.}

The spatial analysis revealed the underlying mechanism: transmission-informed initialization creates local transmission hotspots where infections become concentrated in highly connected social units, leading to rapid local saturation. In contrast, uniform initialization distributes infections across more social units, creating broader initial seeding that ultimately drives faster epidemic expansion. 
This fundamental trade-off between infection clustering and dispersion explains why seemingly minor initialization differences compound into substantial prediction errors over time.

These results have important implications for epidemic modeling practice during emerging outbreaks. 
When detailed transmission chain information is unavailable, as is typically the case during rapidly evolving health emergencies, modelers should recognize that uniform distribution of reported cases likely overestimates epidemic growth rates and intermediate outbreak outcomes. We revealed that the magnitude of this bias depends critically on the social mixing of (initial) transmission events \MK{and it is thus of high importance to learn about the mixing patterns present at the time of outbreak events. I}n particular outbreak settings or in the case of NPIs where only a limited number of local networks, e.g., households, meet, simplified initializations might be sufficient and, otherwise, reverse-engineered model adaptations \MK{which avoid random, uniform initializations} might be needed. \MK{A dynamic reexamination of observed outbreak dynamics on a small local scale can then help to retroactively correct for informed initial data.} Future research should investigate initialization strategies that better approximate realistic transmission clustering when only aggregate surveillance data is available.

\MK{With the quantification given for four realistic outbreak settings in restaurant and workplace settings, the established framework is now ready to be transferred to outbreak settings in health facilities. Future extensions to care homes or hospital wards will help to guide critical policy decisions during health emergencies. In addition, our framework builds the base to study the important effects of aerosol transmission with the consideration of ventilation included, thus, providing highly detailed insights for the location-specific mitigation of disease dynamics, always integrating the larger-scale and complex disease dynamics outside these locations.}


\section*{Acknowledgments}
This study was performed as part of the Helmholtz School for Data Science in Life, Earth and Energy (HDS-LEE) and received funding from the Helmholtz Association of German Research Centres. It furthermore has received funding from the Initiative and Networking Fund of the Helmholtz Association (grant agreement number KA1-Co-08, Project LOKI-Pandemics) and from the German Research Foundation (project PanVadere, project number 515675334). The funders had no role in study design, data collection and analysis, decision to publish, or preparation of the manuscript.

\section*{Declaration of competing interest}
The authors declare to not have any conflict of interest.

\section*{Data availability}

The MEmilio code is publicly available on GitHub under \url{https://github.com/SciCompMod/memilio}~\cite{Bicker_MEmilio_v2_0_0}, the actual simulation code, which is to be found on the branch \textit{abmXpanvadere}. Given this, reproduction of our results is directly possible.
The Vadere code is publicly available on GitLab under \url{https://gitlab.lrz.de/vadere/vadere}. The simulation code, scenario files, and results of this work are located on the branch \textit{memilioXvadere}, ensuring reproducibility.

\section*{CRediT authorship contribution statement}

\noindent\textbf{Conceptualization:} Martin Kühn\\
\noindent\textbf{Data Curation:} Sascha Korf, Sophia Wagner\\
\noindent\textbf{Formal Analysis:} Sascha Korf, Sophia Wagner\\
\noindent\textbf{Funding Acquisition:} Martin Kühn, Gerta Köster\\
\noindent\textbf{Investigation:} Sascha Korf, Sophia Wagner, Martin Kühn\\
\noindent\textbf{Methodology:} Sascha Korf, Sophia Wagner, Martin Kühn\\
\noindent\textbf{Project Administration:} Martin Kühn, Gerta Köster\\
\noindent\textbf{Resources:} Martin Kühn, Gerta Köster\\
\noindent\textbf{Software:} Sascha Korf, Sophia Wagner\\
\noindent\textbf{Supervision:} Martin Kühn, Gerta Köster\\
\noindent\textbf{Validation:} All authors \\
\noindent\textbf{Visualization:} Sascha Korf, Sophia Wagner\\
\noindent\textbf{Writing – Original Draft:} Sascha Korf, Sophia Wagner, Martin Kühn\\
\noindent\textbf{Writing – Review \& Editing:} All authors

\bibliographystyle{elsarticle-num-short-korf}

{

\FloatBarrier
\clearpage
\pagenumbering{roman}
\appendix

\setcounter{figure}{0}
\setcounter{table}{0}

\begin{center}
    \Large{Appendix to Korf et al., On the Effect of Missing Transmission Chain Information in Agent-Based Models: Outcomes of Superspreading Events and Workplace Transmission (2025).}\\
\end{center}

\section{Parameter values}\label{sec:Parameters}

In this section, we provide the parameter values as used for the Vadere outbreak simulation in~\cref{tab:parameters_vadere} as well as for the MEmilio-ABM in~\cref{tab:parameters_mem} and~\cref{fig:viral_load_parameters}.

\begin{table}[!h]
\caption{\textbf{Vadere AirTransmissionModel Parameters.}}
    \begin{adjustbox}{width=\textwidth}
        \begin{tabular}{l l l l}
            \toprule
            \textbf{Variable}                & \textbf{Description}                                                              & \textbf{Value(s) or Distribution} & \textbf{Source and explanation}             \\
            \midrule
            pedestrianRespiratoryCyclePeriod & Duration of one breathing cycle in seconds                                        & \MK{$3.0$}                             & Default value in~\cite{rahn_modelling_2022} \\
            initialPathogenLoad              & Number of pathogens released during exhalation                                    & $10^4$                            & Default value in~\cite{rahn_modelling_2022} \\
            halfLife                         & Time it takes for the airborne pathogen concentration to decay by half in seconds & $600$                             & Default value in~\cite{rahn_modelling_2022} \\
            initialRadius                    & Radius of the aerosol cloud at exhalation in meters                               & $1.5$                             & Default value in~\cite{rahn_modelling_2022} \\
            airDispersionFactor              & Pathogen dispersion factor caused by ambient air flow                             & $0.0$                             & Default value in~\cite{rahn_modelling_2022} \\
            pedestrianDispersionWeight       & Weighting factor for additional pathogen dispersion caused by pedestrian movement & $0.0125$                          & Default value in~\cite{rahn_modelling_2022} \\
            absorptionRate                   & Pathogen absorption rate per inhalation                                           & $5.0 \times 10^{-4}$              & Default value in~\cite{rahn_modelling_2022} \\
            \bottomrule
        \end{tabular}
    \end{adjustbox}
    \label{tab:parameters_vadere}
\end{table}

\begin{table}[!h]
    \caption{\textbf{MEmilio-ABM Parameters.}
    For the first four parameters of the second section, values are given for the different age groups in ascending order. We denote LogNormal distributions with dependency on ({mean, std}) as parameters. Contact Scaling of the baseline contact matrices are provided for the five location types: Home, School, Work, SocialEvent, and BasicsShop.}
    \begin{adjustbox}{width=\textwidth}
        \begin{tabular}{l l l l}
            \toprule
            \textbf{Variable}                   & \textbf{Description}                                                        & \textbf{Value(s) or Distribution}             & \textbf{Source and explanation}                         \\
            \midrule
            $v_{\textrm P}$                     & Viral load peak                                                             & $8.1$                                         & Motivated by~\cite{jones_estimating_2021}               \\
            $v_{\textrm I}$                     & Viral load incline                                                          & $2$                                           & Motivated by~\cite{jones_estimating_2021}               \\
            $v_{\textrm D}$                     & Viral load decline                                                          & $-0.17$                                       & Motivated by~\cite{jones_estimating_2021}               \\
            $\alpha$                            & Viral shed parameter                                                        & $-7$                                          & Motivated by~\cite{jones_estimating_2021}               \\
            $\beta$                             & Viral shed parameter                                                        & $1$                                           & Motivated by~\cite{jones_estimating_2021}               \\
            $s_{\textrm f}$                     & Viral shed factor                                                           & Gamma($1.6, 1/22$)                            & Motivated by~\cite[Fig. 3]{ke_daily_2022}               \\
            $\lambda$                           & Infection rate from viral shed                                              & $22.6$                                        & Motivated by Workplace Scenario~W1                      \\
            $s$                           & Contact Matrix Scaling                                              & $\{21.0, 4.8, 1.5, 14.4,3.16\}$                                        & Motivated by \cite{noauthor_zeitverwendungserhebung_nodate}  and manual assumption                    \\
            \midrule
            $p_{\textrm {Sym}}$                 & Chance to develop symptoms from an infection                                & $\{0.5, 0.55, 0.6, 0.7, 0.83, 0.9\}$          & ~\cite[Tab. 2]{kerr_covasim_2021}                       \\
            $p_{\textrm {Sev}}$                 & Chance to develop severe symptoms from a symptomatic infection              & $\{0.02, 0.03, 0.04, 0.07, 0.17, 0.24\}$      & ~\cite{nyberg_risk_2021}                                \\
            $p_{\textrm C}$                     & Chance to develop critical symptoms from a severe infection                 & $\{0.1, 0.11, 0.12, 0.14, 0.33, 0.62\}$       & ~\cite[Tab. 2]{zali_mortality_2022}                     \\
            $p_{\textrm D}$                     & Chance to die from a critical infection                                     & $\{0.12, 0.13, 0.15, 0.26, 0.4, 0.48\}$       & ~\cite[Tab. 2]{zali_mortality_2022}                     \\
            $t_{\textrm E}^{\textrm N}$         & Time from Exposed to nonsymptomatic                                         & LogNormal(\text{MEAN=}$4.5,\text{STD=} 1.5$)  & ~\cite[Tab. 1]{kerr_covasim_2021} and references within \\
            $t_{\textrm N}^{\textrm {Sym}}$     & Time to develop symptoms after infection in case of a symptomatic infection & LogNormal(\text{MEAN=}$1.1,\text{STD=} 0.9$)  & ~\cite[Tab. 1]{kerr_covasim_2021} and references within \\
            $t_{\textrm N}^{\textrm R}$         & Time to recover in case of an asymptomatic infection                        & LogNormal(\text{MEAN=}$8.0, \text{STD=}2.0$)  & ~\cite[Tab. 1]{kerr_covasim_2021} and references within \\
            $t_{\textrm {Sym}}^{\textrm {Sev}}$ & Time to develop severe symptoms in case of a severe infection               & LogNormal(\text{MEAN=}$6.6,\text{STD=} 4.9$)  & ~\cite[Tab. 1]{kerr_covasim_2021} and references within \\
            $t_{\textrm {Sym}}^{\textrm R}$     & Time to recover in case of a symptomatic infection                          & LogNormal(\text{MEAN=}$8.0,\text{STD=} 2.0$)  & ~\cite[Tab. 1]{kerr_covasim_2021} and references within \\
            $t_{\textrm {Sev}}^{\textrm C}$     & Time to develop critical symptoms in case of a critical infection           & LogNormal(\text{MEAN=}$1.5,\text{STD=} 2.0$)  & ~\cite[Tab. 1]{kerr_covasim_2021} and references within \\
            $t_{\textrm {Sev}}^{\textrm R}$     & Time to recover in case of a severe infection                               & LogNormal(\text{MEAN=}$18.1, \text{STD=}6.3$) & ~\cite[Tab. 1]{kerr_covasim_2021} and references within \\
            $t_{\textrm C}^{\textrm D}$         & Time to die in case of death                                                & LogNormal(\text{MEAN=}$10.7, \text{STD=}4.8$) & ~\cite[Tab. 1]{kerr_covasim_2021} and references within \\
            $t_{\textrm C}^{\textrm R}$         & Time to recover in case of a critical infection                             & LogNormal(\text{MEAN=}$18.1, \text{STD=}6.3$) & ~\cite[Tab. 1]{kerr_covasim_2021} and references within \\
            \midrule
            $q_{\textrm d}$                     & Quarantine Length                                                           & 10 days                            & Manual assumption                                       \\
            $q_{\textrm e}$                     & Quarantine efficiency                                                       & $0.0$                                         & Manual assumption                                       \\
            \bottomrule
        \end{tabular}
    \end{adjustbox}
    \label{tab:parameters_mem}
\end{table}

\FloatBarrier

\begin{figure}[htbp]
    \centering
    \includegraphics[width=\textwidth]{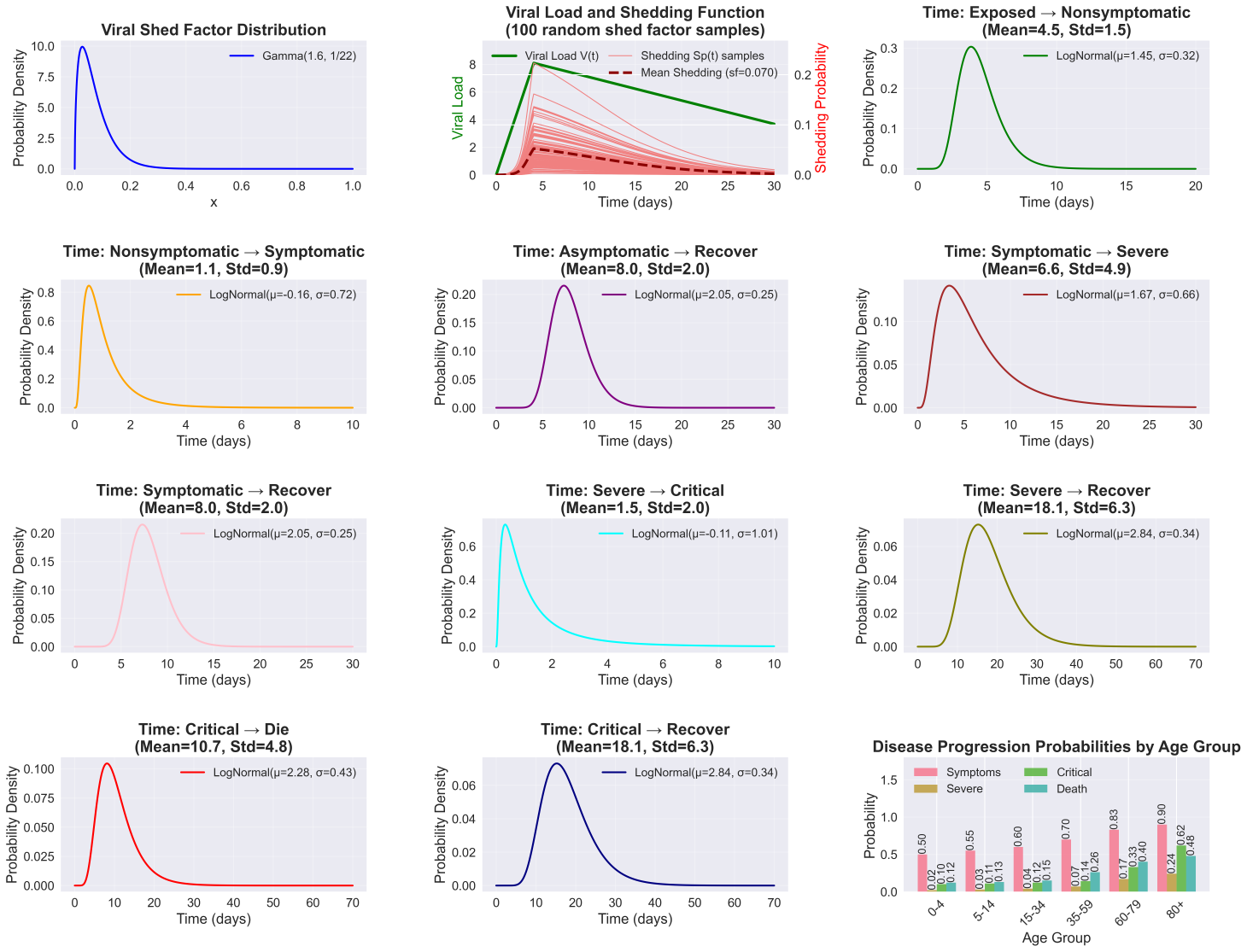}
    \caption{\textbf{Parameter Distributions and Functions for the Viral Load Epidemiological Model.} From left to right, top to bottom: Viral shed factor distribution following Gamma(1.6, 1/22).  Viral load dynamics V(t) over time (green) with 100 individual shedding probability curves Sp(t) (light red) sampled from the shed factor distribution, showing population variability; dashed line represents mean shedding behavior.  Time-to-event distributions for disease progression stages, showing both actual distribution parameters (mean, std) and converted log-normal parameters $(\mu, \sigma)$ used in the model. Age-stratified disease progression probabilities across six age groups (0-4, 5-14, 15-34, 35-59, 60-79, 80+ years) for developing symptoms, severe disease, critical illness, and death.}
    \label{fig:viral_load_parameters}
\end{figure}

\FloatBarrier

\section{Scenario description}\label{app:scen_overview}

\subsection{Restaurant scenarios}

\begin{figure}[htbp]
    \centering
    \includegraphics[width=0.35\textwidth]{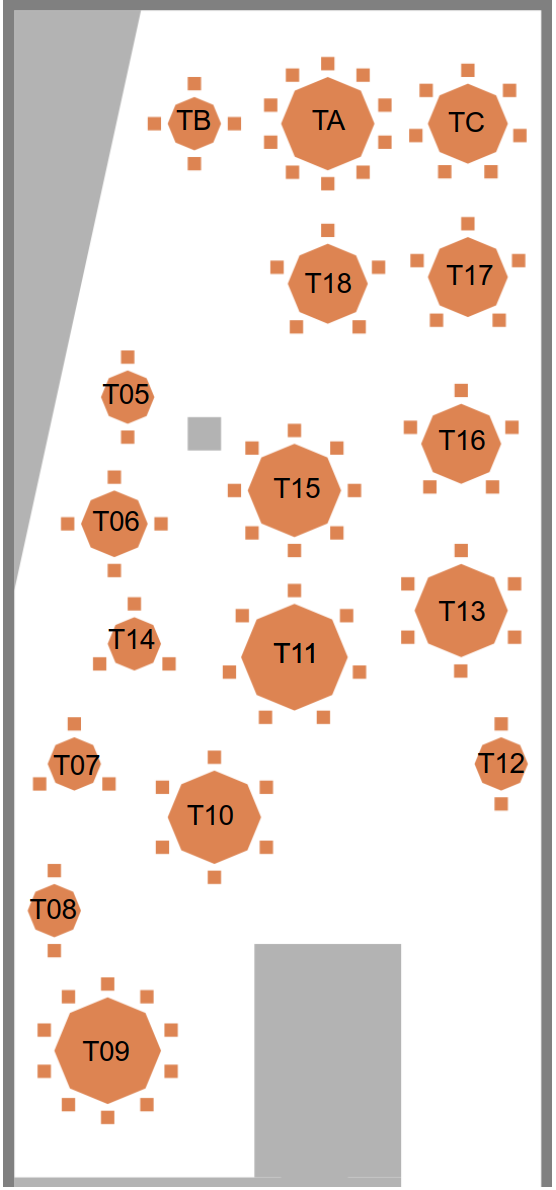}
    \caption{\textbf{Restaurant Scenario Layout with Table Names.} }
    \label{fig:restaurant_seating}
\end{figure}

The restaurant scenarios are based on the COVID-19 outbreak reported in~\cite{lu_restaurant_2020, li_restaurant_2021}. From this setting, we created two scenarios by defining two different household distributions with correspondingly selected individuals from our city setup that are mapped to the table and chair positions as depicted in~\cref{fig:restaurant_seating}. For more details on how the scenarios are defined, see Section 2.4. 

In the setting (independently of the scenario), 88 agents are susceptible and 1 agent is infectious. Our simulation results in 8 new infections, compared to 9 in the original outbreak.

The dimensions of the restaurant are based on~\cite{lu_restaurant_2020} ($8.3m \times 17.5m$), while table layout and naming conventions follow~\cite{li_restaurant_2021} and are shown in Figure~\ref{fig:restaurant_seating}, with the infectious agent seated on table TA.
The time schedule is also adapted from~\cite{li_restaurant_2021}. The simulation starts at 12:00 PM, shortly before the infectious agent arrives. Earlier times are omitted, as the infectious agent is not yet present. The simulation ends at 2:00 PM, while the infectious agent, sitting on table TA, leaves at 1:23 PM.

\begin{itemize}[noitemsep, topsep=0pt, label=\tiny\textbullet]
    \item TA comes in 1 min after start, stays for 82 mins.
    \item TB comes at start, stays for 54 mins.
    \item TC comes in 3 mins after start, stays for 75 mins.
    \item T05 comes at start, stays for 53 minutes.
    \item T06 comes at start, stays for 83 minutes.
    \item T07 comes at start, stays for 70 minutes.
    \item T08 comes in 28 minutes after start, stays for  69 minutes.
    \item T09 comes at start, stays for 76 minutes.
    \item T10 comes at start, stays for 88 minutes.
    \item T11 comes at start, stays for 71 minutes.
    \item T12 comes in 13 minutes after start, stays for 64 minutes.
    \item T13 comes in at start, stays for 51 minutes.
    \item T14 comes in at start, stays for 62 minutes.
    \item T15 comes in at start, stays for 90 minutes.
    \item T16 comes in at start, stays for 49 minutes.
    \item T17 comes in 60 minutes after start, stays until the end.
    \item T18 comes in at start, stays for 78 minutes.
\end{itemize}

The original outbreak scenario included airflow. For simplicity, we neglect airflow here, as our goal is not model validation but rather a realistic scenario for analysis.

\subsection{Workplace scenarios}

We consider two different workplace scenarios, one with few meetings (W1) and one with more meetings (W2). Both scenarios are custom-designed. In Scenario~W2, there is one initial meeting in which all agents participate. During all subsequent meetings in both W1 and W2, the agents maintain physical distance from one another.

The spatial layout of the workplace, including the corresponding room numbers, workplace assignments, and conference seating positions, is illustrated in Figure \ref{fig:office_seating}. The exact schedule of the meetings is as follows:

\begin{figure}[htbp]
    \centering
    \includegraphics[width=\textwidth]{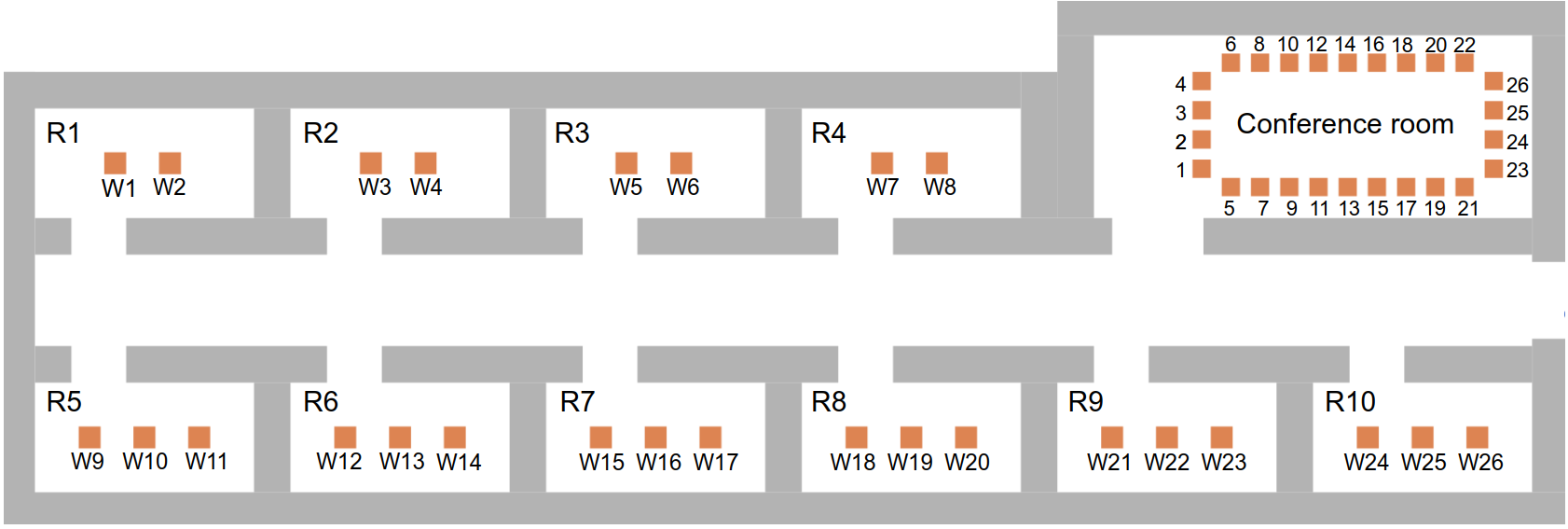}
    \caption{\textbf{Layout of the Workplace Scenario with Room Numbers, Workplace Assignments, and Conference Seating Positions.}}
    \label{fig:office_seating}
\end{figure}

\textbf{Scenario~W1 (few meetings)}

\begin{itemize}[noitemsep, topsep=0pt, label=\tiny\textbullet]
    \item Person 1 (room R1) goes to workplace W1 at start, goes to conference seat C2 at 120 min, goes to workplace W1 at 180 min.
    \item Person 2 (room R1) goes to workplace W2 at start, goes to conference seat C8 at 120 min, goes to workplace W2 at 180 min.
    \item Person 3 (room R2) goes to workplace W3 at start, goes to conference seat C2 at 180 min, goes to workplace W3 at 240 min.
    \item Person 4 (room R2) goes to workplace W4 at start, goes to conference seat C8 at 180 min, goes to workplace W4 at 240 min.
    \item Person 5 (room R3) goes to workplace W5 at start.
    \item Person 6 (room R3) goes to workplace W6 at start.
    \item Person 7 (room R4) goes to workplace W7 at start.
    \item Person 8 (room R4) goes to workplace W8 at start.
    \item Person 9 (room R5) goes to workplace W9 at start, goes to conference seat C9 at 120 min, goes to workplace W9 at 180 min.
    \item Person 10 (room R5) goes to workplace W10 at start, goes to conference seat C12 at 120 min, goes to workplace W10 at 180 min.
    \item Person 11 (room R5) goes to workplace W11 at start, goes to conference seat C13 at 120 min, goes to workplace W11 at 180 min.
    \item Person 12 (room R6) goes to workplace W12 at start, goes to conference seat C9 at 180 min, goes to workplace W12 at 240 min.
    \item Person 13 (room R6) goes to workplace W13 at start, goes to conference seat C12 at 180 min, goes to workplace W13 at 240 min.
    \item Person 14 (room R6) goes to workplace W14 at start, goes to conference seat C13 at 180 min, goes to workplace W14 at 240 min.
    \item Person 15 (room R7) goes to workplace W15 at start.
    \item Person 16 (room R7) goes to workplace W16 at start.
    \item Person 17 (room R7) goes to workplace W17 at start.
    \item Person 18 (room R8) goes to workplace W18 at start.
    \item Person 19 (room R8) goes to workplace W19 at start.
    \item Person 20 (room R8) goes to workplace W20 at start.
    \item Person 21 (room R9) goes to workplace W21 at start, goes to conference seat C2 at 240 min, goes to workplace W21 at 300 min.
    \item Person 22 (room R9) goes to workplace W22 at start, goes to conference seat C8 at 240 min, goes to workplace W22 at 300 min.
    \item Person 23 (room R9) goes to workplace W23 at start, goes to conference seat C9 at 240 min, goes to workplace W23 at 300 min.
    \item Person 24 (room R10) goes to workplace W24 at start, goes to conference seat C12 at 240 min, goes to workplace W24 at 300 min.
    \item Person 25 (room R10) goes to workplace W25 at start, goes to conference seat C13 at 240 min, goes to workplace W25 at 300 min.
    \item Person 26 (room R10) goes to workplace W26 at start, goes to conference seat C16 at 240 min, goes to workplace W26 at 300 min.
\end{itemize}

\textbf{Scenario~W2 (more meetings)}

\begin{itemize}[noitemsep, topsep=0pt, label=\tiny\textbullet]
    \item Person 1 (room R1) goes to conference seat C1 at 0 min, returns to workplace W1 at 60 min, goes to conference seat C2 at 240 min, and returns to workplace W1 at 300 min.
    \item Person 2 (room R1) goes to conference seat C2 at 0 min, returns to workplace W2 at 60 min, goes to conference seat C8 at 240 min, and returns to workplace W2 at 300 min.
    \item Person 3 (room R2) goes to conference seat C3 at 0 min, returns to workplace W3 at 60 min, goes to conference seat C9 at 240 min, and returns to workplace W3 at 300 min.
    \item Person 4 (room R2) goes to conference seat C4 at 0 min, returns to workplace W4 at 60 min, goes to conference seat C12 at 240 min, and returns to workplace W4 at 300 min.
    \item Person 5 (room R3) goes to conference seat C5 at 0 min, then to conference seat C2 at 60 min, and returns to workplace W5 at 120 min.
    \item Person 6 (room R3) goes to conference seat C6 at 0 min, then to conference seat C8 at 60 min, and returns to workplace W6 at 120 min.
    \item Person 7 (room R4) goes to conference seat C7 at 0 min, then to conference seat C9 at 60 min, and returns to workplace W7 at 120 min.
    \item Person 8 (room R4) goes to conference seat C8 at 0 min, then to conference seat C12 at 60 min, and returns to workplace W8 at 120 min.
    \item Person 9 (room R5) goes to conference seat C9 at 0 min, returns to workplace W9 at 60 min, goes to conference seat C13 at 240 min, and returns to workplace W9 at 300 min.
    \item Person 10 (room R5) goes to conference seat C10 at 0 min, returns to workplace W10 at 60 min, goes to conference seat C16 at 240 min, and returns to workplace W10 at 300 min.
    \item Person 11 (room R5) goes to conference seat C11 at 0 min, returns to workplace W11 at 60 min, goes to conference seat C17 at 240 min, and returns to workplace W11 at 300 min.
    \item Person 12 (room R6) goes to conference seat C12 at 0 min, returns to workplace W12 at 60 min, goes to conference seat C20 at 240 min, and returns to workplace W12 at 300 min.
    \item Person 13 (room R6) goes to conference seat C13 at 0 min, returns to workplace W13 at 60 min, goes to conference seat C21 at 240 min, and returns to workplace W13 at 300 min.
    \item Person 14 (room R6) goes to conference seat C14 at 0 min, returns to workplace W14 at 60 min, goes to conference seat C25 at 240 min, and returns to workplace W14 at 300 min.
    \item Person 15 (room R7) goes to conference seat C15 at 0 min, then to conference seat C13 at 60 min, and returns to workplace W15 at 120 min.
    \item Person 16 (room R7) goes to conference seat C16 at 0 min, then to conference seat C16 at 60 min, and returns to workplace W16 at 120 min.
    \item Person 17 (room R7) goes to conference seat C17 at 0 min, then to conference seat C17 at 60 min, and returns to workplace W17 at 120 min.
          -\item Person 18 (room R8) goes to conference seat C18 at 0 min, then to conference seat C20 at 60 min, and returns to workplace W18 at 120 min.
          -\item Person 19 (room R8) goes to conference seat C19 at 0 min, then to conference seat C21 at 60 min, and returns to workplace W19 at 120 min.
          -\item Person 20 (room R8) goes to conference seat C20 at 0 min, then to conference seat C25 at 60 min, and returns to workplace W20 at 120 min.
    \item Person 21 (room R9) goes to conference seat C21 at 0 min, returns to workplace W21 at 60 min, goes to conference seat C2 at 300 min, and returns to workplace W21 at 360 min.
    \item Person 22 (room R9) goes to conference seat C22 at 0 min, returns to workplace W22 at 60 min, goes to conference seat C8 at 300 min, and returns to workplace W22 at 360 min.
    \item Person 23 (room R9) goes to conference seat C23 at 0 min, returns to workplace W23 at 60 min, goes to conference seat C9 at 300 min, and returns to workplace W23 at 360 min.
    \item Person 24 (room R10) goes to conference seat C24 at 0 min, returns to workplace W24 at 60 min, goes to conference seat C12 at 300 min, and returns to workplace W24 at 360 min.
    \item Person 25 (room R10) goes to conference seat C25 at 0 min, returns to workplace W25 at 60 min, goes to conference seat C13 at 300 min, and returns to workplace W25 at 360 min.
    \item Person 26 (room R10) goes to conference seat C26 at 0 min, returns to workplace W26 at 60 min, goes to conference seat C16 at 300 min, and returns to workplace W26 at 360 min.
\end{itemize}

\section{Simulation Results for Different Seeds}
We present the outcomes of infection counts for \SK{50} different seeds, creating \SK{50} different city districts, for all four scenarios.

\begin{figure}[H]
    \centering
    \includegraphics[width=\textwidth]{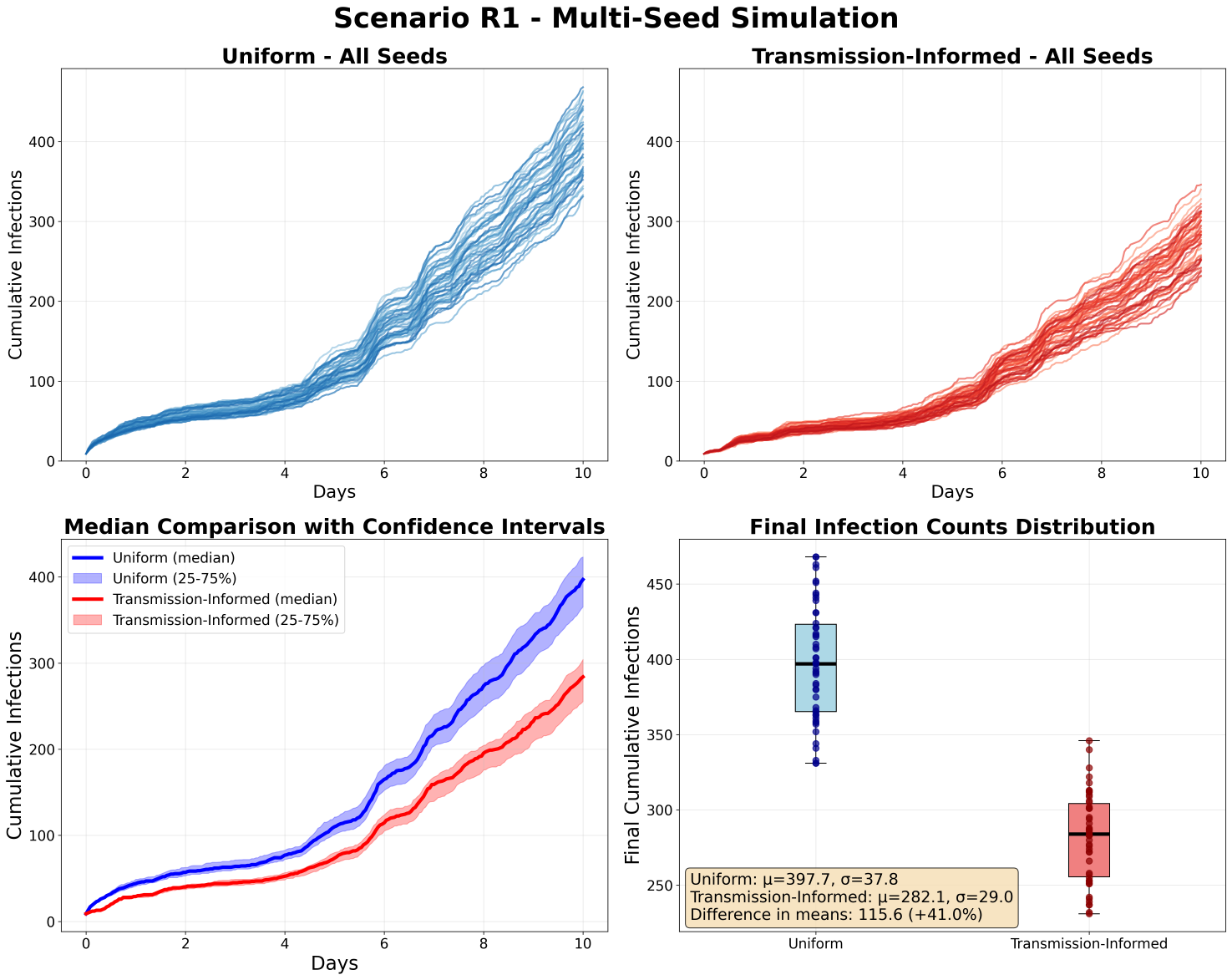}
    \caption{\textbf{Multi-Seed Simulation Comparison for the Scenario~R1 with Uniform (blue) and Transmission-Informed (red) Initializations.} Upper panels display median cumulative infection values of \SK{50} different seed initializations over a 10-day simulation period. Lower left panel presents median trajectories of the \SK{50} seed runs with 25th-75th percentile confidence intervals aggregated across all seed runs, while the lower right panel shows the distribution of final infection counts at day 10.}
    \label{fig:multi_seed_comparison_R1}
\end{figure}

\begin{figure}[H]
    \centering
    \includegraphics[width=\textwidth]{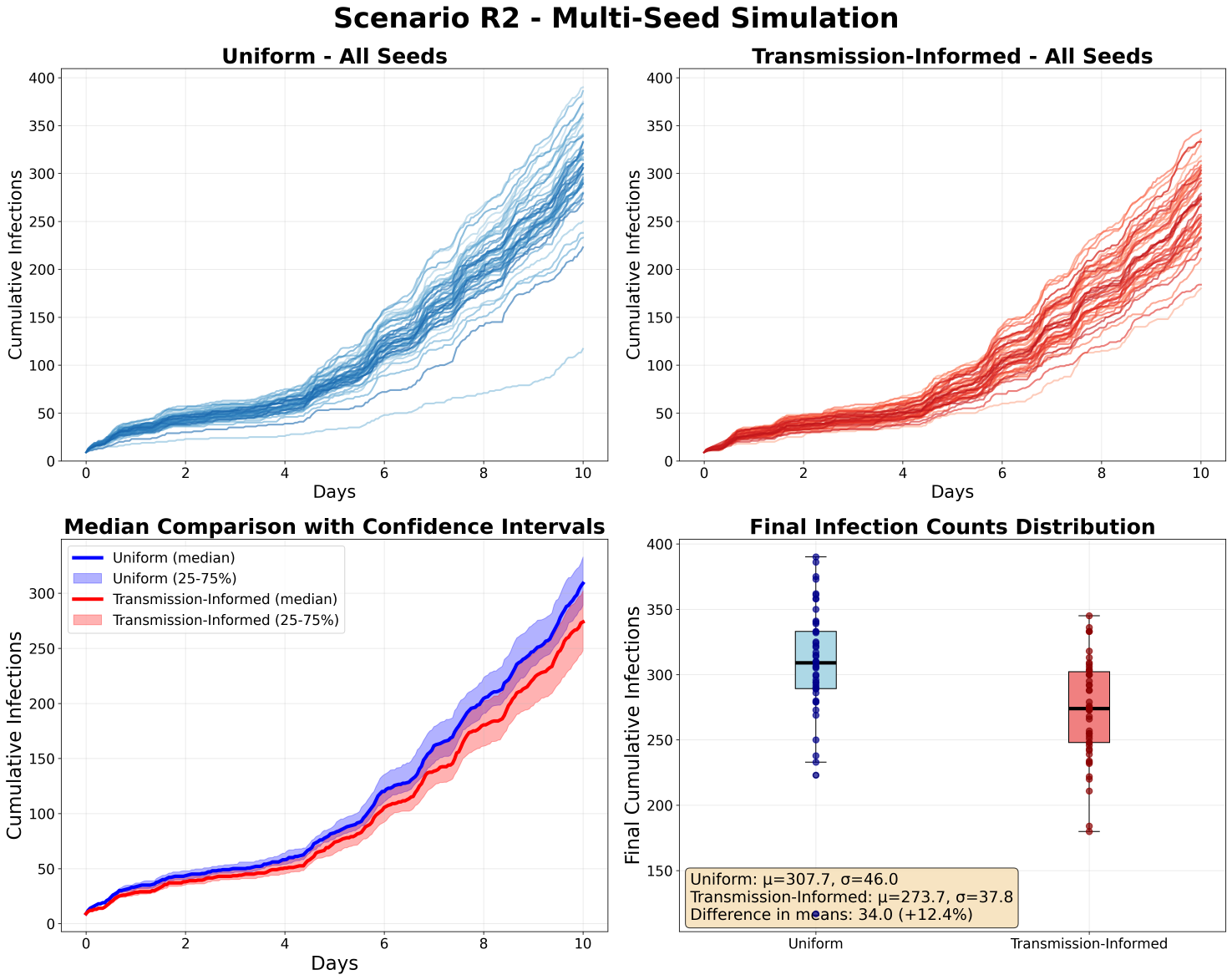}
    \caption{\textbf{Multi-Seed Simulation Comparison for the Scenario~R2 with Uniform (blue) and Transmission-Informed (red) Initializations.} Upper panels display median cumulative infection values of \SK{50} different seed initializations over a 10-day simulation period. Lower left panel presents median trajectories of the \SK{50} seed runs with 25th-75th percentile confidence intervals aggregated across all seed runs, while the lower right panel shows the distribution of final infection counts at day 10.}
    \label{fig:multi_seed_comparison_R2}
\end{figure}

\begin{figure}[H]
    \centering
    \includegraphics[width=\textwidth]{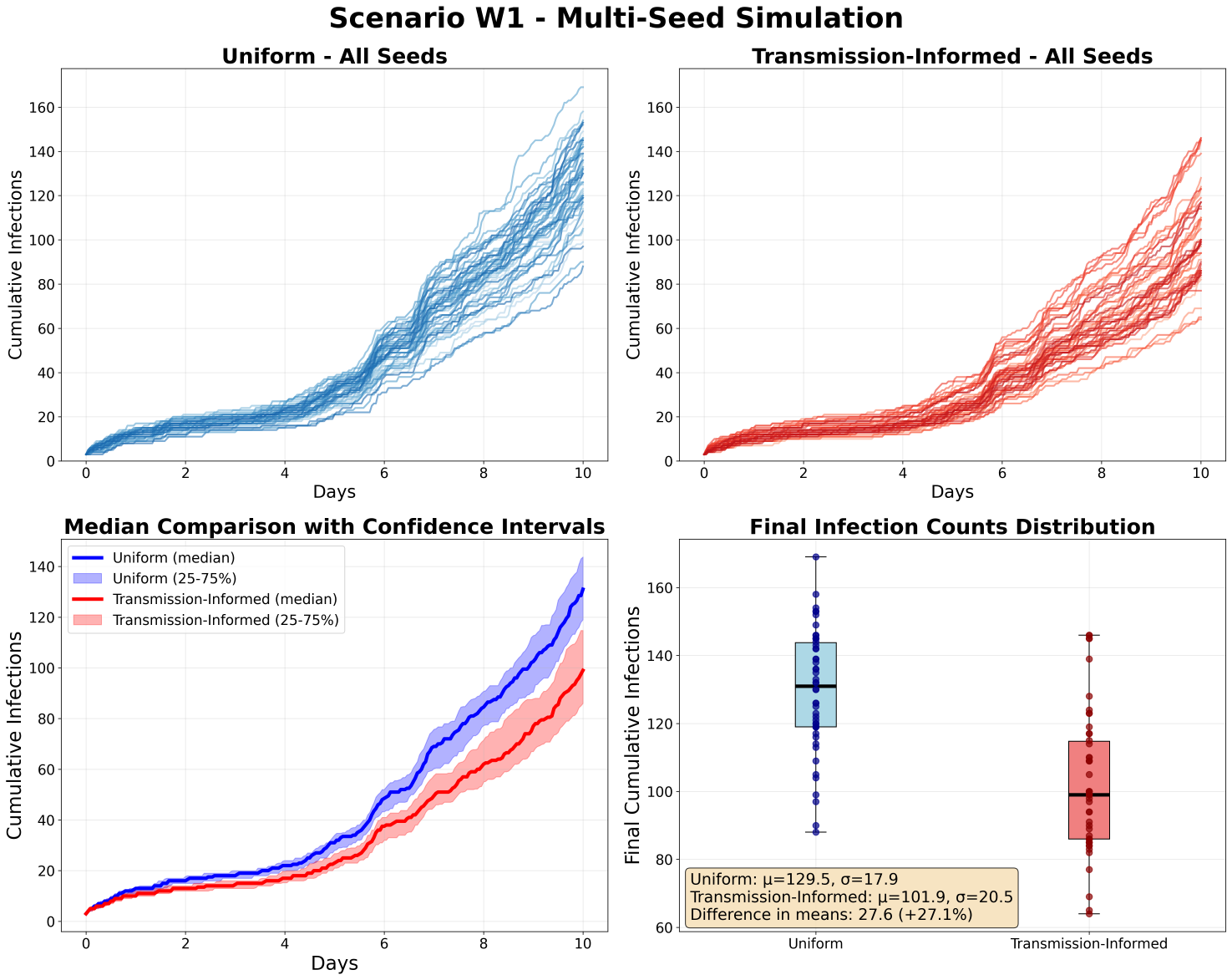}
    \caption{\textbf{Multi-Seed Simulation Comparison for the Scenario~W1 with Uniform (blue) and Transmission-Informed (red) Initializations.} Upper panels display median cumulative infection values of \SK{50} different seed initializations over a 10-day simulation period. Lower left panel presents median trajectories of the \SK{50} seed runs with 25th-75th percentile confidence intervals aggregated across all seed runs, while the lower right panel shows the distribution of final infection counts at day 10.}
    \label{fig:multi_seed_comparison_W1}
\end{figure}

\begin{figure}[H]
    \centering
    \includegraphics[width=\textwidth]{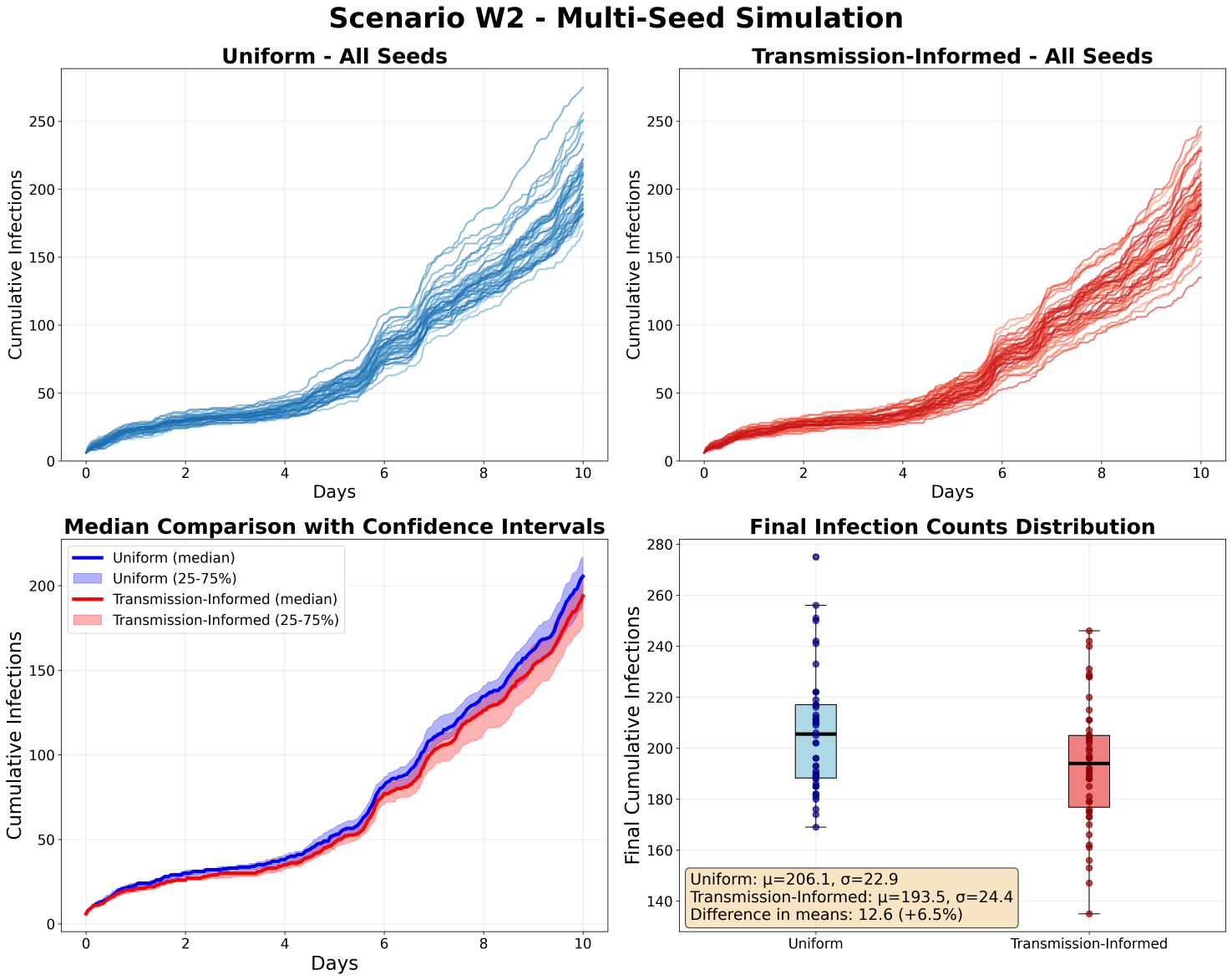}
    \caption{\textbf{Multi-Seed Simulation Comparison for the Scenario~W2 with Uniform (blue) and Transmission-Informed (red) Initializations.} Upper panels display median cumulative infection values of \SK{50} different seed initializations over a 10-day simulation period. Lower left panel presents median trajectories of the \SK{50} seed runs with 25th-75th percentile confidence intervals aggregated across all seed runs, while the lower right panel shows the distribution of final infection counts at day 10.}
    \label{fig:multi_seed_comparison_W2}
\end{figure}

\section{Infection by Location Type}
\begin{figure}[H]
    \centering
    
    \begin{subfigure}[b]{\textwidth}
        \centering
        \begin{subfigure}[b]{0.48\textwidth}
            \centering
            \includegraphics[width=\textwidth]{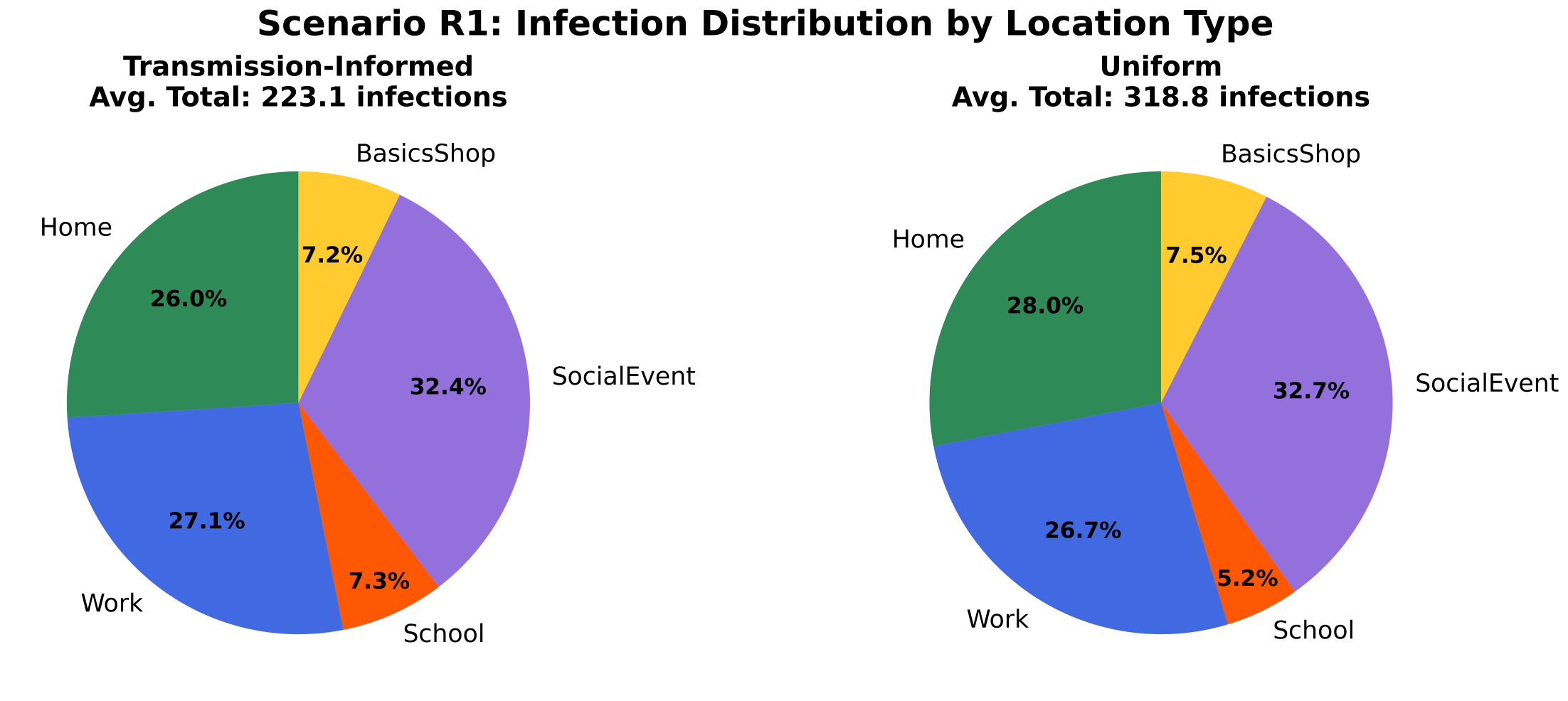}
            \caption{\textbf{R1}}
        \end{subfigure}
        \hfill
        \begin{subfigure}[b]{0.48\textwidth}
            \centering
            \includegraphics[width=\textwidth]{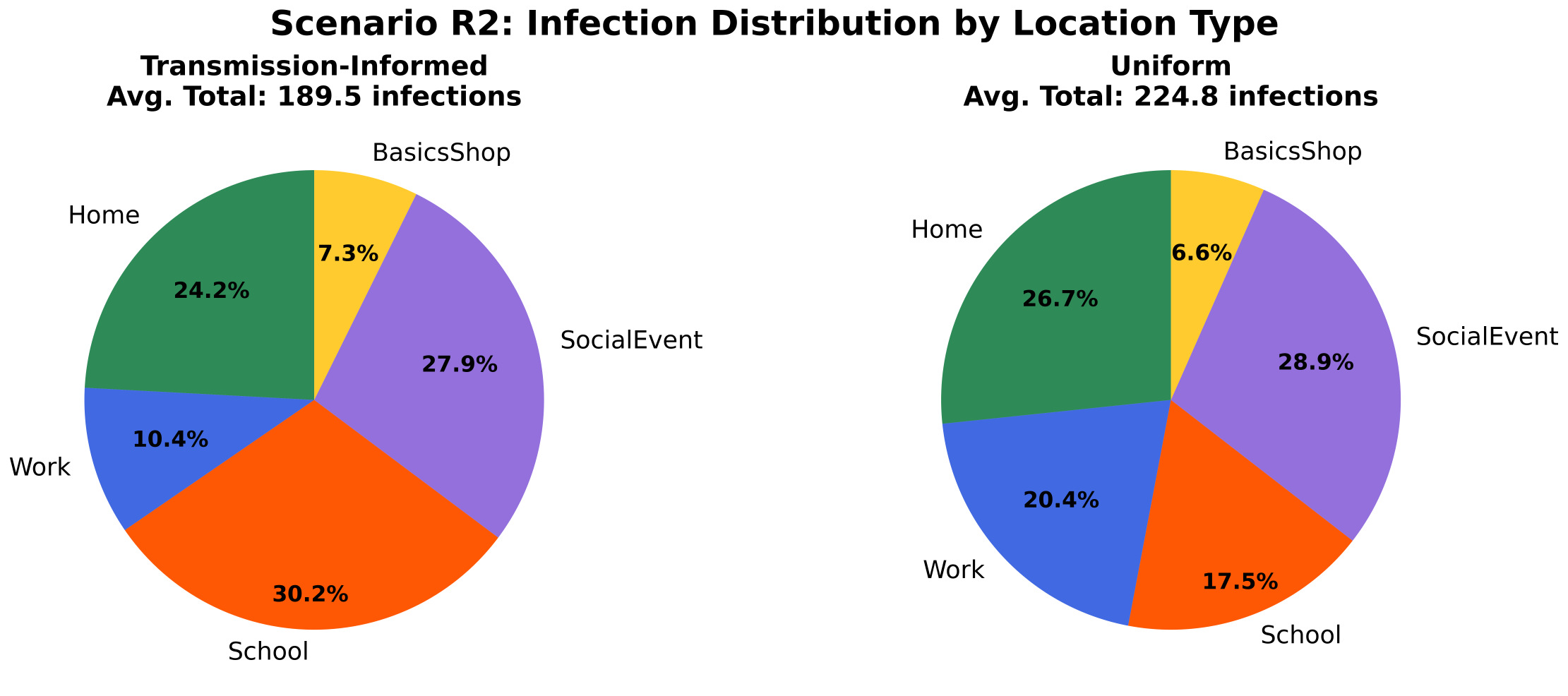}
            \caption{\textbf{R2}}
        \end{subfigure}
        \label{fig:restaurant_scenarios}
    \end{subfigure}
    
    \vspace{1cm}
    
    \begin{subfigure}[b]{\textwidth}
        \centering
        \begin{subfigure}[b]{0.48\textwidth}
            \centering
            \includegraphics[width=\textwidth]{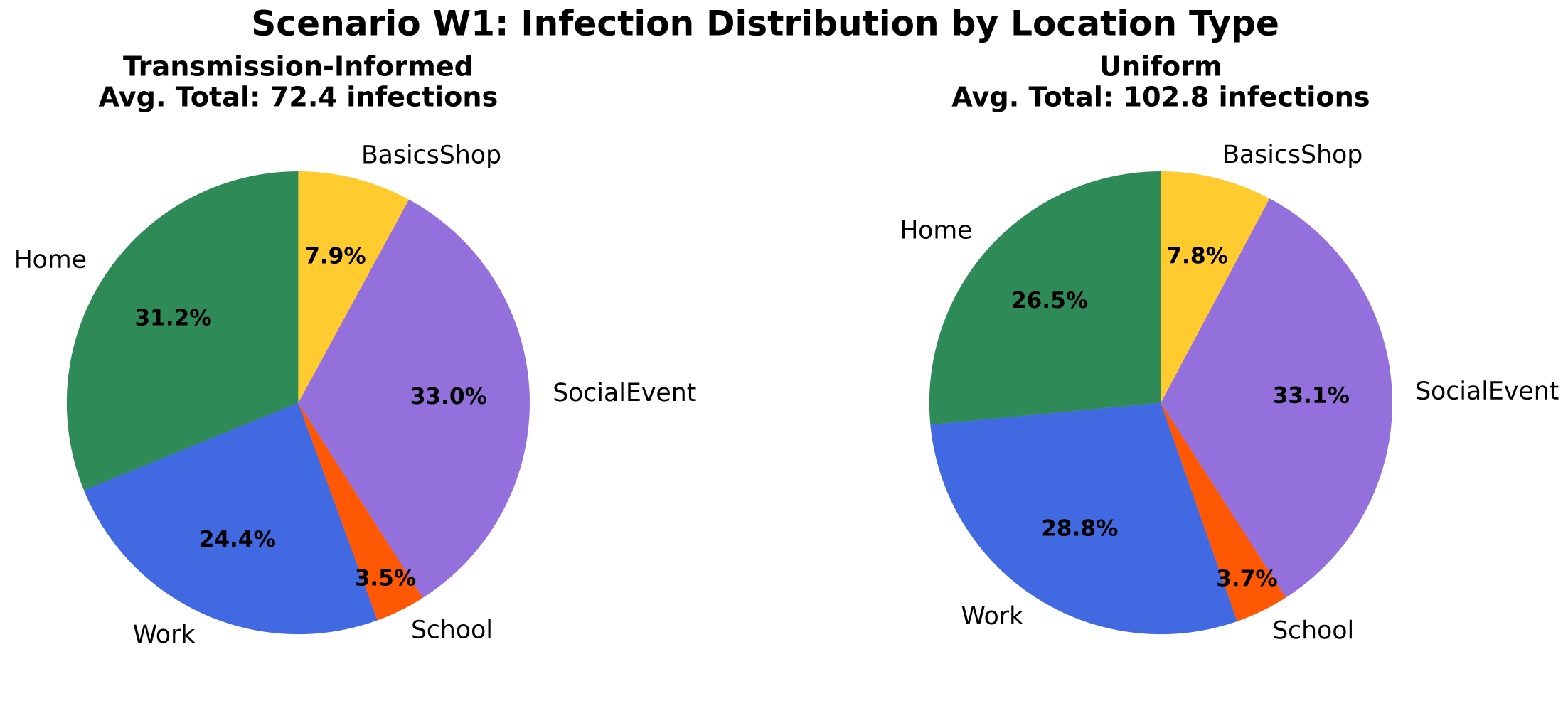}
            \caption{\textbf{W1}}
        \end{subfigure}
        \hfill
        \begin{subfigure}[b]{0.48\textwidth}
            \centering
            \includegraphics[width=\textwidth]{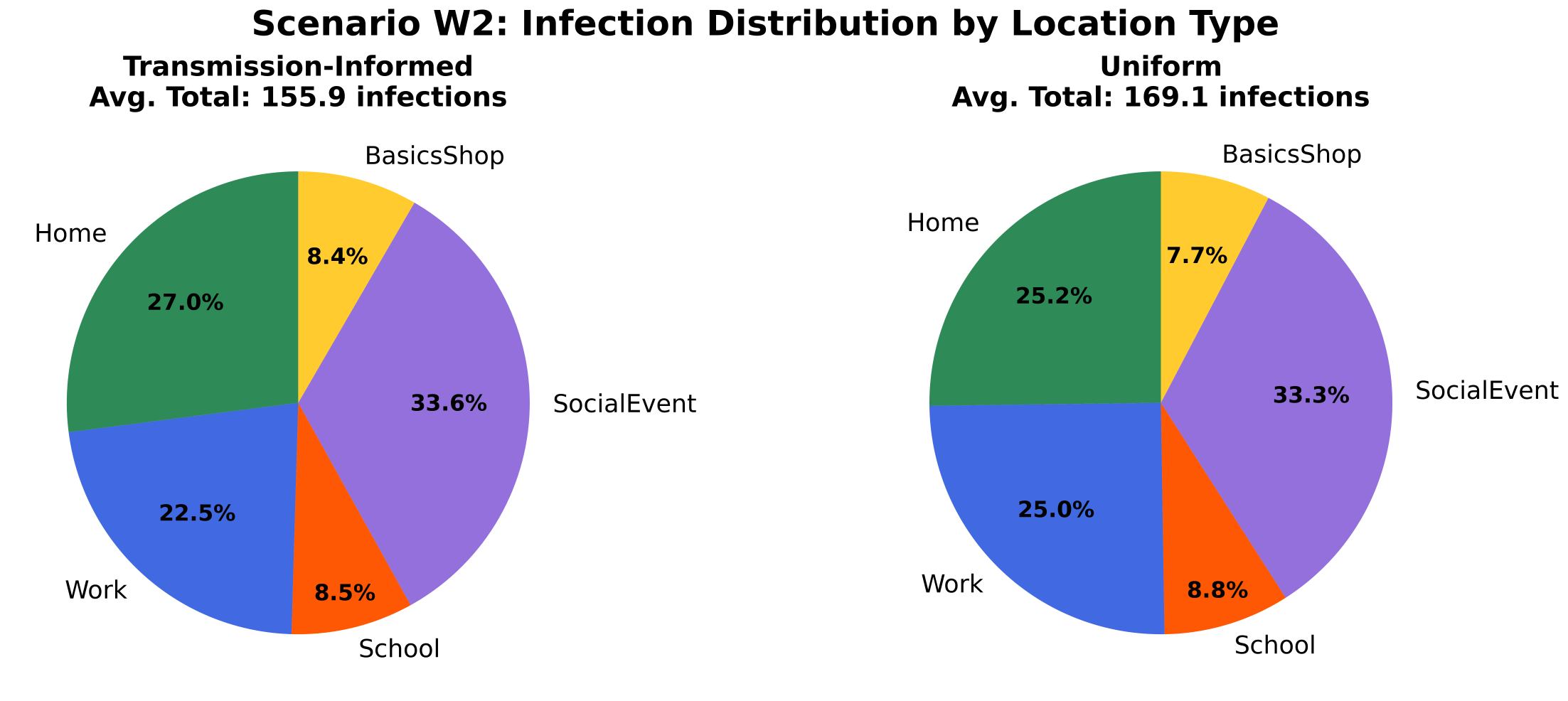}
            \caption{\textbf{W2}}
        \end{subfigure}
        \label{fig:workplace_scenarios}
    \end{subfigure}
    
    \caption{\textbf{Location-based Infection Distribution Analysis across Epidemic Initialization Strategies.} We compare transmission-informed versus uniform initialization across four distinct scenarios: restaurant Scenarios R1 and R2 (top row) and workplace Scenarios W1 and W2 (bottom row). Each pie chart pair shows the average distribution of infections by location type (Home, Work, School, BasicsShop, SocialEvent) across 100 simulation runs for one representative seed.}
    \label{fig:location_infection_analysis}
\end{figure}

\section{Household Sizes of Initially Infected for Different Population Sizes}

\begin{table}[!h]
\caption{\MK{\textbf{Averaged Absolute Household Sizes of the Infected Individuals with Uniform and Transmission-Informed Initializations Applied to the Initial Outbreak.}}}
 \vspace{2mm} 

\begin{adjustwidth}{-1in}{-1in}
\centering
\resizebox{1.3\textwidth}{!}{%
\begin{tabular}{lccccccccc}
\toprule
\footnotesize
\textbf{Scenario}& 1k & 2k& 5k & 7k & 10k & 12k& 15k & 20k & 50k \\
\midrule
R1 &  4.49/4.79&4.49/4.79&4.48/4.79&4.48/4.79&4.47/4.78&4.50/4.79&4.49/4.78&4.49/2.78&4.49/4.78 \\
R2 & 2.79/3.07&2.75/3.05&2.74/3.03&2.78/3.05&2.73/2.95&2.75/3.08&2.79/3.06&2.82/3.09&2.78/3.13 \\
W1 &  2.69/2.63&2.79/2.89&2.72/2.79&2.61/2.71&2.74/\textbf{2.47}&2.64/2.70&2.69/2.77&2.58/2.80&2.80/3.02 \\
W2 & 2.85/2.71&2.89/3.01&2.77/2.86&2.83/2.82&2.69/2.64&2.87/2.84&2.71/2.86&2.66/2.82&3.06/3.15 \\
\bottomrule
\end{tabular}
}
\end{adjustwidth}
\label{tab:hh_size}
\end{table}

\section{Saturation and Infection by Location Type for Reduced $\lambda$}\label{sec:satur_lambda025}

\begin{figure}[H]
    \centering
    
    \begin{subfigure}[b]{\textwidth}
        \centering
        \begin{subfigure}[b]{0.48\textwidth}
            \centering
            \includegraphics[width=\textwidth]{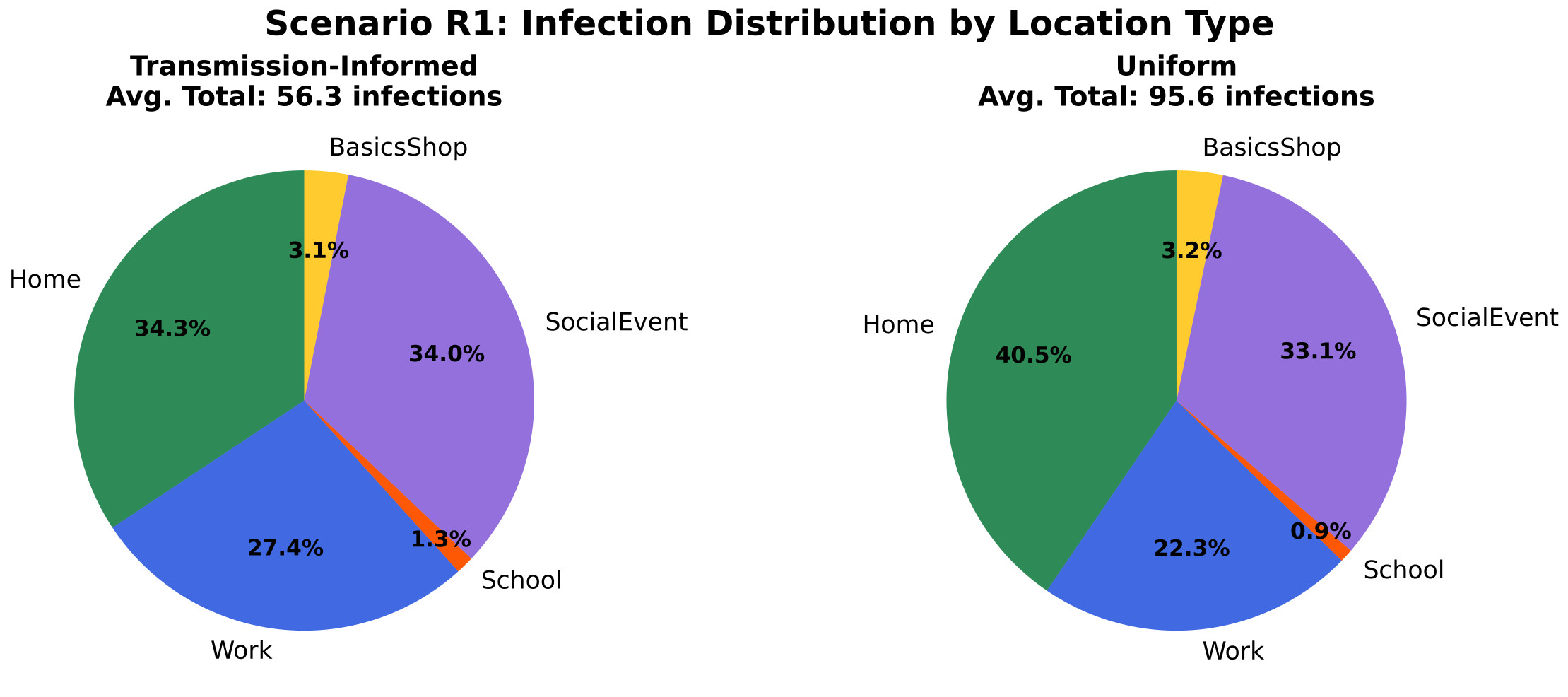}
            \caption{\textbf{R1}}
        \end{subfigure}
        \hfill
        \begin{subfigure}[b]{0.48\textwidth}
            \centering
            \includegraphics[width=\textwidth]{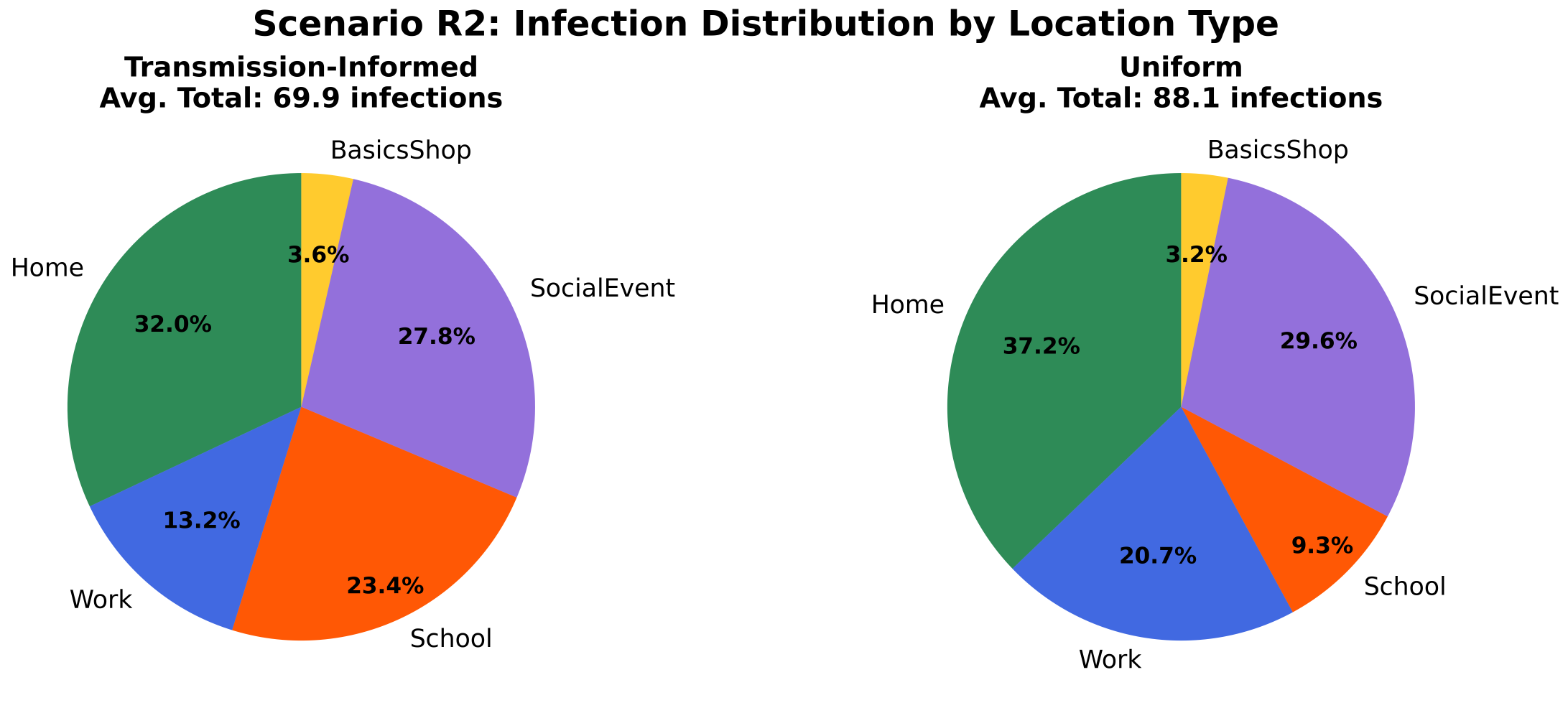}
            \caption{\textbf{R2}}
        \end{subfigure}
        \label{fig:restaurant_scenarios_ql}
    \end{subfigure}
    
    \vspace{1cm}
    
    \begin{subfigure}[b]{\textwidth}
        \centering
        \begin{subfigure}[b]{0.48\textwidth}
            \centering
            \includegraphics[width=\textwidth]{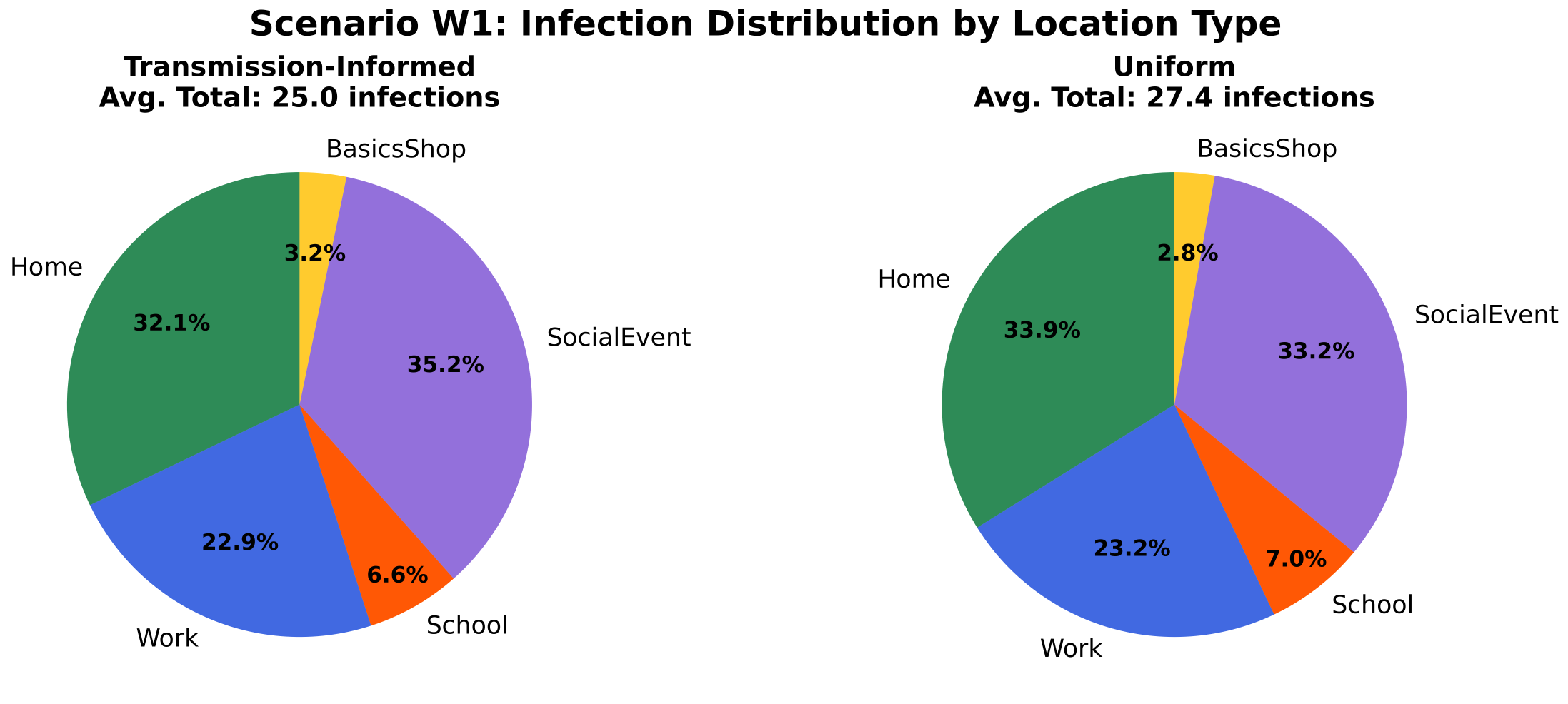}
            \caption{\textbf{W1}}
        \end{subfigure}
        \hfill
        \begin{subfigure}[b]{0.48\textwidth}
            \centering
            \includegraphics[width=\textwidth]{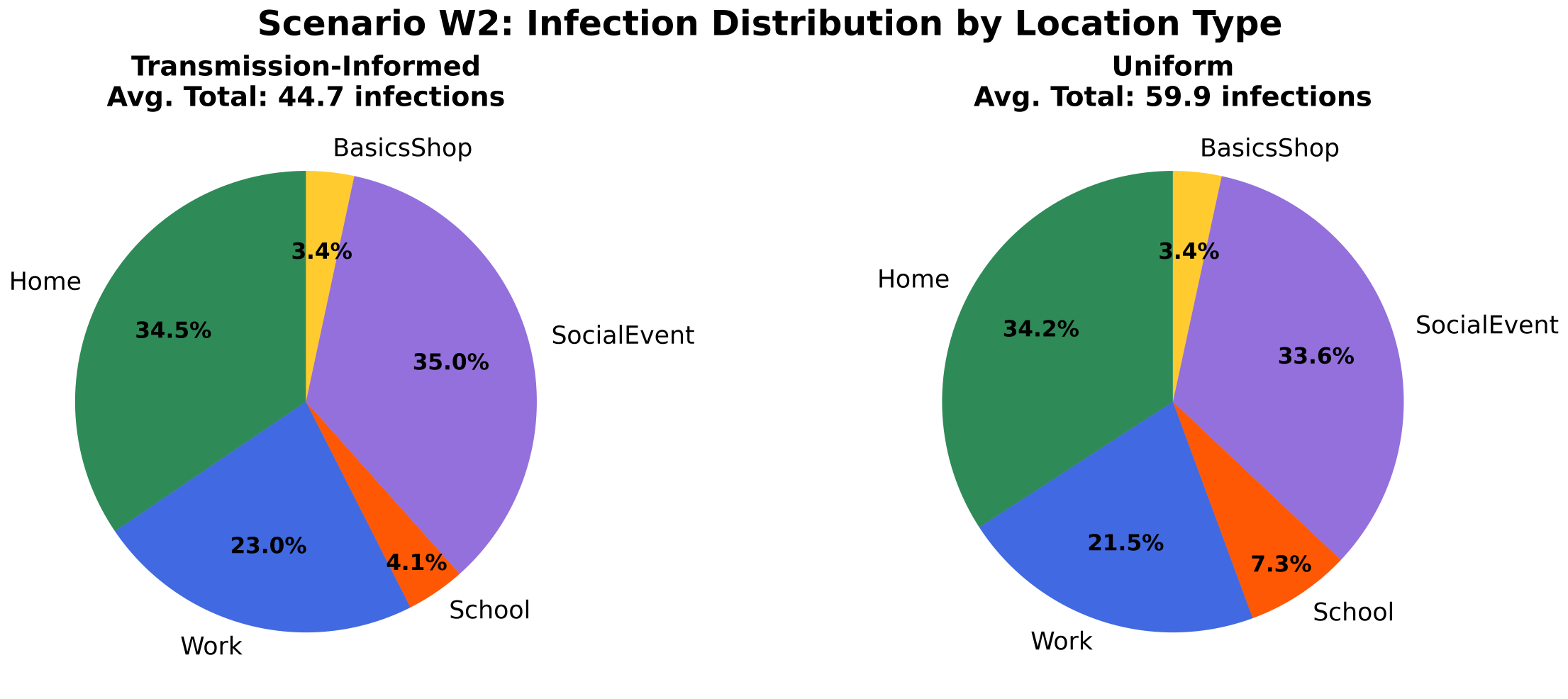}
            \caption{\textbf{W2}}
        \end{subfigure}
        \label{fig:workplace_scenarios_ql}
    \end{subfigure}
    
    \caption{\MK{\textbf{Location-based Infection Distribution Analysis across Epidemic Initialization Strategies with $\frac{1}{4}\lambda$.} We compare transmission-informed versus uniform initialization across four distinct scenarios: restaurant Scenarios R1 and R2 (top row) and workplace Scenarios W1 and W2 (bottom row). Each pie chart pair shows the average distribution of infections by location type (Home, Work, School, BasicsShop, SocialEvent) across 100 simulation runs for one representative seed.}}
    \label{fig:location_infection_analysis_ql}
\end{figure}

\begin{figure}[!h]
    \centering
    \includegraphics[width=\textwidth]{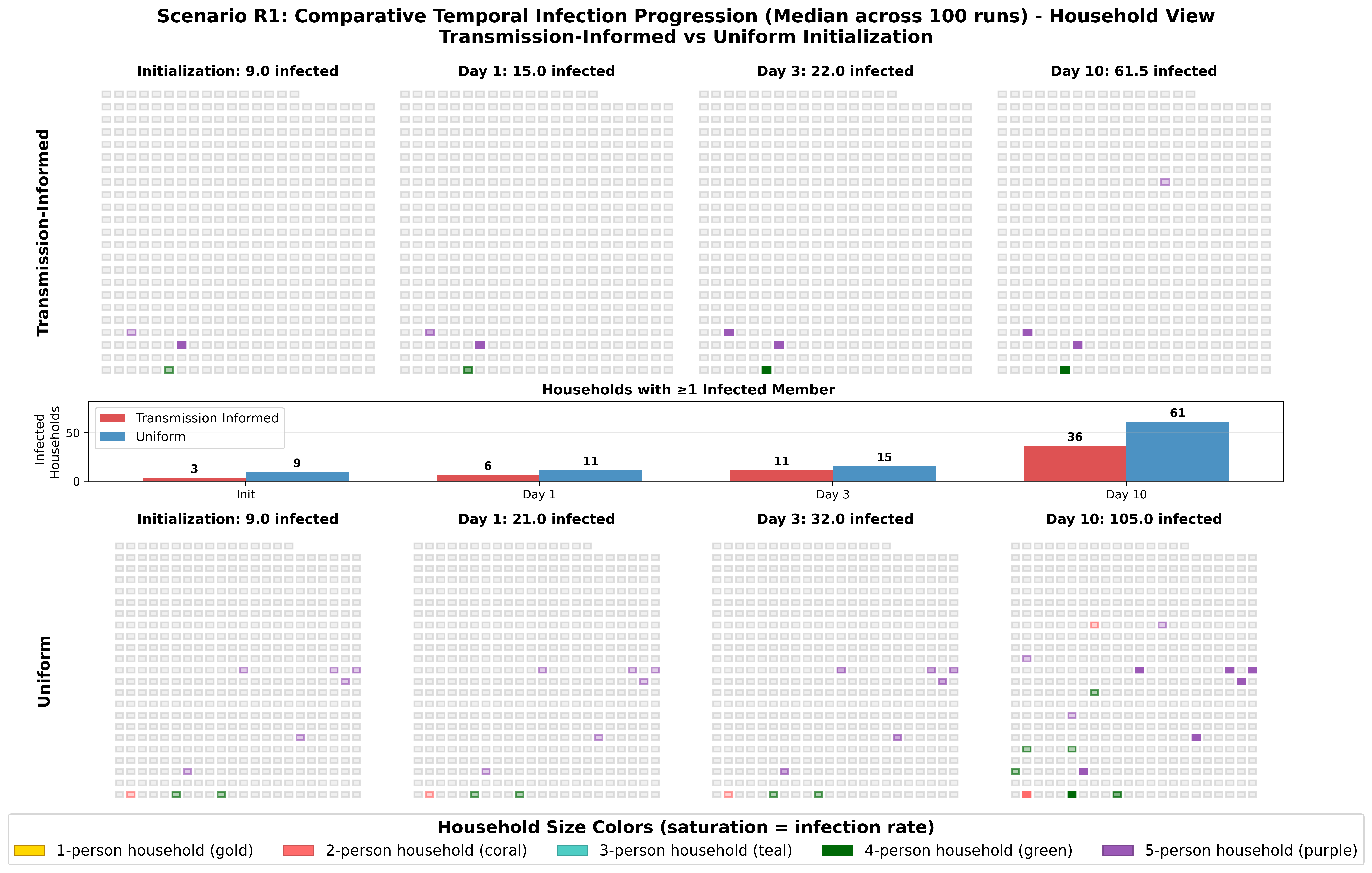}
    \caption{\MK{\textbf{Heatmap of Spatially Distributed Infections for Restaurant Scenario~R1 (Limited mixing) with $\frac{1}{4}\lambda$}: We compare transmission-informed initialization (top row) versus uniform initialization (bottom row) across four time points: initialization, day 1, day 3, and day 10. Rectangular shapes represent households, with color intensity indicating the median number of infected members per household across 100 simulation runs (one color per household size, light shading for a low percentage of infected, dark shading for a high percentage). The center panel shows the median number of affected households over time.}}
    \label{fig:heatmap_R1_ql}
\end{figure}

\begin{figure}[!h]
    \centering
    \includegraphics[width=\textwidth]{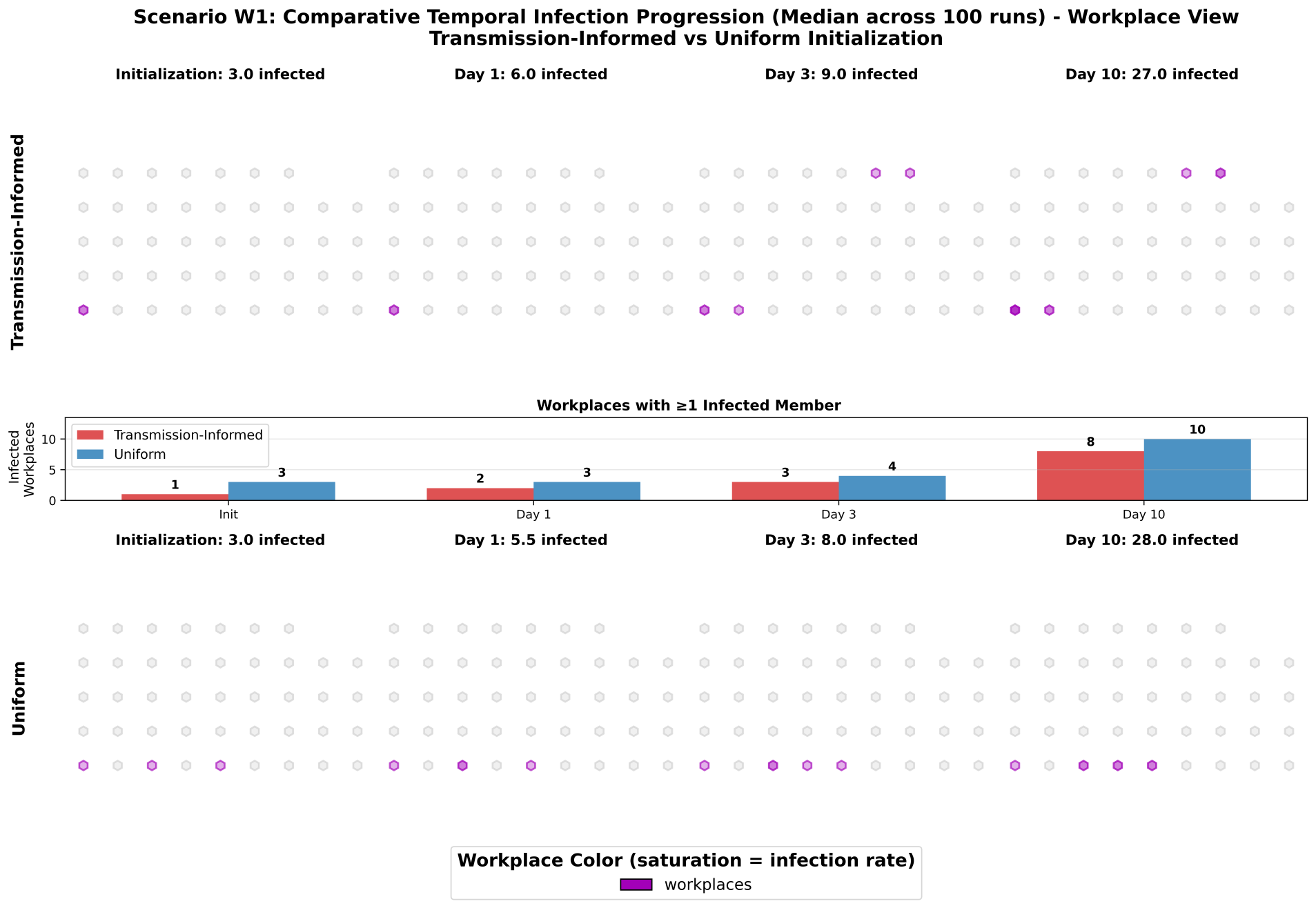}
    \caption{\MK{\textbf{Heatmap of Spatially Distributed Infections for Workplace Scenario~W1 (Limited mixing) with $\frac{1}{4}\lambda$}: We compare transmission-informed initialization (top row) versus uniform initialization (bottom row) across four time points: initialization, day 1, day 3, and day 10. Hexagonal shapes represent workplaces, with color intensity indicating the median number of infected members per workplace across 100 simulation runs (light shading for a low percentage of infected, dark shading for a high percentage). The center panel shows the number of affected workplaces over time.}}
    \label{fig:heatmap_W1_ql}
\end{figure}

}

\end{document}